\documentstyle[preprint,aps,eqsecnum]{revtex}

\begin{document}

\baselineskip 20pt

\newcommand{\beq}{\begin{equation}}
\newcommand{\eeq}{\end{equation}}
\newcommand{\bea}{\begin{eqnarray}}
\newcommand{\eea}{\end{eqnarray}}
\newcommand{\cir}{{\buildrel \circ \over =}}

\title{Generalized Eulerian Coordinates for Relativistic Fluids:
Hamiltonian Rest-Frame Instant Form, Relative Variables,
Rotational Kinematics.}

\author{David Alba}

\address
{Dipartimento di Fisica\\
Universita' di Firenze\\
Via G. Sansone 1\\
50019 Sesto Fiorentino (FI), Italy\\
E-mail: ALBA@FI.INFN.IT}

\author{and}

\author{Luca Lusanna}

\address
{Sezione INFN di Firenze\\
Via G. Sansone 1\\
50019 Sesto Fiorentino (FI), Italy\\
E-mail: LUSANNA@FI.INFN.IT}

\maketitle

\begin{abstract}

We study the rest-frame instant form of a new formulation of
relativistic perfect fluids in terms of new generalized Eulerian
configuration coordinates. After the separation of the
relativistic center of mass from the relative variables on the
Wigner hyper-planes, we define orientational and shape variables
for the fluid, viewed as a relativistic extended deformable body,
by introducing dynamical body frames. Finally we define Dixon's
multipoles for the fluid.

\today

\end{abstract}

\newpage


\section{Introduction}

Both theoretical and numerical investigations concerning
relativistic hydrodynamics are becoming very important for
astrophysics, cosmology and also for heavy-ions collisions.
Therefore it is important to develop analytical methods able to
describe the various aspects of the theory of relativistic perfect
fluids, in particular their properties as isolated extended
relativistic systems.
\medskip

Usually relativistic fluids are described by assigning i) a unit
four-velocity field ${\tilde U}^\mu(z)$ [${\tilde U}^{\mu}\,
{\tilde U}_{\mu} = 1$]; ii) the {\it local} thermodynamical
functions {\em internal energy} $\tilde \rho(z)$, {\em particle
density} $\tilde n(z)$, {\em local pressure} $\tilde p(z)$, {\em
local temperature} $\tilde T(z)$, {\em entropy per particle}
$\tilde s(z)$); iii) an equation of state. Here $z^{\mu}$ are the
coordinates of a fixed point in Minkowski space-time and sometimes
this description is named Eulerian. However some clarification is
needed about this terminology at the relativistic level, where a
notion of simultaneity has to be introduced to give sense to the
relativistic equations of motion, which very often are deduced
from an action principle.

\medskip

In the standard approach to non-relativistic fluid dynamics, not
based on variational principles, two distinct points of view are
usually used \footnote{See for instance Section 3.2 of
Ref.\cite{1}.}: the {\it Lagrangian (or material)} point of view
and the {\it Eulerian (or local)} point of view. The equations of
motion, i) the continuity equation or mass conservation; ii)
Euler-Newton equations or balance of linear momentum; iii) the
conservation of energy; look different in the two points of view.
\medskip

A) In the {\it Lagrangian point of view} the fluid is described by
the {\it flux lines} $\vec x(t, {\vec x}_o)$ with $\vec x(0, {\vec
x}_o) = {\vec x}_o$, defined as the integral lines of the
3-velocity field $\vec u(t, {\vec x}_o) = {{\partial \vec x(t,
{\vec x}_o)}\over {\partial t}}$. Each integral line is labeled
with its initial coordinate ${\vec x}_o$. The coordinates ${\vec
x}_o$'s are the {\it Lagrangian (or comoving) coordinates of the
Lagrangian point of view}. The flux lines have the role to
describe the {\em mechanical} aspect of the flow of the fluid. If
we think that associated with each flux line there is a {\it
material particle} defined by an infinitesimal volume of fluid
around the point ${\vec x}_o$, then the flux line $\vec x(t, {\vec
x}_o)$ is also the trajectory followed by the material particle,
so that also the name {\it material point of view} is used. For a
fixed value of ${\vec x}_o$, $\vec x(t, {\vec x}_o)$ specifies the
path of the mass element which was at ${\vec x}_o$ at $t = 0$; for
a fixed value of $t$, $\vec x(t, {\vec x}_o)$ determines the
transformation of the region initially occupied by the whole mass
of the fluid. For every local thermodynamical function evaluated
on the flux lines ${\tilde {\cal G}}(t, \vec x(t, {\vec x}_o))$
its expression in the Lagrangian point of view is ${\hat {\cal
G}}(t, {\vec x}_o) = {\tilde {\cal G}}(t, \vec x(t, {\vec x}_o))$.
Therefore the equations of motion in the Lagrangian point of view
are written using only the ordinary time derivative,
${{\partial}\over {\partial t}}\, {\hat {\cal G}}(t, {\vec x}_o)$.
\medskip

B) Instead in the {\it Eulerian (or local) point of view} the
fluid is described by taking as {\it Eulerian coordinates}  a set
of coordinates $\vec x$ referring to a fixed location in space and
not to a moving mass element of fluid and by using the 3-velocity
field $\vec v(t, \vec x)$, which is the velocity of the fluid
particle that happens to be at the location $\vec  x$ at time $t$.
If at time $t$ we make the identification $\vec x = \vec x(t,
{\vec x}_o)$, namely the fixed coordinate is seen as the
coordinate of the flux line through that point, we recover the
3-velocity field of the other point of view: $\vec v(t, \vec
x){|}_{\vec x = \vec x(t, {\vec x}_o)} = \vec u(t,{\vec x}_o)$.
Now a local thermodynamical function is described by the local
function ${\tilde {\cal G}}(t, \vec x)$ with ${\tilde {\cal G}}(t,
\vec x) {|}_{\vec x = \vec x(t, {\vec x}_o)} = {\hat {\cal G}}(t,
{\vec x}_o)$, so that this description of the fluid is also named
the {\it local point of view}. The equations of motion in the
Eulerian point of view involve the so-called {\it total (or
material) derivative} ${D\over {Dt}}\, {\tilde {\cal G}}(t, \vec
x) = ({{\partial}\over {\partial t}} + \vec u(t, \vec x) \cdot
{{\partial}\over {\partial \vec x}})\, {\tilde {\cal G}}(t, \vec
x)$, since we have ${{\partial}\over {\partial t}}\, {\hat {\cal
G}}(t, {\vec x}_o) = {{\partial}\over {\partial t}}\, {\tilde
{\cal G}}(t, \vec x(t, {\vec x}_o)) = \Big( {{\partial}\over
{\partial t}}\, + \vec u(t, {\vec x}_o) \cdot {{\partial}\over
{\partial \vec x}}\Big)\, {\tilde {\cal G}}(t, \vec x(t, {\vec
x}_o)) = \Big( {D\over {Dt}}\, {\tilde {\cal G}}(t, \vec x)\Big)
{|}_{\vec x = \vec x(t, {\vec x}_o)}$.

\bigskip

To extend these descriptions to the relativistic level in the
Minkowski space-time $M^4$ with coordinates $z^{\mu}$ without
introducing an explicit breaking of covariance like the one
implied by the decomposition $z^{\mu} = ( z^o = ct; \vec z)$, it
is convenient to work in the context of {\em Dirac's parametrized
Minkowski theories} \cite{2,3} on arbitrary (simultaneity and
Cauchy) space-like hyper-surfaces, leaves of the foliation
associated to a 3+1 splitting of Minkowski space-time. If $\tau$
is the scalar parameter (mathematical time) which labels the
leaves $\Sigma_{\tau}$ of the foliation, $\vec \sigma$ are
curvilinear coordinates on them (with respect to an arbitrary
centroid $x^{\mu}(\tau ) = z^{\mu}(\tau ,\vec 0)$ chosen as
origin) and $z^{\mu}(\tau ,\vec \sigma )$ are the embeddings of
the hyper-surfaces $\Sigma_{\tau}$ in Minkowski space-time, then
every local thermodynamical function has an {\it equal time}
re-formulation: ${\tilde {\cal G}}(z(\tau ,\vec \sigma )) = {\cal
G}(\tau ,\vec \sigma )$. By using these adapted coordinates we
have the following natural relativistic generalization of the two
points of view.
\medskip

A) In the {\it Lagrangian (or comoving) point of view} the fluid
is described by the flux lines ${\tilde \zeta}^\mu(z_o,\tilde
\tau)$, defined as the integral lines of the four-velocity field
with {\em initial condition} $z_o^{\mu}$ and parametrized by their
proper time $\tilde \tau$

\begin{equation}
 \frac{d}{d\tilde \tau}{\tilde \zeta}^\mu(z_o,\tilde \tau) =
 {\tilde U}^\mu({\tilde \zeta}^\mu(z_o,\tilde \tau)),
\qquad {\tilde \zeta}^\mu(z_o,0) = z^\mu_o.
 \label{1.1}
 \end{equation}

The points $z_o^{\mu}$ used for the initial conditions must belong
to a {\em space-like Cauchy hyper-surface} in Minkowski
space-time. If we use the embeddings $z^{\mu}(\tau ,\vec \sigma )$
of the hyper-surfaces $\Sigma_{\tau}$ of a foliation, the flux
world-lines are described by functions $\zeta^{\mu}(\tau ,{\vec
\sigma}_o ) = {\tilde \zeta}^{\mu}(z_o, \tilde \tau (\tau ,{\vec
\sigma}_o)) = z^{\mu}(\tau, \vec \Sigma (\tau ,{\vec \sigma}_o ))$
with $\tilde \tau (0, {\vec \sigma}_o) = 0$, $\vec \Sigma (0,
{\vec \sigma}_o) = {\vec \sigma}_o$, $\zeta^{\mu}(0, {\vec
\sigma}_o) = z^{\mu}(\tau_o, {\vec \sigma}_o) = z^{\mu}_o$.  Since
$\tau$ is not the proper time of any flux line, on $\Sigma_{\tau}$
we have ${{d \zeta^{\mu}(\tau ,{\vec \sigma}_o)}\over {d\tau}} /
\sqrt{({{d \zeta (\tau ,{\vec \sigma}_o)}\over {d\tau}})^2} =
U^{\mu}(\tau ,{\vec \sigma}_o)$ for the flux line through ${\vec
\sigma}_o$ at $\tau = 0$.

In these adapted coordinates the coordinates ${\vec \sigma}_o$ on
the Cauchy surface $\Sigma_{\tau_o}$ are the {\it Lagrangian (or
comoving) coordinates of the Lagrangian point of view}, replacing
the non-relativistic ${\vec x}_o$, while the functions $\vec
\Sigma (\tau ,{\vec \sigma}_o)$, describing the flux lines,
replace the non-relativistic $\vec x(t, {\vec x}_o)$ and $\vec u
(\tau ,{\vec \sigma}_o) = {{\partial \vec \Sigma (\tau ,{\vec
\sigma}_o)}\over {\partial \tau}}$ is the 3-velocity field
replacing $\vec u(t, {\vec x}_o)$. The local thermodynamical
functions evaluated along the flux lines ${\tilde {\cal
G}}(\zeta^{\mu}(\tau , {\vec \sigma}_o)) = {\tilde {\cal
G}}(z^{\mu}(\tau , \vec \Sigma (\tau , {\vec \sigma}_o))) = {\cal
G}(\tau , \vec \Sigma (\tau , {\vec \sigma}_o))$ have the
expression ${\hat {\cal G}}(\tau , {\vec \sigma}_o) = {\cal
G}(\tau , \vec \sigma ) {|}_{\vec \sigma = \vec \Sigma (\tau ,
{\vec \sigma}_o)} = {\cal G}(\tau , \vec \Sigma (\tau , {\vec
\sigma}_o))$. The equations of motion involve only the ordinary
time derivative ${{\partial}\over {\partial \tau}}\, {\hat {\cal
G}}(\tau , {\vec \sigma}_o)$. To each flux line there is
associated a material particle at ${\vec \sigma}_o$ on the Cauchy
surface $\Sigma_{\tau = 0}$.

\medskip

B) In the {\it Eulerian point of view} the fluid is described by
taking the fixed 3-coordinates $\vec \sigma$ (replacing the
non-relativistic $\vec x$) as {\it Eulerian coordinates} and
$\tau$ as the time in the coordinatization dictated by the
embedding. The 3-velocity field is $\vec v(\tau ,\vec \sigma )$
and it can be shown that its expression in terms of the 4-velocity
field $U^A(\tau ,\vec \sigma )$, written in adapted coordinates,
is $\vec v(\tau ,\vec \sigma ) = {\vec U}(\tau ,\vec \sigma ) /
U^{\tau}(\tau ,\vec \sigma )$ [see Eq.(\ref{3.16})]. When the
Eulerian coordinate is identified with the flux line
$\zeta^{\mu}(\tau , {\vec \sigma}_o) = z^{\mu}(\tau , \vec \Sigma
(\tau , {\vec \sigma}_o))$ we get $\vec v(\tau ,\vec \sigma
){|}_{\vec \sigma = \vec \Sigma (\tau , {\vec \sigma}_o)} =
{{\partial \vec \Sigma (\tau ,{\vec \sigma}_o)}\over {\partial
\tau}}$. The local thermodynamical functions are described by
local functions ${\cal G}(\tau ,\vec \sigma )$ with ${\cal G}(\tau
, \vec \Sigma (\tau , {\vec \sigma}_o)) = {\hat {\cal G}}(\tau ,
{\vec \sigma}_o)$. Their equations of motion in the Eulerian point
of view involve a total $\tau$-derivative, ${D\over {D\tau}}\,
{\cal G}(\tau , \vec \sigma)  = ({{\partial}\over {\partial \tau}}
+ \vec v(\tau ,\vec \sigma ) \cdot {{\partial}\over {\partial \vec
\sigma}})\, {\cal G}(\tau , \vec \sigma)$, since ${{\partial}\over
{\partial \tau}}\, {\hat {\cal G}}(\tau , {\vec \sigma}_o) =
({{\partial}\over {\partial \tau}} + {{\partial \vec \Sigma (\tau
, {\vec \sigma}_o)}\over {\partial \tau}} \cdot {{\partial}\over
{\partial \vec \sigma}})\, {\cal G}(\tau ,\vec \Sigma (\tau ,{\vec
\sigma}_o)) =\Big( {D\over {D\tau}}\, {\cal G}(\tau , \vec \sigma)
\Big) {|}_{\vec \sigma = \vec \Sigma (\tau ,{\vec \sigma}_o)} $.
\medskip

Therefore, while in the Lagrangian (or material) point of view we
follow the evolution of a thermodynamic function evaluated in a
material particle (namely along the flux line physically
determined by the average particle motion), in the Eulerian (or
local) point of view we follow the evolution of the same function
evaluated in a given space-time point coinciding with the
associated material particle only on the Cauchy surface.

Till now only in the non-relativistic framework of the
Euler-Newton equations  there has been a study of the
transformation from Eulerian to Lagrangian coordinates \cite{4}.

\bigskip

The next problem is how to derive the equations of motion of the
Lagrangian and Eulerian point of views as the Euler-Lagrange (EL)
equations of an action principle.

\medskip

An important development in relativistic hydrodynamics has been
given by Brown \cite{5}, who built a general framework
encompassing all the known variational principles for relativistic
perfect fluids and clarifying the inter-connections among very
different Lagrangian approaches.

In Brown's paper the fluid is described by means of a set of
scalar 3-dimensional configuration variables ${\tilde
\alpha}^i(z)$, $i = 1,2,3,$ of the space-time coordinates
$z^{\mu}$, interpreted as {\it Lagrangian (or comoving)
configuration coordinates} for the fluid labeling the fluid flow
lines.

In  Ref. \cite{6} one of the action principles of Ref. \cite{5}
has been re-formulated in the contest of {\em Dirac's parametrized
Minkowski theories} \cite{2,3} on arbitrary (simultaneity and
Cauchy) space-like hyper-surfaces, leaves of the foliation
associated to a 3+1 splitting of Minkowski space-time for
arbitrary equations of state of the type $\tilde \rho (z) = \tilde
\rho (\tilde n(z), \tilde s(z))$. Now the Lagrangian (or comoving)
coordinates of the fluid are $\alpha^i(\tau ,\vec \sigma ) =
{\tilde \alpha}^i(z(\tau , \vec \sigma ))$ \footnote{The fluid is
supposed to have compact support $V_{\alpha}(\tau ) \subset
\Sigma_{\tau }$, whose boundary $\partial V_{\alpha}(\tau )$ is
dynamically determined as the 2-dimensional surface in each
$\Sigma_{\tau}$ where the pressure vanishes, $\tilde p(z(\tau
,\vec \sigma )) = 0$ for $\vec \sigma
\in
\partial V_{\alpha}(\tau )$. In the case of $N$ disjoint fluid
sectors, we can use the same description with $V_{\alpha}(\tau ) =
\cup_i\, V_{\alpha\, i}(\tau )$ till when the compact supports
$V_{\alpha\, i}(\tau )$ do not overlap.}. For each value of
$\tau$, we can invert $\alpha^i = \alpha^i(\tau ,\vec \sigma )$ to
$\vec \sigma = \vec \sigma (\tau ,\alpha^i)$ and use the
$\alpha^i$'s as a special coordinate system on $\Sigma_{\tau}$
inside the fluid support $V_{\alpha}(\tau )\subset \Sigma_{\tau}$:
$z^{\mu}(\tau ,\vec \sigma (\tau ,\alpha^i))={\check z}^{\mu}(\tau
,\alpha^i)$. This approach is reviewed in Section II.

Since it can be shown [see Eqs.(\ref{2.16}) and(\ref{3.2})] that
we have $\alpha^i(\tau , \vec \sigma ) {|}_{\vec \sigma = \vec
\Sigma (\tau , {\vec \sigma}_o)} = \alpha^i(0, {\vec \sigma}_o) =
\alpha^i_o({\vec \sigma}_o)$, namely that the coordinates
$\alpha^i(\tau ,\vec \sigma )$ are constant along the flux lines,
this explains why these coordinates are a possible set of
Lagrangian (comoving) coordinates for the fluid in alternative to
the ${\vec \sigma}_o$'s . Therefore every action principle studied
in Ref.\cite{5} generates Euler-Lagrange equations [like
Eq.(\ref{2.19})] which describe what happens at the fixed location
$\vec \sigma$ when $\tau$ changes, i.e. these EL equations
generate the equations of motion of the thermodynamical functions
${\cal G}(\tau ,\vec \sigma )$ in the Eulerian (or local) point of
view. This is a consequence of the necessity that the
configuration variables of an action principle be
$\tau$-dependent, even when, like in this case, they are used to
simulate the fixed comoving coordinates of the fluid.

\bigskip

In this paper we will show that it is possible to define a
variational approach whose  configuration variables are the
adapted  3-coordinates $\vec \Sigma (\tau ,{\vec \sigma}_o)$ of
the flux lines, which evolve from the Lagrangian coordinates
${\vec \sigma}_o = \vec \Sigma (0, {\vec \sigma}_o)$ on
$\Sigma_{\tau = 0}$, instead of the Lagrangian (or comoving)
variables $\vec \alpha (\tau ,\vec \sigma)$ used in Ref.\cite{5}.
The configuration 3-coordinates $\vec \Sigma (\tau ,{\vec
\sigma}_o)$ are strictly speaking neither Lagrangian nor Eulerian
coordinates. However they can be considered as {\it generalized
Eulerian  configuration coordinates}, because through the position
$\vec \sigma = \vec \Sigma (\tau ,{\vec \sigma}_o)$ (connecting
the two points of view) they allow to obtain the Eulerian
description of what happens in the fixed point ${\vec \sigma}_o$
when $\tau$ changes. Indeed in this case the resulting EL
equations (\ref{3.17}) [replacing the Eulerian ones (\ref{2.19})]
will correspond to the equations of motion of the thermodynamical
functions ${\hat {\cal G}}(\tau ,{\vec \sigma}_o)$ in the
Lagrangian (or material) point of view. It will be shown (see
footnote 7) how we can transform these EL equations into the
equations of motion in the Eulerian point of view for ${\cal
G}(\tau ,\vec \sigma )$. This will allow to obtain a Hamiltonian
formulation with Eulerian coordinates using Poisson brackets
instead of the Lie-Poisson brackets of Ref.\cite{7}.

\medskip

We shall study the Hamiltonian  first class constraints of the
fluid on arbitrary space-like hyper-surfaces and, then, the
restriction to the rest frame foliation (Wigner hyper-planes), in
order to obtain its {\em rest-frame instant form}, already used
for particles and fields in Refs. \cite{8,9,10,11,12}. Then,
following the methods of Refs. \cite{13,14,15}, we shall study the
problem of the center-of-mass and relative variables, the
separation of relative variables in orientational and vibrational
ones by means of the introduction of dynamical body frames and
Dixon's multipoles \cite{16,17} of the fluid.

\medskip

In this way we get a complete control on the relativistic
kinematics of perfect fluids considered as extended deformable
objects. The next step will be to couple the perfect fluid to
metric and tetrad canonical gravity, whose rest-frame instant form
has been developed in Refs.\cite{18,19,20}, both to develop a
scheme of Hamiltonian numerical gravity in accord with constraint
theory and to study the linearized theory in a completely fixed
3-orthogonal Hamiltonian gauge following the scheme of
Ref.\cite{21}. Another future development \cite{22} will be to
study the non-relativistic limit of this approach and, then, after
the addition of Newton gravitational potential, to recover the
ellipsoidal equilibrium configurations \cite{23} in this
kinematics in the case of incompressible fluids.

\bigskip

While in Section II we review the description of perfect fluids
with Lagrangian coordinates, in Section III we introduce the new
formulation with Eulerian coordinates. In Section IV we derive the
Hamiltonian formulation associated with the action written in the
previous Section and we get the usual constraints of  {\em Dirac's
parametrized Minkowski theories} and their Dirac Poisson algebra
\cite{2}. In Section V the {\em rest frame instant form} of the
dynamic is constructed. This form of the dynamic is such that we
can discuss the problem of the separation of the relative
variables from the center of mass-like variables and the analogous
problem for the rotational and shape variables in the same way as
it has already been done for relativistic particles. This is done
in Sections VI and VII. In Section VIII we discuss the various
either exact or approximate forms in which the invariant mass of
the fluid, i.e. the Hamiltonian in the rest-frame instant form,
may be presented as a function of the orientational and shape
variables. Dixon's multipoles for the fluid are defined in Section
IX. In the final Section there are some concluding remarks.

Appendix A reviews notations on space-like hyper-surfaces. In
Appendix B there is a list of the equations of state for which we
can obtain a closed form of the fluid invariant mass. Appendix C
contains remarks on Poisson brackets. Appendix D describes the
Gartenhaus-Schwartz transformation. Finally in Appendix E there
are some solutions for the kernels associated with the relative
variables.

\vfill\eject

\section{Lagrangian formulation of the dynamics of relativistic perfect fluids
in parametrized Minkowski space-time with Lagrangian comoving
configuration coordinates.}

In this Section we review some of the results of Refs. \cite{5,6}.
One of the many action principles for a Lagrangian description of
relativistic perfect fluid dynamics described in Ref.\cite{5} has
been re-formulated in Ref.\cite{6} in the context of the {\em
parametrized Minkowski theories} \cite{8,3}. As said in the
Introduction the starting point of these theories \cite{2} is the
foliation of the Minkowski space-time by a family of space-like
hyper-surfaces $\Sigma_{\tau}$ defined by the {\em embedding}
$z^\mu(\tau,\vec{\sigma})$ ($R\times\Sigma\rightarrow M^4$)
\footnote{We use the notation
$\sigma^{\check{A}}=(\tau,\sigma^{\check{r}})$ where
$\check{A}=(\tau,\check{r})$, $\check{r}=1,2,3$, for these
coordinates adapted to the foliation. The notation $r$ will be
reserved in Section V for the 3-vectors on the {\em Wigner
hyper-planes}}. The fields $z^\mu(\tau,\vec{\sigma})$ define a
coordinates transformation $z^{\mu} \mapsto \sigma^{\check A}$ on
the pseudo-Riemannian manifold $M^4$, a field of {\em cotetrads}
($\mu$ are the flat indices)

\begin{equation}
z^\mu_{\check{A}}(\tau,\vec{\sigma})= \frac{\partial
z^\mu(\tau,\vec{\sigma})} {\partial \sigma^{\check{A}}},
 \label{2.1}
\end{equation}

\noindent and the induced metric

\begin{equation}
g_{\check{A}\check{B}}(\tau,\vec{\sigma})=
z^\mu_{\check{A}}(\tau,\vec{\sigma})\eta_{\mu\nu}z^\nu
_{\check{B}}(\tau,\vec{\sigma}).
 \label{2.2}
 \end{equation}

\noindent See Appendix A for other properties of space-like
hyper-surfaces.

In parametrized Minkowski theories the Lagrangian of every
isolated system is written as a functional of the Lagrangian
coordinates $\alpha^i(\tau ,\vec \sigma )$ of the system, adapted
to the foliation, and of the embedding $z^\mu(\tau,\vec{\sigma})$
interpreted as the Lagrangian coordinates describing the
hyper-surface in this enlarged configuration space. This
functional is determined by considering the Lagrangian of the
system coupled to an external gravitational field and replacing
the 4-metric $g_{\mu\nu}(z)$ with the induced metric (\ref{2.2})
in the adapted coordinates.

\medskip

As said in the Introduction in the Eulerian point of view, the
relativistic perfect fluid is characterized by the 4-velocity
field ${\tilde U}^\mu(z)$ (${\tilde U}_\mu {\tilde U}^\mu=1$) on
Minkowski space-time $M^4$ and by a set of local thermodynamical
functions. After the foliation of $M^4$ with the hyper-surfaces
$\Sigma_{\tau}$, the 4-velocity field has the adapted covariant
components

\begin{equation}
U_{\check{A}}(\tau,\vec{\sigma})=z^\mu_{\check{A}}(\tau,
\vec{\sigma}){\tilde U}_\mu(z(\tau,\vec{\sigma})).
 \label{2.3}
 \end{equation}

With the adopted parametrization of Minkowski space-time the local
thermodynamical functions can be seen as functions of
$(\tau,\vec{\sigma})$ by means of the replacement $z^{\mu} =
z^{\mu}(\tau,\vec{\sigma})$ \footnote{From now on we shall denote
with $f(\tau ,\vec \sigma )$ the functions $f(\tau ,\vec \sigma )
= \tilde f(z(\tau ,\vec \sigma ))$.} . In particular let us
consider the {\em numerical density of particles}
$n(\tau,\vec{\sigma})$. Together with the 4-velocity field
$U^{\check{A}}(\tau, \vec{\sigma})$ it defines the {\em numerical
density current} $n(\tau,\vec{\sigma})\,
U^{\check{A}}(\tau,\vec{\sigma})$. {\em The conservation of the
total particle number } is the following constraint on this
current ("$;$" and "$,$" denote the covariant and ordinary
derivative, respectively)

\begin{equation}
\left[ n(\tau,\vec{\sigma})\, U^{\check{A}}(\tau,\vec{\sigma})
\right]_{;\check{A}} = \frac{1}{\sqrt{g(\tau,\vec{\sigma})}}\,
\frac{\partial}{\partial\sigma^{\check{A}}}\, \Big[
\sqrt{g(\tau,\vec{\sigma})}\, n(\tau,\vec{\sigma})\,
U^{\check{A}}(\tau,\vec{\sigma})\Big] = 0.
 \label{2.4}
 \end{equation}

\medskip

We can also to consider the {\em energy density function}
$\rho(\tau,\vec{\sigma})$, the local {\em pressure}
$p(\tau,\vec{\sigma})$, the local {\em temperature}
$T(\tau,\vec{\sigma})$, the {\em entropy per particles}
$s(\tau,\vec{\sigma})$ and the {\em chemical potential}

\begin{equation}
\mu(\tau,\vec{\sigma}) = \frac{\rho(\tau,\vec{\sigma}) +
p(\tau,\vec{\sigma})}{n(\tau,\vec{\sigma})}.
 \label{2.5}
 \end{equation}

The {\em first principle of thermodynamics} is given by the
following differential relation

\begin{equation}
d\rho(\tau,\vec{\sigma}) = \mu(\tau,\vec{\sigma})\,
dn(\tau,\vec{\sigma}) + n(\tau,\vec{\sigma})\,
T(\tau,\vec{\sigma})\, ds(\tau,\vec{\sigma}).
 \label{2.6}
 \end{equation}

The {\em equation of state} is given interpreting $\rho$ as a
function of $n$ and $s$

\begin{equation}
\rho(\tau,\vec{\sigma}) =
\rho(n(\tau,\vec{\sigma}),s(\tau,\vec{\sigma})),
 \label{2.7}
 \end{equation}

\noindent and we can obtain the other thermodynamical quantities
as functions of $n$ and $s$, in particular

\begin{eqnarray}
T(\tau,\vec{\sigma}) &\equiv& \frac{1}{n(\tau,\vec{\sigma})}\,
\frac{\partial \rho}{\partial s}
(n(\tau,\vec{\sigma}),s(\tau,\vec{\sigma})),\nonumber \\
 &&{}\nonumber \\
p(\tau,\vec{\sigma}) &\equiv& n(\tau,\vec{\sigma})\,
\frac{\partial \rho}{\partial n}
(n(\tau,\vec{\sigma}),s(\tau,\vec{\sigma})) -
\rho(n(\tau,\vec{\sigma}),s(\tau,\vec{\sigma})).
 \label{2.8}
 \end{eqnarray}

Finally, we have to add {\em the entropy conservation}

\begin{equation}
U^{\check{A}}(\tau,\vec{\sigma})\, \frac{\partial}{\partial
\sigma^{\check{A}}}\, s(\tau,\vec{\sigma}) = 0.
 \label{2.9}
 \end{equation}

\noindent This constraint tells us that we don't have loss of
entropy out the flux tube defined by the fluid's flux lines.

\medskip

Due to the constraints (\ref{2.4}),(\ref{2.9}), the four-velocity
field $U_{\check{A}}(\tau,\vec{\sigma})$ and the independent
thermodynamic functions, $n(\tau,\vec{\sigma})$,
$s(\tau,\vec{\sigma})$ are a set of redundant variables for the
fluid description . In Ref.\cite{5} it is shown that the
constraints (\ref{2.4}),(\ref{2.9}) may be enforced introducing
some {\em Lagrangian (or comoving) variables} $\alpha^i(\tau,
\vec{\sigma})$ for describing the fluid. It is useful to introduce
the fields $J^{\check{A}}(\tau,\vec{\sigma})$

\begin{equation}
 \sqrt{g(\tau,\vec{\sigma})}\, n(\tau,\vec{\sigma})\,
U^{\check{A}}(\tau,\vec{\sigma}) =
J^{\check{A}}(\tau,\vec{\sigma}).
 \label{2.10}
 \end{equation}

\noindent These fields are dependent on the Lagrangian coordinates
$\alpha^i(\tau,\vec{\sigma}),\;i=1,2,3$, according to the
definition

\begin{eqnarray}
J^\tau(\tau,\vec{\sigma}) &=& - \det\, I(\tau,\vec{\sigma}),
\nonumber\\
 J^{\check{r}}(\tau,\vec{\sigma}) &=& \det\,
I(\tau,\vec{\sigma}) \,\,
\left[I^{-1}(\tau,\vec{\sigma})\right]^{\check{r}}_{\;i}\,
\frac{\partial\alpha^i(\tau,\vec{\sigma})}{\partial \tau},
 \label{2.11}
\end{eqnarray}

\noindent where the $3\times 3$ matrix $I(\tau,\vec{\sigma})$ is

\begin{equation}
\left[I(\tau,\vec{\sigma})\right]^i_{\;\check{r}} =
\left(\frac{\partial\alpha^i(\tau,\vec{\sigma})}
{\partial\sigma^{\check{r}}}\right).
 \label{2.12}
\end{equation}

From the definition (\ref{2.11}) it follows that
($\sigma^\tau\equiv\tau$)

\begin{equation}
\frac{\partial}{\partial\sigma^{\check{A}}}\,
J^{\check{A}}(\tau,\vec{\sigma}) = 0.
 \label{2.13}
 \end{equation}

\noindent This equation is equivalent to the constraint
(\ref{2.4}). In particular, it follows from Eq. (\ref{2.13}) that
the {\em total number of particles} on the hyper-surfaces
$\Sigma_{\tau}$ may be defined as ($V_\alpha(\tau)$ is the fluid's
volume on $\Sigma_{\tau}$)

\begin{equation}
{\cal N} = \int_{V_\alpha(\tau)}d^3\sigma\,
J^\tau(\tau,\vec{\sigma}),
 \label{2.14}
 \end{equation}

\noindent and it is conserved by the evolution in the time
parameter $\tau$. In this parametrization the  entropy per
particle is a function of the $\alpha^i(\tau,\vec{\sigma})$ alone

\begin{equation}
s \equiv s(\alpha^i(\tau,\vec{\sigma})).
 \label{2.15}
 \end{equation}

\noindent By construction it satisfy the entropy constraint
(\ref{2.9}) [see also the following equation (\ref{2.16})].

\bigskip

The Lagrangian coordinates $\alpha^i(\tau,\vec{\sigma})$ can be
interpreted as resulting from a coordinate transformation
$\sigma^{\check r} \mapsto \alpha^i$ on the hyper-surface
$\Sigma_{\tau}$. In particular $\alpha^i(0,\vec{\sigma})$ define a
coordinate transformation $\sigma^{\check r} \mapsto
\alpha^i(0,\vec \sigma )$  on the hyper-surface $\Sigma_{\tau=0}$;
if $V_\alpha(0)$ is the fluid total volume in this hyper-surface,
every point in the total volume on the hyper-surface
$\Sigma_{\tau}$, $V_\alpha(\tau)$, is in a one to one
correspondence with a point in $V_\alpha(0)$ by means of the flux
lines. Due to the definitions (\ref{2.10}) and (\ref{2.11}) it
follows that the fields $\alpha^i(\tau,\vec{\sigma})$ are {\it
constant along the flux lines}, since by construction we have

\begin{equation}
U^{\check{A}}(\tau,\vec{\sigma})\, \frac{\partial} {\partial
\sigma^{\check{A}}}\, \alpha^i(\tau,\vec{\sigma}) = 0.
 \label{2.16}
\end{equation}

\noindent Then the fields $\alpha^i(\tau,\vec{\sigma})$ can be
interpreted also as {\em labels} assigned to the flux lines; the
field $\alpha^i(\tau,\vec{\sigma})$ tells us that the flux line
labeled with $\alpha^i$ {\it goes through} the point
$z^\mu(\tau,\vec{\sigma})\in M^4$.

\medskip

With the previous definitions and observations, the action defined
in Section 5 of Ref. \cite{5} has been rewritten in Ref.\cite{6}
in the form

\begin{equation}
S = - \int d\tau\, d^3 \sigma\, \sqrt{g(\tau,\vec{\sigma})}\,
\rho(n[\alpha],s[\alpha]),
 \label{2.17}
\end{equation}

\noindent where $s$ is given by Eq.(\ref{2.15}) and from
Eq.(\ref{2.10}) it follows that

\begin{equation}
 n(\tau ,\vec \sigma ) =
 \frac{\sqrt{g_{\check{A}\check{B}}(\tau,\vec{\sigma})\,
J^{\check{A}}(\tau,\vec{\sigma})\,
J^{\check{B}}(\tau,\vec{\sigma})}} {\sqrt{g(\tau,\vec{\sigma})}}.
 \label{2.18}
 \end{equation}

\medskip

The stationarity of the action with respect to variations of the
$\alpha$'s gives  the fluid equations of motion \footnote{They are
equations of motion in the Eulerian (or local) point of view. }

\begin{equation}
0 = 2\, V_{[\check{E},\check{F}]}(\tau,\vec{\sigma})\,
U^{\check{F}}(\tau,\vec{\sigma}) + T(\tau,\vec{\sigma})\,
s_{,\check{E}}(\tau,\vec{\sigma}),
 \label{2.19}
 \end{equation}

\noindent where

\begin{equation}
V_{\check{A}}(\tau,\vec{\sigma}) = \mu(\tau,\vec{\sigma})\,
U_{\check{A}}(\tau,\vec{\sigma}),
 \label{2.20}
 \end{equation}

\noindent is the {\em Taub vector}, if $\mu (\tau ,\vec \sigma )$
is the chemical potential (\ref{2.5}). In the previous relation we
have used the notation

\begin{equation}
V_{[\check{E},\check{F}]} = V_{[\check{E};\check{F}]} =
\frac{V_{\check{E},\check{F}} - V_{\check{F},\check{E}}}{2}.
 \label{2.21}
 \end{equation}

The variation of the action with respect to the metric variations
$\delta g_{\check A\check B}$ defines the {\em stress-energy
tensor}

\begin{eqnarray}
T^{\check A\check B}(\tau,\vec{\sigma}) &=& -
\frac{2}{\sqrt{g(\tau,\vec{\sigma})}}\, \frac{\delta S}{\delta
g_{\check A\check B}(\tau,\vec{\sigma})} = \nonumber \\
 &&\nonumber\\
  &=& \Big[\rho(\tau,\vec{\sigma}) + p(\tau,\vec{\sigma})\Big]\,
U^{\check{A}}(\tau,\vec{\sigma})\,
U^{\check{B}}(\tau,\vec{\sigma}) - p(\tau,\vec{\sigma})\,
g^{\check{A}\check{B}}(\tau,\vec{\sigma}).
 \label{2.22}
 \end{eqnarray}

In Ref.\cite{5} it is showed that the equations of motion
(\ref{2.19}) are equivalent to the stress-energy tensor
conservation law

\begin{equation}
\left.T^{\check{A}\check{B}}\right._{;\check{B}} = 0.
 \label{2.23}
\end{equation}

\noindent To see this, we have to observe that Eq.(\ref{2.23}) is
equivalent to

\begin{eqnarray}
U_{\check{A}} \left.T^{\check{A}\check{B}}\right._{;\check{B}}&=&
0,\nonumber\\
 (g_{\check{A}\check{B}} - U_{\check{A}}\, U_{\check{B}})
\left.T^{\check{B}\check{C}}\right._{;\check{C}}&=& 0.
 \label{2.24}
\end{eqnarray}

\noindent The first of these equations is equivalent to the
entropy constraint (\ref{2.9}), which is then  satisfied,

\begin{equation}
U_{\check{A}} \left.T^{\check{A}\check{B}}\right._{;\check{B}} = -
n\, T\;s_{,\check{B}}\, U^{\check{B}} = 0.
 \label{2.25}
\end{equation}

\noindent The second equation is equivalent to the {\em Euler
equations}

\begin{equation}
(\rho + p)\, U_{\check{A};\check{B}}\, U^{\check{B}} + (-
\delta^{\check{B}}_{\check{A}} + U^{\check{A}}\, U^{\check{B}})\,
p_{,\check{B}} = 0.
 \label{2.26}
 \end{equation}

\medskip

The stationarity of the action (\ref{2.17}) with respect to the
variations of the $z^{\mu}(\tau,\vec{\sigma})$'s gives us the
equations of motion of the embeddings $z^{\mu}(\tau,
\vec{\sigma})$. Since the action depends on $z^{\mu}(\tau,
\vec{\sigma})$ only through the induced metric, they are

\begin{eqnarray}
\frac{\delta S}{\delta z^\mu(\tau,\vec{\sigma})} &=& 2\,
\eta_{\mu\nu}\, \frac{\partial}{\partial\sigma^{\check A}}\,
\left[ \frac{\delta S}{\delta g_{\check A\check
B}(\tau,\vec{\sigma})}\, z^\nu_{\check
B}(\tau,\vec{\sigma})\right] = \nonumber\\
 &&\nonumber\\
 &=& - \eta_{\mu\nu}\, \frac{\partial}{\partial\sigma^{\check{A}}}\, \left[
\sqrt{g(\tau,\vec{\sigma})}\, T^{\check{A}\check{B}}
(\tau,\vec{\sigma})\, z^\nu_{\check{B}}(\tau,\vec{\sigma})\right]
= \nonumber\\
 &&\nonumber\\
  &=& \sqrt{g(\tau,\vec{\sigma})}\,
  z^{\check{C}}_\mu(\tau,\vec{\sigma})\,
g_{\check{C}\check{D}}(\tau,\vec{\sigma})
\left[T^{\check{D}\check{A}}(\tau,\vec{\sigma})\right]_{;\check{A}}
= 0.
 \label{2.27}
\end{eqnarray}

Due to the stress-energy conservation (\ref{2.23}), following from
the fluid equations of motion (\ref{2.19}), these equations are
always satisfied without any restriction on the embeddings
$z^\mu(\tau,\vec{\sigma})$, which remain arbitrary. In other
words, the equations of motion  (\ref{2.19}) and (\ref{2.27}) are
not independent. This is the Lagrangian manifestation that the
parametrized Minkowski theories are singular theories. In these
theories the $z^\mu(\tau,\vec{\sigma})$ are {\em gauge variables}
and in the Hamiltonian formulation their conjugate momenta are
defined by first class Dirac constraints \cite{3,8} as it will
shown in Section IV .

\vfill\eject

\section{Lagrangian formulation of the dynamics of relativistic perfect fluids
in parametrized Minkowski space-time with Eulerian configuration
coordinates.}

In this Section we introduce a different parametrization of the
action (\ref{2.17}) using a new set of configuration coordinates.
The geometrical interpretation of the {\em old} Lagrangian
(comoving) coordinates $\alpha^i(\tau , \vec \sigma )$ as labels
for the flux lines suggests that it is possible to parametrize the
action (\ref{2.17}) with new adapted 3-coordinates
$\vec{\Sigma}(\tau, \vec{\sigma}_o)$, which describe the flux
lines as the integral curves of the 4-velocity field starting from
the Lagrangian coordinates ${\vec \sigma}_o = \vec \Sigma (0,
{\vec \sigma}_o)$. As explained in the Introduction they can be
named generalized Eulerian configuration coordinates. The derived
variables $\zeta^\mu(\tau, {\vec \sigma}_o) =
z^\mu(\tau,\vec{\Sigma}(\tau,\vec{\sigma}_o))$ define the
four-dimensional {\it flux line} going through ${\vec \sigma}_o$
at $\tau = 0$

\begin{equation}
{{\frac{d}{d\tau}\,\zeta^\mu(\tau, {\vec \sigma}_o)}\over {
\sqrt{\eta_{\alpha\beta}\, \frac{d}{d\tau}\,\zeta^\alpha(\tau,
{\vec \sigma}_o)\, \frac{d}{d\tau}\,\zeta^\beta(\tau, {\vec
\sigma}_o) }}} = {\tilde U}^\mu(\zeta(\tau ,{\vec \sigma}_o )),
 \label{3.1}
 \end{equation}

\noindent with the initial condition $\zeta^\mu(0, {\vec
\sigma}_o) = z^\mu(0,\vec{\sigma}_o)$, namely
$\vec{\Sigma}(0,\vec{\sigma}_o)=\vec{\sigma}_o$.

\medskip

Moreover, due to Eq.(\ref{2.16}), we get consistently

\begin{equation}
\alpha^i(\tau,\vec{\Sigma}(\tau,\vec{\sigma}_o)) =
\alpha^i(0,\vec{\sigma}_o) \equiv \alpha_o^i(\vec{\sigma}_o),
 \label{3.2}
 \end{equation}

\noindent i.e. the functions $\alpha_o^i(\vec{\sigma}_o)$ (a
possible set of Lagrangian coordinates replacing the ${\vec
\sigma}_o$'s) are constant along the flux line through each point
${\vec \sigma}_o$. Then the inverse function theorem implies $\vec
\Sigma (\tau ,{\vec \sigma}_o) = \vec F(\tau, \vec \alpha (0,{\vec
\sigma}_o))$ and ${\vec \sigma}_o = \vec \Sigma (0,{\vec
\sigma}_o) = \vec F(0, \vec \alpha (0,{\vec \sigma}_o))$.
Therefore at $\tau = 0$ the generalized Eulerian coordinates of
the fluid are just the Lagrangian 3-coordinates ${\vec \sigma}_o$
of the points where the flux lines intersect the hyper-surface
$\Sigma_{\tau = 0}$ and they are connected to the Lagrangian
comoving coordinates $\alpha^i(0 ,\vec \sigma )$ at $\tau = 0$ by
the change of coordinates $\vec \sigma \mapsto \vec \alpha (0,
\vec \sigma )$. By inverting $\vec \sigma = \vec \Sigma (\tau ,
{\vec \sigma}_o)$ to $ {\vec \sigma}_o = {\vec g}_{\vec
\Sigma}(\tau ,\vec \sigma )$, from Eq.(\ref{3.2}) we get
$\alpha^i(\tau ,\vec \sigma ) = \alpha^i(0, {\vec g}_{\vec
\Sigma}(\tau , \vec \sigma ))$. While on $\Sigma_{\tau}$ with
$\tau > 0$ the position of the flux lines is identified by the
Eulerian coordinates $\vec \sigma = \vec \Sigma (\tau , {\vec
\sigma}_o)$ in the Eulerian point of view, in the Lagrangian one
this position is identified by the Lagrangian coordinates
$\alpha^i(\tau ,\vec \sigma ) = \alpha^i(0, {\vec g}_{\vec
\Sigma}(\tau , \vec \sigma ))$ [see also Eq.(\ref{3.9})].

\medskip

If we remember the definition (\ref{2.12}), we have

\begin{equation}
\frac{d\alpha^i(\tau,\vec{\Sigma}(\tau,\vec{\sigma}_o))}{d\tau} =
\frac{\partial\alpha^i(\tau,\vec{\Sigma}(\tau,\vec{\sigma}_o))}{\partial\tau}
+ [I(\tau,\vec{\Sigma}(\tau,\vec{\sigma}_o))]^i_{\;\check{r}}\,
\frac{\partial\Sigma^{\check{r}}(\tau,\vec{\sigma}_o)}{\partial\tau}
= 0,
 \label{3.3}
 \end{equation}

\noindent so that

\begin{equation}
\frac{\partial\Sigma^{\check{r}}(\tau,\vec{\sigma}_o)}
{\partial\tau} = -
[I^{-1}(\tau,\vec{\Sigma}(\tau,\vec{\sigma}_o))]^{\check{r}}_{\;i}\,
\frac{\partial\alpha^i(\tau,\vec{\Sigma}(\tau,\vec{\sigma}_o))}{\partial\tau}
= + \frac{J^{\check{r}}(\tau,\vec{\Sigma}(\tau,\vec{\sigma}_o))}
{J^\tau(\tau,\vec{\Sigma}(\tau,\vec{\sigma}_o))}.
 \label{3.4}
 \end{equation}

Since we have

\begin{equation}
[I(\tau,\vec{\Sigma}(\tau,\vec{\sigma}_o))]^i_{\;\check{r}} \,
\frac{\partial\Sigma^{\check{r}}(\tau,\vec{\sigma}_o)}
{\partial\sigma^{\check{s}}_o} =
\frac{\partial\alpha_o^i(\vec{\sigma}_o)}
{\partial\sigma^{\check{s}}_o},
 \label{3.5}
 \end{equation}

\noindent then we get

\begin{eqnarray}
J^\tau(\tau,\vec{\Sigma}(\tau,\vec{\sigma}_o)) &=& - \det\,
(I(\tau,\vec{\Sigma}(\tau,\vec{\sigma}_o))) = -
{\det}^{-1}\left(\frac{\partial\Sigma}{\partial\sigma_o}\right)\,
\det\left(\frac{\partial\alpha_o(\vec{\sigma}_o)}{\partial\sigma_o}\right)
=\nonumber\\
 &&\nonumber\\
  &=& n_o(\vec{\sigma}_o)\, {\det}^{-1}
\left(\frac{\partial\Sigma}{\partial\sigma_o}\right).
 \label{3.6}
 \end{eqnarray}

The function

\begin{equation}
n_o(\vec{\sigma}_o) = -
\det\left(\frac{\partial\alpha_o(\vec{\sigma}_o)}{\partial\sigma_o}\right),
 \label{3.7}
 \end{equation}

\noindent is {\it the particle numerical density on the Cauchy
hyper-surface  $\Sigma_{\tau=0}$ and is known from the fluid
initial conditions}. With the previous results, we can write

\begin{eqnarray}
J^\tau(\tau,\vec{\Sigma}(\tau,\vec{\sigma}_o)) &=&
n_o(\vec{\sigma}_o)\, {\det}^{-1}
\left(\frac{\partial\Sigma}{\partial\sigma_o}\right), \nonumber\\
&&\nonumber\\
J^{\check{r}}(\tau,\vec{\Sigma}(\tau,\vec{\sigma}_o)) &=&
n_o(\vec{\sigma}_o)\, {\det}^{-1}
\left(\frac{\partial\Sigma}{\partial\sigma_o}\right) \,
\frac{\partial\Sigma^{\check{r}}}{\partial\tau}.
 \label{3.8}
 \end{eqnarray}

\medskip

Let us notice that, as said in the Introduction,  any functional
of the thermodynamical functions admits many expressions ${\cal
G}(\tau , \vec \sigma ) = {\tilde {\cal G}}(z(\tau , \vec \sigma
)) = {\cal G}(\tau , \vec \Sigma (\tau , {\vec \sigma}_o)) = {\hat
{\cal G}}(\tau , {\vec \sigma}_o) = {\cal G}_{\alpha}(\tau , \vec
\alpha (\tau , \vec \sigma ))$ (we have added the expression in
terms of the Lagrangian comoving coordinates) \footnote{Note that
in general each of these functions is actually a functional of the
associated coordinates and of their time and spatial gradients.}.
In particular $\widehat{\cal G}(\tau,\vec{\sigma}_o)$ and ${\cal
G}(\tau ,\vec \sigma )$ are the expressions in the Lagrangian (or
material) and Eulerian (or local) point of view respectively.
While ${\hat {\cal G}}(\tau , {\vec \sigma}_o)$ is defined only on
$\Sigma_{\tau = 0}$, ${\cal G}(\tau , \vec \sigma )$ gives the
expression on arbitrary hyper-surfaces $\Sigma_{\tau \not= 0}$. We
have ${\cal G}(\tau ,\vec \sigma ) {|}_{\vec \sigma = \vec \Sigma
(\tau , {\vec \sigma}_o)} = {\hat {\cal G}}(\tau , {\vec
\sigma}_o)$ and ${\cal G}(\tau , \vec \sigma ) = {\hat {\cal
G}}(\tau , {\vec \sigma}_o) {|}_{{\vec \sigma}_o = {\vec
g}_{\Sigma}(\tau , \vec \sigma )}$. Therefore we get the following
equation connecting them

\begin{equation}
{\cal G}(\tau,\vec{\sigma})=\int d^3\sigma_o\,
\det\left(\frac{\partial\Sigma}{\partial\sigma_o}\right)
\delta^3(\vec{\sigma}-\vec{\Sigma}(\tau,\vec{\sigma}_o)) \hat{\cal
G}(\tau,\vec{\sigma}_o).
 \label{3.9}
 \end{equation}

\medskip

The particle density $\hat{n}$ is still given by Eq.(\ref{2.18})
with the $J^{\check{A}}$ given by Eqs. (\ref{3.8}). Then,
introducing the notation

\begin{equation}
{\cal
R}(\tau,\vec{\sigma}_o)=\sqrt{g_{\tau\tau}(\tau,\vec{\Sigma}) +
2\, g_{\tau \check{r}}(\tau,\vec{\Sigma})\,
\frac{\partial\Sigma^{\check{r}}}{\partial\tau} +
g_{\check{r}\check{s}}(\tau,\vec{\Sigma}) \,
\frac{\partial\Sigma^{\check{r}}}{\partial\tau} \,
\frac{\partial\Sigma^{\check{s}}}{\partial\tau}}\, {|}_{\vec
\Sigma = \vec \Sigma (\tau,\vec{\sigma}_o)},
 \label{3.10}
 \end{equation}

\noindent we get

\begin{equation}
\hat{n}(\tau ,{\vec \sigma}_o) = n(\tau ,\vec \sigma ){|}_{\vec
\sigma = \vec \Sigma (\tau ,{\vec \sigma}_o)} =
n_o(\vec{\sigma}_o)\, \frac{\det^{-1} \left(\frac{\partial\Sigma
(\tau,\vec{\sigma}_o) }{\partial\sigma_o}\right)}{\sqrt{g(\tau,
\vec{\Sigma}(\tau,\vec{\sigma}_o))}}\, {\cal
R}(\tau,\vec{\sigma}_o).
 \label{3.11}
\end{equation}

\medskip

Instead the entropy per particle $s$ can be rewritten as a
function dependent on $\vec{\sigma}_o$ alone

\begin{equation}
s(\alpha(\tau,\vec{\Sigma}(\tau,\vec{\sigma}_o))) =
s(\alpha_o(\vec{\sigma}_o)) = s_o(\vec{\sigma}_o),
 \label{3.12}
\end{equation}

\noindent and it is given with the initial conditions. In this
case the constraint (\ref{2.9}) is trivially satisfied being
equivalent to

\begin{equation}
\frac{\partial}{\partial\tau}\, s_o(\vec{\sigma}_o) = 0.
 \label{3.13}
 \end{equation}

Analogously the particle number constraint is satisfied by
construction being in particular

\begin{equation}
{\cal N} = \int_{V_\alpha(\tau)}d^3\sigma\,
J^\tau(\tau,\vec{\sigma}) = \int_{V_{\alpha}(0)} d^3\sigma_o\,
n_o(\vec{\sigma}_o).
 \label{3.14}
 \end{equation}

\bigskip

The action in the new coordinates is

\begin{eqnarray}
S &=& - \int d\tau\, \int_{\Sigma_{\tau}} d^3\sigma\,
\sqrt{g(\tau,\vec{\sigma})}\nonumber \\
 && \int_{\Sigma_{\tau = 0}} d^3\sigma_o\,
 \delta^3(\vec{\sigma}-\vec{\Sigma}(\tau,\vec{\sigma}_o))
\, \det\left(\frac{\partial\Sigma
(\tau,\vec{\sigma}_o)}{\partial\sigma_o}\right)\,
\rho(\hat{n}(\tau , {\vec \sigma}_o),s_o({\vec \sigma}_o)\, )
 =\nonumber \\
&&\nonumber\\
 &=& - \int_{\Sigma_{\tau = 0}}\, d\tau\,d^3\sigma_o\,
 \sqrt{g(\tau,\vec{\Sigma}(\tau,\vec{\sigma}_o))}\,
\det\left(\frac{\partial\Sigma
(\tau,\vec{\sigma}_o)}{\partial\sigma_o}\right)\,
\rho(\hat{n}(\tau ,{\vec \sigma}_o),s_o({\vec \sigma}_o) ),
 \label{3.15}
 \end{eqnarray}

\noindent with the spatial integral restricted to the volume
$V_{\alpha}(0)$ in $\Sigma_{\tau = 0}$.

\medskip

The equations of motion of  the fluid variables derive from the
stationarity of the new action with respect to the variations
$\delta\vec{\Sigma}(\tau,\vec{\sigma}_o)$. If the four-velocity
$\widehat{U}^{\check{A}}(\tau,\vec{\sigma}_o)$ is defined by Eqs.
(\ref{2.10}) with the new expression for the
$J^{\check{A}}(\tau,\vec{\sigma}_o)$ given by Eqs. (\ref{3.8}),
that is

\begin{equation}
\widehat{U}^{\check{A}}(\tau,\vec{\sigma}_o) = \frac{1}{{\cal
R}(\tau,\vec{\sigma}_o)}\,
\frac{\partial\Sigma^{\check{A}}(\tau,\vec{\sigma}_o)}{\partial
\tau},\;\;\;where\;\;\;\Sigma^{\check{A}}=(\tau,\Sigma^{\check{r}}),
 \label{3.16}
 \end{equation}

\medskip

\noindent then the equations of motion are

\begin{equation}
- \widehat{U}^\tau(\tau,\vec{\sigma}_o)\,
\frac{\partial\Sigma^{\check{r}}(\tau,\vec{\sigma}_o)}
{\partial\sigma_o^{\check{s}}}\, \frac{\partial}{\partial\tau}\,
\widehat{V}_{\check{r}} (\tau,\vec{\sigma}_o) +
\widehat{U}^{\check{A}}(\tau,\vec{\sigma}_o)\,
\frac{\partial}{\partial\sigma_o^{\check{s}}}\,
\widehat{V}_{\check{A}}(\tau,\vec{\sigma}_o) -
\widehat{T}(\tau,\vec{\sigma}_o)\, \frac{\partial
s_o(\vec{\sigma}_o)}{\partial\sigma_o^{\check{s}}} = 0,
 \label{3.17}
 \end{equation}

\noindent with $\widehat{V}^{\check{A}}(\tau,\vec{\sigma}_o)=
\widehat{\mu}(\tau,\vec{\sigma}_o)\,\widehat{U}^{\check{A}}(\tau,\vec{\sigma}_o)$.
Let us prove that these equations are equivalent to Eqs.
(\ref{2.19}). First we can observe that from Eq.(\ref{3.9}) we get

\begin{eqnarray}
\frac{\partial}{\partial\sigma^{\check{r}}}{\cal
G}(\tau,\vec{\sigma}) &=& \int d^3\sigma_o\,
\det\left(\frac{\partial\Sigma
(\tau,\vec{\sigma}_o)}{\partial\sigma_o}\right)\,
\delta^3(\vec{\sigma} - \vec{\Sigma}(\tau,\vec{\sigma}_o))\,
K_{\check{r}}^{\;\check{s}}(\tau,\vec{\sigma}_o)\,
\frac{\partial}{\partial\sigma_o^{\check{s}}}\hat{\cal
G}(\tau,\vec{\sigma}_o),\nonumber\\
 &&\nonumber\\
 \frac{\partial}{\partial\tau}{\cal G}(\tau,\vec{\sigma}) &=& \int
d^3\sigma_o\, \det\left(\frac{\partial\Sigma (\tau,\vec{\sigma}_o)
}{\partial\sigma_o}\right)\, \delta^3(\vec{\sigma} -
\vec{\Sigma}(\tau,\vec{\sigma}_o)) \times\nonumber\\
 &&\nonumber\\
  &\times& \left(\frac{\partial}{\partial\tau} -
\frac{\partial\Sigma^{\check
r}(\tau,\vec{\sigma}_o)}{\partial\tau}\,
K_{\check{r}}^{\;\check{s}}(\tau,\vec{\sigma}_o)\,
\frac{\partial}{\partial\sigma_o^{\check{s}}}\right)\, \hat{\cal
G}(\tau,\vec{\sigma}_o),
 \label{3.18}
 \end{eqnarray}

\noindent where we have adopted the notation

\begin{equation}
K_{\check{r}}^{\;\check{s}}(\tau,\vec{\sigma}_o)\,
\frac{\partial\Sigma^{\check{u}}(\tau,\vec{\sigma}_o)}
{\partial\sigma_o^{\check{s}}}
= \delta^{\check{u}}_{\check{r}}.
 \label{3.19}
 \end{equation}

Using the rules (\ref{3.18}) on the equations of motion
(\ref{2.19}) we can verify that Eqs. (\ref{3.17}) are the material
(Lagrangian) representation of the local (Eulerian) equations
(\ref{2.19}). \footnote{Let us observe that we get

\begin{eqnarray*}
 {D\over {D\tau}}\, {\cal G}(\tau ,\vec \sigma ) &=&
  {{\partial {\cal G}(\tau ,\vec \sigma )}\over {\partial \tau}}
 - \vec v(\tau ,\vec \sigma ) \cdot {{\partial}\over {\partial \vec
\sigma}}\, {\cal G}(\tau ,\vec \sigma ) =\nonumber \\
 &&{}\nonumber \\
 &=& \int_{\Sigma_{\tau = 0}}\, d^3\sigma_o\,
 \det\, \Big( {{\partial \Sigma (\tau,\vec{\sigma}_o)}\over
  {\partial \sigma_o}} \Big)\, \delta^3(\vec \sigma
 - \vec \Sigma (\tau , {\vec \sigma}_o))\, {{\partial {\hat {\cal G}}(\tau ,
 {\vec \sigma}_o)}\over {\partial \tau}},
  \end{eqnarray*}

 \noindent with $\vec v(\tau ,\vec \sigma ) = \Big( {{\partial \vec \Sigma}\over
{\partial \tau}} (\tau , {\vec \sigma}_o)\Big) {|}_{{\vec
\sigma}_o = {\vec g}_{\Sigma}(\tau ,\vec \sigma)}$ and with
${D\over {D\tau}}$ denoting the {\it material temporal
derivative}.} Therefore the action (\ref{3.15}) defines the
correct fluid equations of motion in the Lagrangian (material)
point of view.

\bigskip

For the stress-energy tensor we get

\begin{eqnarray}
T^{\check{A}\check{B}}(\tau,\vec{\sigma}) &=& -
\frac{2}{\sqrt{g(\tau,\vec{\sigma})}}\, \frac{\delta S}{\delta
g_{\check{A}\check{B}}(\tau,\vec{\sigma})} =\nonumber\\
&&\nonumber\\
 &=& \int d^3\sigma_o\,
\det\left(\frac{\partial\Sigma
(\tau,\vec{\sigma}_o)}{\partial\sigma_o}\right)\,
\delta^3(\vec{\sigma} - \vec{\Sigma}(\tau,\vec{\sigma}_o))\,
\widehat{T}^{\check{A}\check{B}}(\tau,\vec{\sigma}_o),
 \label{3.20}
 \end{eqnarray}

\noindent where

\begin{equation}
\widehat{T}^{\check{A}\check{B}}(\tau,\vec{\sigma}_o) =
\left(\widehat{\rho}(\tau,\vec{\sigma}_o) +
\widehat{p}(\tau,\vec{\sigma}_o)\right)\,
\widehat{U}^{\check{A}}(\tau,\vec{\sigma}_o)\,
\widehat{U}^{\check{B}}(\tau,\vec{\sigma}_o) -
\widehat{p}(\tau,\vec{\sigma}_o)\,
g^{\check{A}\check{B}}(\tau,\vec{\Sigma}(\tau,\vec{\sigma}_o)).
 \label{3.21}
 \end{equation}

Obviously Eqs.(\ref{3.17}), being equivalent to Eqs.(\ref{2.19}),
imply the conservation of the stress-energy tensor (\ref{3.20}).
Again there are no equations of motion for the
$z^{\mu}(\tau,\vec{\sigma})$'s, so that they remain gauge
variables also for the action (\ref{3.15}).

\vfill\eject

\section{The Hamiltonian formulation.}

Let us study the Hamiltonian formulation of relativistic perfect
fluid dynamics implied by the action (\ref{3.15}). In this Section
we use the notation $f(\Sigma)$ instead of $f(\tau,
\vec{\Sigma}(\tau, \vec{\sigma}_o))$ for the sake of simplicity.
The action (\ref{3.15}) depends on the generalized Eulerian
coordinates $\vec \Sigma(\tau, \vec \sigma )$ of the fluid and on
the embeddings $z^{\mu}(\tau ,\vec \sigma )$ as configurational
variables.

\subsection{The First Class Constraints.}

We can define the following canonical momenta: the {\em momentum
density for the fluid}

\begin{equation}
K_{\check{r}}(\tau,\vec{\sigma}_o)=-\frac{\delta S}{\delta\left(
\frac{\partial\Sigma^{\check{r}}(\tau,\vec{\sigma}_o)}{\partial\tau}\right)}
=n_o(\vec{\sigma}_o)\,\frac{\partial\rho}{\partial\hat{n}}
\left[\frac{g_{\check{r}\tau}(\Sigma)+g_{\check{r}\check{s}}
(\Sigma) \frac{\partial\Sigma^{\check{s}}}{\partial\tau}}{\cal
R}(\tau,\vec{\sigma}_o)\right],
 \label{4.1}
 \end{equation}

\noindent and the {\em momentum density of the embedding}

\begin{eqnarray}
\rho_\mu(\tau,\vec{\sigma}) &=& - \frac{\delta S}{\delta
z^\mu_\tau(\tau,\vec{\sigma})} =\nonumber\\
 &&\nonumber\\
  &=&\int d^3\sigma_o\,
\det\left(\frac{\partial\Sigma
(\tau,\vec{\sigma}_o)}{\partial\sigma_o}\right)
\delta^3(\vec{\sigma}-\vec{\Sigma}(\tau,\vec{\sigma}_o)) \,\left(
\rho(\hat{n},s)-\frac{\partial\rho}{\partial\hat{n}}\hat{n}
\right)\frac{\partial\sqrt{g(\Sigma)}} {\partial
z^\mu_\tau(\Sigma)}+\nonumber\\ &&\nonumber\\ &+&\int
d^3\sigma_o\,
\delta(\vec{\sigma}-\vec{\Sigma}(\tau,\vec{\sigma}_o))\,
n_o(\vec{\sigma}_o)\,\frac{\partial\rho}{\partial\hat{n}}\,
\frac{z_{\mu\tau}(\Sigma)+z_{\mu\,\check{r}}(\Sigma)
\frac{\partial\Sigma^{\check{r}}}{\partial\tau}}{\cal
R}(\tau,\vec{\sigma}_o).
 \label{4.2}
 \end{eqnarray}

\medskip

The Darboux canonical basis of phase space is represented in the
following table

\begin{equation}
\begin{array}{|c|c|}
\hline
z^\mu(\tau,\vec{\sigma})&\Sigma^{\check{r}}(\tau,\vec{\sigma}_o)\\
\rho_\mu(\tau,\vec{\sigma})&K_{\check{r}}(\tau,\vec{\sigma}_o)\\
\hline
\end{array}
 \label{4.3}
 \end{equation}

\noindent and we assume the following non null Poisson Brackets

\begin{eqnarray}
\{z^\mu(\tau,\vec{\sigma}),\rho_\nu(\tau,\vec{\sigma}')\}&=&
-\eta^\mu_\nu\,\delta^3(\vec{\sigma}-\vec{\sigma}'), \nonumber\\
 &&\nonumber\\
 \{\Sigma^{\check{r}}(\tau,\vec{\sigma}_o),
K_{\check{s}}(\tau,\vec{\sigma}'_o)\}&=&
-\delta^{\check{r}}_{\check{s}}\,\delta^3(\vec{\sigma}_o-\vec{\sigma}'_o).
 \label{4.4}
 \end{eqnarray}

\medskip

As we said at the end of Sections II and III, we expect that the
momentum density (\ref{4.2}) is equivalent to four first class
constraints. Since we have

\begin{equation}
\frac{\partial\sqrt{g(\Sigma)}} {\partial
z^\mu_\tau(\Sigma)}\,z^\mu_{\check{r}}(\Sigma)=0,\qquad
  \frac{\partial\sqrt{g(\Sigma)}} {\partial
z^\mu_\tau(\Sigma)}\,l^\mu(\Sigma)=\sqrt{\gamma(\Sigma)},
 \label{4.5}
 \end{equation}

\noindent then, using Eq.(\ref{4.1}),  Eq.(\ref{4.2}) implies

\begin{eqnarray}
\rho_\mu(\tau,\vec{\sigma})z^\mu_{\check{r}}(\tau,\vec{\sigma})&=&
\int d^3\sigma_o\,
\delta^3(\vec{\sigma}-\vec{\Sigma}(\tau,\vec{\sigma}_o))
K_{\check{r}}(\tau,\vec{\sigma}_o),\nonumber\\
 &&\nonumber\\
 \rho_\mu(\tau,\vec{\sigma})l^\mu(\tau,\vec{\sigma})&=& \int
d^3\sigma_o\,
\delta^3(\vec{\sigma}-\vec{\Sigma}(\tau,\vec{\sigma}_o))\times\nonumber\\
 &&\nonumber\\
 &\times& \Big[ \det\left(\frac{\partial\Sigma
(\tau,\vec{\sigma}_o) }{\partial\sigma_o}\right)
\sqrt{\gamma(\Sigma)} \,\left(
\rho(\hat{n},s)-\frac{\partial\rho}{\partial\hat{n}}\hat{n}
\right)+\nonumber \\
 &&{}\nonumber \\
 &+& \frac{n^2_o(\vec{\sigma}_o)}{\sqrt{\gamma(\Sigma)}
\det\left(\frac{\partial\Sigma}{\partial\sigma_o}\right)}
\,\frac{1}{\hat{n}}\,\frac{\partial\rho}{\partial\hat{n}}\,
\Big](\tau ,{\vec \sigma}_o).\nonumber \\
 &&{}
 \label{4.6}
 \end{eqnarray}

\medskip

The first expression is already in a constraint form. In the
right-hand side of the second one the only dependence on the
velocities is in the density $\hat{n}(\tau ,{\vec \sigma}_o)$ [see
the definition (\ref{3.11})]. Nevertheless we observe that from
the definition (\ref{4.1}) we obtain

\begin{eqnarray}
\gamma^{\check{r}\check{s}}(\Sigma)K_{\check{r}}(\tau,\vec{\sigma}_o)
K_{\check{s}}(\tau,\vec{\sigma}_o) &=&
\left(n_o(\vec{\sigma}_o)\,\frac{\partial\rho}{\partial\hat{n}}(\hat{n},s)\right)^2
\nonumber \\
 &&{}\nonumber \\
 &&\left[-\frac{n_o^2(\vec{\sigma}_o)}{\hat{n}^2(\tau ,{\vec \sigma}_o)} \left( \frac{1}
{\det\left(\frac{\partial\Sigma}{\partial\sigma_o}\right)
\sqrt{\gamma(\Sigma)}}\right)^2+1 \right].
 \label{4.7}
 \end{eqnarray}

Then we can replace $\hat{n}(\tau ,{\vec \sigma}_o)$ with the
(implicit) solution $X(\tau ,{\vec \sigma}_o)$ of the equation

\begin{eqnarray}
\gamma^{\check{r}\check{s}}(\Sigma)K_{\check{r}}(\tau,\vec{\sigma}_o)
K_{\check{s}}(\tau,\vec{\sigma}_o) &=&
\left(n_o(\vec{\sigma}_o)\,\frac{\partial\rho}{\partial X}(X(\tau
,{\vec \sigma}_o),s)\right)^2\nonumber \\
 &&{}\nonumber \\
 &&\left[-\frac{n_o^2(\vec{\sigma}_o)}{X^2(\tau ,{\vec \sigma}_o)} \left(
\frac{1}{\det\left(\frac{\partial\Sigma}{\partial\sigma_o}\right)
\sqrt{\gamma(\Sigma)}}\right)^2+1 \right].
 \label{4.8}
 \end{eqnarray}

It is evident by inspection that $X(\tau ,{\vec \sigma}_o)$ is a
function of the canonical variables alone. In other words $X(\tau
,{\vec \sigma}_o)$ is independent from the $\tau$-derivative of
the $z^{\mu}$'s and $\Sigma$'s. It depends on the initial particle
density $n_o({\vec \sigma}_o)$, on the Eulerian coordinates
through ${{n_o({\vec \sigma}_o)}\over
{\det\left(\frac{\partial\Sigma}{\partial\sigma_o}\right)
\sqrt{\gamma(\Sigma)} }}$ and on the Eulerian momenta through
$\gamma^{\check{r}\check{s}}(\Sigma)K_{\check{r}}(\tau,\vec{\sigma}_o)
K_{\check{s}}(\tau,\vec{\sigma}_o)$.

\bigskip

Then we have the following Dirac constraints:

\begin{eqnarray}
{\cal H}_{\check r}(\tau,\vec{\sigma})&=&
\rho_\mu(\tau,\vec{\sigma})z^\mu_{\check r}(\tau,\vec{\sigma})-
\int d^3\sigma_o\cdot
\delta^3(\vec{\sigma}-\vec{\Sigma}(\tau,\vec{\sigma}_o)) K_{\check
r}(\tau,\vec{\sigma}_o)\approx 0,\nonumber\\
 &&\nonumber\\
  {\cal H}_\perp(\tau,\vec{\sigma})&=&\rho_\mu(\tau,\vec{\sigma})l^\mu(\tau,\vec{\sigma})-
\int d^3\sigma_o\cdot
\delta^3(\vec{\sigma}-\vec{\Sigma}(\tau,\vec{\sigma}_o))\times\nonumber\\
 &&\nonumber\\
 && \left[
\det\left(\frac{\partial\Sigma}{\partial\sigma_o}\right)
\sqrt{\gamma(\Sigma)} \,\left(
\rho(X,s)-\frac{\partial\rho}{\partial X}\,X \right)+
\frac{n^2_o(\vec{\sigma}_o)}{\sqrt{\gamma(\Sigma)}
\det\left(\frac{\partial\Sigma}{\partial\sigma_o}\right)}
\,\frac{1}{X}\,\frac{\partial\rho}{\partial X}\, \right](\tau
,{\vec \sigma}_o)\approx 0.\nonumber \\
 &&{}
 \label{4.9}
 \end{eqnarray}

\medskip

As shown in Appendix B, following Ref.\cite{6}, only in few cases,
including the dust, the photon gas and a polytropic with $n =
{1\over 2}$ or $\gamma = 1 + {1\over n} = 3$, a closed form of
Eq.(\ref{4.9}) can be obtained.

Working on the implicit definition of $X(\tau ,{\vec \sigma}_o)$
given by  Eq.(\ref{4.8}) and using the results of Appendix C, it
is possible to prove that the previous constraints satisfy the
Dirac algebra

\begin{eqnarray}
\{{\cal H}_{\check{r}}(\tau,\vec{\sigma}),{\cal
H}_{\check{s}}(\tau,\vec{\sigma}')\}&=& {\cal
H}_{\check{r}}(\tau,\vec{\sigma}')
\frac{\partial}{\partial\sigma^{\prime\,\check{s}}}\,
\delta(\vec{\sigma}-\vec{\sigma}')- {\cal
H}_{\check{s}}(\tau,\vec{\sigma})
\frac{\partial}{\partial\sigma^{\check{r}}}\,
\delta^3(\vec{\sigma}-\vec{\sigma}'),\nonumber\\
 &&\nonumber\\
 \{{\cal H}_\perp(\tau,\vec{\sigma}),{\cal
H}_\perp(\tau,\vec{\sigma}')\}&=& \left[{\cal
H}_{\check{r}}(\tau,\vec{\sigma})
\gamma^{\check{r}\check{s}}(\tau,\vec{\sigma})+ {\cal
H}_{\check{r}}(\tau,\vec{\sigma}')
\gamma^{\check{r}\check{s}}(\tau,\vec{\sigma}')\right]\,
\frac{\partial}{\partial\sigma^{\check{s}}}\,\delta^3(\vec{\sigma}-\vec{\sigma}'),\nonumber\\
 &&\nonumber\\
  \{{\cal H}_\perp(\tau,\vec{\sigma}),{\cal
H}_{\check{r}}(\tau,\vec{\sigma}')\}&=& {\cal
H}_\perp(\tau,\vec{\sigma}')
\frac{\partial}{\partial\sigma^{\prime\,\check{r}}}\,
\delta^3(\vec{\sigma}-\vec{\sigma}'),
 \label{4.10}
 \end{eqnarray}

\noindent so that they are first class constraint.

It is convenient to rewrite the constraints in the form

\begin{eqnarray}
{\cal H}^\mu(\tau,\vec{\sigma}) &=& {\cal
H}_\perp(\tau,\vec{\sigma}) l^\mu(\tau,\vec{\sigma})+{\cal
H}_{\check{r}}(\tau,\vec{\sigma})
\gamma^{\check{r}\check{s}}(\tau,\vec{\sigma})
z^\mu_{\check{s}}(\tau,\vec{\sigma})=\nonumber \\
 &&{}\nonumber \\
 &=& \rho^{\mu}(\tau ,\vec \sigma ) - l^{\mu}(\tau ,\vec \sigma
 )\, \int d^3\sigma_o\cdot
\delta^3(\vec{\sigma}-\vec{\Sigma}(\tau,\vec{\sigma}_o))\times\nonumber\\
 &&\nonumber\\
 && \left[
\det\left(\frac{\partial\Sigma}{\partial\sigma_o}\right)
\sqrt{\gamma(\Sigma)} \,\left(
\rho(X,s)-\frac{\partial\rho}{\partial X}\,X \right)+
\frac{n^2_o(\vec{\sigma}_o)}{\sqrt{\gamma(\Sigma)}
\det\left(\frac{\partial\Sigma}{\partial\sigma_o}\right)}
\,\frac{1}{X}\,\frac{\partial\rho}{\partial X}\, \right](\tau
,{\vec \sigma}_o) -\nonumber \\
 &&{}\nonumber \\
 &-& z^{\mu}_{\check r}(\tau ,\vec \sigma )\,
\int d^3\sigma_o\cdot
\delta^3(\vec{\sigma}-\vec{\Sigma}(\tau,\vec{\sigma}_o)) K_{\check
r}(\tau,\vec{\sigma}_o)\approx 0,
 \label{4.11}
 \end{eqnarray}

\noindent because then we get

\begin{equation}
\{{\cal H}^\mu(\tau,\vec{\sigma}), {\cal
H}^\nu(\tau,\vec{\sigma}')\}=0.
 \label{4.12}
 \end{equation}

The Hamiltonian gauge transformations generated by these
constraints change the form and the coordinatization of the
space-like hyper-surfaces $\Sigma_{\tau}$. Therefore the
embeddings $z^{\mu}(\tau ,\vec \sigma )$ are the {\it gauge
variables} of this special relativistic general covariance
according to which the description of isolated systems (here the
perfect fluid) does not depend from the choice of the 3+1
splitting of Minkowski space-time.

\medskip

The Dirac Hamiltonian is $H_D = \int d^3\sigma\,
\lambda_{\mu}(\tau ,\vec \sigma )\, {\cal H}^{\mu}(\tau ,\vec
\sigma )$, where the $\lambda_{\mu}$'s are arbitrary Dirac
multipliers. Since only the embedding carries Lorentz indices, the
generators of the Poincare' group are $p_{\mu} = \int d^3\sigma\,
\rho_{\mu}(\tau ,\vec \sigma )$ and $J^{\mu\nu} = \int d^3\sigma\,
\Big( z^{\mu}\, \rho^{\nu} - z^{\nu}\, \rho^{\mu} \Big)(\tau ,\vec
\sigma )$.

\subsection{The Restriction to Space-Like Hyper-Planes.}

Following Refs.\cite{8,10,3} let us restrict ourselves to
foliations whose leaves are space-like hyper-planes by adding the
gauge-fixings:

\begin{equation}
\zeta^\mu(\tau,\vec{\sigma})=z^\mu(\tau,\vec{\sigma})-
x^\mu(\tau)-b^\mu_{\check{r}}(\tau)\sigma^{\check{r}}\approx 0
 \label{4.13}
 \end{equation}

With this condition many geometrical quantities take a trivial
expression, in particular we have:

\begin{eqnarray}
z^\mu_{\check{r}}(\tau,\vec{\sigma})&\approx&b^\mu_{\check{r}}(\tau),\nonumber
\\
 z^\mu_{\check{r}}(\tau,\vec{\sigma})&\approx&\dot{x}^\mu(\tau)+
\dot{b}^\mu_{\check{r}}(\tau)\sigma^{\check{r}},\nonumber \\
 g_{\check{r}\check{s}}(\tau,\vec{\sigma})&\approx&-\delta_{\check{r}\check{s}},\;\;
\gamma^{\check{r}\check{s}}(\tau,\vec{\sigma})\approx-\delta^{\check{r}\check{s}},\;\;
\gamma(\tau,\vec{\sigma})\approx 1.
 \label{4.14}
 \end{eqnarray}

We also introduce the following notation for the unit normal to
the hyper-planes defined in Eq.(\ref{a.3})

\begin{equation}
b^\mu_\tau(\tau)=l^\mu(\tau) \approx l^\mu(\tau,\vec{\sigma})
 \label{4.15}
 \end{equation}

The hyper-planes define a {\em true global foliation} only if the
normal $l_\mu$ is $\tau$-independent, because only in this case
the hyper-planes are parallel and not intersecting and there is a
one to one global correspondences between points $z^\mu$ and
coordinates $\tau,\vec{\sigma}$. We ignore from now on this
observation, but we shall return on the consequences of the
$l^\mu=constant$ request at the end of this Section.

\medskip

Since we have

\begin{equation}
\{\zeta^\mu(\tau,\vec{\sigma}),{\cal H}_\nu(\tau,\vec{\sigma}')\}=
-\eta^\mu_\nu\delta^3(\vec{\sigma}-\vec{\sigma}'),
 \label{4.16}
 \end{equation}

\noindent it is possible to define the following Dirac brackets

\begin{eqnarray}
&&\{A,B\}^*= \{A,B\}-\int d^3\sigma\cdot\left[
\{A,\zeta^\mu(\tau,\vec{\sigma})\}\{{\cal
H}_\mu(\tau,\vec{\sigma}),B\}- \{A,{\cal
H}^\mu(\tau,\vec{\sigma})\}\{\zeta_\mu(\tau,\vec{\sigma}),B\}
\right].\nonumber \\
 &&{}\nonumber \\
 \label{4.17}
 \end{eqnarray}

In the reduced phase space the hyper-surface canonical variables
$z^{\mu}(\tau ,\vec \sigma )$, $\rho_{\mu}(\tau ,\vec \sigma )$
are reduced to only ten canonical pairs and the first class
constraints (\ref{4.11}) are reduced to only ten \cite{8,6}.

\medskip

We can verify that four pairs of canonical variables are the
centroid $x^\mu(\tau)$, used as origin of the 3-coordinates on the
hyper-planes, and the conjugate momentum $p^\mu(\tau)$

\begin{equation}
\{x^\mu(\tau),p^\nu(\tau)\}^*=-\eta^{\mu\nu}.
 \label{4.18}
 \end{equation}

Since $p^{\mu}$ is the Poincare' generator of the translations, it
describes the total 4-momentum of the system. As a consequence we
have the following decomposition for the canonical generators of
Lorentz transformations inside the Poincare' algebra

 \begin{equation}
J^{\mu\nu}(\tau)=x^\mu(\tau)p^\nu(\tau)-x^\nu(\tau)p^\mu(\tau)
+S^{\mu\nu}(\tau).
 \label{4.19}
 \end{equation}

The remaining canonical variables defining the hyper-planes are
the variables $\phi_\lambda(\tau)$ ( $\lambda=1,...,6$) that
parametrize the orthonormal tetrad $b^\mu_{\check{A}}(\tau)$ such
that

\begin{equation}
b^\mu_{\check{A}}(\tau)\eta_{\mu\nu}b^\nu_{\check{B}}(\tau)=
\eta_{\check{A}\check{B}},
 \label{4.20}
 \end{equation}

\noindent and the associated conjugate variables
$T_\lambda(\tau)$. Nevertheless it is possible \cite{8} to use a
set of redundant, non independent and non canonical variables.
These are the tetrads $b^\mu_{\check{A}}(\tau)$ and the spin
tensor $S^{\mu\nu}(\tau)$ of Eq.(\ref{4.19}), if they satisfy the
following Poisson brackets \footnote{They are the Dirac brackets
enforcing the fulfillment  \cite{24} of Eqs.(\ref{4.20}). }

\begin{eqnarray}
\{S^{\mu\nu}(\tau),b^\rho_{\check{A}}(\tau)\}^*&=&
\eta^{\rho\nu}b^\mu_{\check{A}}(\tau)-\eta^{\rho\mu}b^\nu_{\check{A}}(\tau),
\nonumber\\ \{S^{\mu\nu}(\tau),S^{\rho\sigma}(\tau)\}^*&=&
C^{\mu\nu\rho\sigma}_{\alpha\beta}S^{\alpha\beta}(\tau),
 \label{4.21}
 \end{eqnarray}

\noindent with the Lorentz group constant structures:

\begin{equation}
C^{\mu\nu\rho\sigma}_{\alpha\beta}=
\eta^\nu_\alpha\eta^\rho_\beta\eta^{\mu\sigma}-
\eta^\mu_\alpha\eta^\rho_\beta\eta^{\nu\sigma}-
\eta^\nu_\alpha\eta^\sigma_\beta\eta^{\mu\rho}+
\eta^\mu_\alpha\eta^\sigma_\beta\eta^{\nu\rho}.
 \label{4.22}
 \end{equation}

Finally we have the unchanged fluids variables:

\begin{equation}
\{\Sigma^{\check{r}}(\tau,\vec{\sigma}_o),
K^{\check{s}}(\tau,\vec{\sigma}'_o)\}^*=
\delta(\vec{\sigma}_o-\vec{\sigma}'_o)\,
\delta^{\check{r}\check{s}}.
 \label{4.23}
 \end{equation}

\medskip

In conclusion this non-Darboux canonical basis of the reduced
phase space can be represented in the table:

\begin{equation}
\begin{array}{|cc|c|}
\hline
x^\mu(\tau)&S^{\mu\nu}(\tau)&\Sigma^{\check{r}}(\tau,\vec{\sigma}_o)\\
p^\mu(\tau)&b^\mu_{\check{A}}(\tau)&K^{\check{r}}(\tau,\vec{\sigma}_o)\\
\hline
\end{array}.
 \label{4.24}
 \end{equation}

The stability in time of the gauge fixings (\ref{4.13}) force the
arbitrary Dirac multiplier to take the reduced form:

\begin{equation}
\frac{\partial}{\partial\tau}\zeta(\tau,\vec{\sigma})\approx 0
\Rightarrow \lambda^\mu(\tau,\vec{\sigma})\approx
\lambda^\mu(\tau)+\lambda^{\mu\nu}(\tau)
\eta_{\nu\rho}b^\rho_{\check{r}}(\tau)\sigma^{\check{r}}.
 \label{4.25}
 \end{equation}

\medskip

The Dirac Hamiltonian become

\begin{equation}
H_D=\lambda^\mu(\tau)H_\mu(\tau)+\lambda^{\mu\nu}(\tau)H_{\mu\nu}(\tau),
 \label{4.26}
 \end{equation}

\noindent where, if we define

\begin{eqnarray}
{\cal M}(\tau)&=&\int d^3\sigma_o\, \left[
\det\left(\frac{\partial\Sigma}{\partial\sigma_o}\right) \,\left(
\rho(X,s)-\frac{\partial\rho}{\partial X}\,X \right)+
\frac{n^2_o(\vec{\sigma}_o)}{
\det\left(\frac{\partial\Sigma}{\partial\sigma_o}\right)}
\,\frac{1}{X}\,\frac{\partial\rho}{\partial X}\, \right](\tau
,{\vec \sigma}_o), \nonumber\\
 &&\nonumber\\
  {\cal P}^{\check{r}}(\tau)&=&\int
d^3\sigma_o\, K^{\check{r}}(\tau,\vec{\sigma}_o), \nonumber\\
 &&\nonumber\\
  {\cal J}^{\tau\check{r}}(\tau)&=&{\cal
K}^{\check{r}}(\tau)=-\int d^3\sigma_o\,
\Sigma^{\check{r}}(\tau,\vec{\sigma}_o)\times\nonumber\\
 &&\nonumber\\
  &\times& \left[
\det\left(\frac{\partial\Sigma}{\partial\sigma_o}\right) \,\left(
\rho(X,s)-\frac{\partial\rho}{\partial X}\,X \right)+
\frac{n^2_o(\vec{\sigma}_o)}{
\det\left(\frac{\partial\Sigma}{\partial\sigma_o}\right)}
\,\frac{1}{X}\,\frac{\partial\rho}{\partial X}\, \right](\tau
,{\vec \sigma}_o), \nonumber\\
 &&\nonumber\\
  {\cal J}^{\check{r}\check{s}}(\tau)&=&\int d^3\sigma_o\,\left[
\Sigma^{\check{r}}(\tau,\vec{\sigma}_o)K^{\check{s}}(\tau,\vec{\sigma}_o)-
\Sigma^{\check{s}}(\tau,\vec{\sigma}_o)K^{\check{r}}(\tau,\vec{\sigma}_o)\right],
 \label{4.27}
 \end{eqnarray}

\medskip

\noindent we have the following form of the ten remaining first
class constraints

 \begin{eqnarray}
H^\mu(\tau)&=& p^\mu(\tau)-l^\mu(\tau){\cal M}(\tau)
+b^\mu_{\check{r}}(\tau){\cal P}^{\check{r}}(\tau)\approx
0,\nonumber\\
 &&\nonumber\\
  H^{\mu\nu}(\tau)&=&S^{\mu\nu}(\tau) +
\left[b^\mu_{\check{r}}(\tau)l^\nu(\tau)-b^\nu_{\check{r}}(\tau)l^\mu(\tau)\right]
{\cal J}^{\tau\check{r}}(\tau)+\nonumber\\
 &&\nonumber\\
 &-& {1\over 2}
\left[b^\mu_{\check{r}}(\tau)b^\nu_{\check{s}}(\tau)-b^\mu_{\check{s}}(\tau)b^\mu_{\check{r}}(\tau)\right]
{\cal J}^{\check{r}\check{s}}(\tau)\approx 0.
 \label{4.28}
 \end{eqnarray}

\noindent They satisfy the Poisson algebra

\begin{eqnarray}
\{H^\mu(\tau),H^\nu(\tau)\}&=&\{H^\mu(\tau),H^{\alpha\beta}(\tau)\}=0,
\nonumber\\
 &&\nonumber\\
 \{H^{\mu\nu}(\tau),H^{\rho\sigma}(\tau)\}&=&
C^{\mu\nu\rho\sigma}_{\alpha\beta}H^{\alpha\beta}(\tau)\approx 0.
 \label{4.29}
 \end{eqnarray}

These ten first class constraints imply that $x^{\mu}(\tau )$,
$b^{\mu}_{\check r}(\tau )$ are {\it gauge variables} and that the
description is independent from the choice of the space-like
hyper-planes.

\medskip

We can discuss now  the condition $l_\mu=constant$.

By using the brackets (\ref{4.21}), the Hamiltonian (\ref{4.20})
implies the following equations of motion for the normal
$l_\mu(\tau)$

\begin{equation}
\frac{d}{d\tau}l_\mu(\tau)=2\lambda_{\mu\nu}(\tau)l^\nu(\tau).
 \label{4.30}
 \end{equation}

Then the condition $l_\mu=constant$ restricts the arbitrariness of
the $\lambda_{\mu\nu}(\tau)$'s with the condition

\begin{equation}
\lambda_{\mu\nu}(\tau)l^\nu=0,\quad \Rightarrow
\lambda^{\mu\nu}(\tau ) = b^{\mu}_{\check r}(\tau )\,
b^{\nu}_{\check s}(\tau )\, \lambda^{\check r\check s}(\tau
),\quad  \lambda^{\check r\check s}(\tau ) =-  \lambda^{\check
s\check r}(\tau ).
 \label{4.31}
 \end{equation}

In other words the condition $l_\mu=constant$ have to be
interpreted as 3 gauge fixing conditions for the Lorentz
transformations generated by the constraints $H^{\mu\nu}(\tau )
\approx 0$, leaving only the possibility of Hamiltonian gauge
rotations generated by $H_{\check r\check s}(\tau ) =
b^{\mu}_{\check r}(\tau )\, b^{\nu}_{\check s}(\tau )\,
H_{\mu\nu}(\tau )$. This implies that a foliation with space-like
hyper-planes has 7 gauge variables: $x^{\mu}(\tau )$ and 3 angles
inside $b^{\mu}_{\check r}(\tau )$ describing the linear
acceleration of the origin and the arbitrary rotation of the
spatial axes with respect to an inertial frame, respectively.
Correspondingly only 7 of the first class constraints (\ref{4.28})
are independent, namely $H^{\mu}(\tau )$ and $H_{\check r\check
s}(\tau )$.

\vfill\eject

\section{The rest frame instant form.}

In this Section, following Ref. \cite{8}, we define a  new instant
form of dynamics \cite{24} called the {\em rest frame instant
form}. This is done by selecting all the configurations of the
isolated system with time-like conserved total 4-momentum and, for
each of them, by choosing the foliation whose space-like
hyper-planes are orthogonal to this 4-momentum. Physically these
hyper-planes, named {\it Wigner hyper-planes}, correspond to the
intrinsic rest frame of the configuration of the isolated system.
Moreover we have to fix the four acceleration degrees of freedom
of the centroid $x^{\mu}(\tau )$, reducing it to the world-line of
an inertial observer.

\medskip

Before doing this we have to recall the notion of {\em standard
Wigner boost} and {\em Wigner rotation}. It is known that in the
rest frame a timelike four-vector $p^\mu$ assume the standard form
$\bar{p}^\mu=(\sqrt{p^2},\vec{0})$. In the general theory of the
induced representation of the Poincar\'e group a {\em standard
Wigner boost} $L^\mu_{\;\;\nu}(p, \bar p)$ for time-like Poincare'
orbits is defined such that

\begin{equation}
L^\mu_{\;\;\nu}(p, \bar p)\, \bar{p}^\nu=p^\mu.
 \label{5.1}
 \end{equation}

We can use the rows of the {\em standard Wigner boost} matrix to
define a orthonormal tetrad (we define $n^\mu=p^\mu/\sqrt{p^2}$)

\begin{eqnarray}
\epsilon^\mu_o(p)&=&L^\mu_{\;\;o}(p, \bar p)=n^\mu,\nonumber\\
\epsilon^\mu_r(p)&=&L^\mu_{\;\;r}(p, \bar p )=\Big(-n_r;
\delta^i_r-n^in_r(1+n_o)^{1/2}\Big).
 \label{5.2}
 \end{eqnarray}

We use the notation

\begin{equation}
\epsilon^\mu_A(p)=(\epsilon^\mu_o(p),\epsilon^\mu_r(p)).
 \label{5.3}
 \end{equation}

The inverse of standard boost defines naturally the inverse tetrad

\begin{equation}
\epsilon^A_\mu(p)=L_\mu^{\;\;A}(p, \bar p),
 \label{5.4}
 \end{equation}

\noindent such that

\begin{eqnarray}
\epsilon^o_\mu(p)&=&n_\mu = {{p_{\mu}}\over
{\sqrt{p^2}}},\nonumber\\
 \epsilon^r_\mu(p)&=&\Big(\delta^{rs}
u_s; \delta^r_s-\delta^{rs}\delta_{jh} n^hn_s(1+n_o)^{1/2}\Big),
 \label{5.5}
 \end{eqnarray}

\noindent and

\begin{eqnarray}
\epsilon^A_\mu(p)\epsilon^\nu_A(p)&=&\eta^\nu_\mu,\qquad
\epsilon^A_\mu(p)\epsilon^\mu_B(p)=\eta^A_B,\nonumber\\
 &&{}\nonumber \\
 \eta^{\mu\nu}&=& n^\mu
n^\nu-\sum_r\epsilon^\mu_r(p)\epsilon^\nu_r(p),\qquad
\eta_{AB}=\epsilon^\mu_A(p)\eta_{\mu\nu}\epsilon^\nu_B(p).
 \label{5.6}
 \end{eqnarray}

Moreover we can verify that

\begin{eqnarray}
p_\mu\epsilon^\mu_r(p)&=&p^\mu\epsilon^r_\mu(p)=0,\nonumber\\
 &&\nonumber\\
  p^\mu\frac{\partial}{\partial
p^\mu}\epsilon^A_\nu(p)&=& p^\mu\frac{\partial}{\partial
p^\mu}\epsilon^\nu_A(p)=0.
 \label{5.7}
 \end{eqnarray}

The standard boost can be used to define the {\em Wigner rotation}
$R(\Lambda,p)$ associated to any Lorentz transformation $\Lambda$

\begin{equation}
L^{-1}(p, \bar p)\Lambda^{-1}L(\Lambda p, \bar p)=\left(
\begin{array}{cc}
1&0\\
&\\
0&R(\Lambda,p)
\end{array}
\right).
 \label{5.8}
 \end{equation}

Then we have the following property

\begin{equation}
\epsilon^\mu_r(\Lambda p)=\Lambda^\mu_{\;\;\nu}\epsilon^\nu_s(p)
R^s_{\;\; r}(\Lambda,p).
 \label{5.9}
 \end{equation}

\bigskip

We can use the tetrad (\ref{5.3}) to define a gauge fixing for the
$b^\mu_{\check{A}}(\tau)$. Before doing this, it is useful to
replace the centroid $x^\mu(\tau)$ with the new one

\begin{equation}
q^\mu(\tau)=x^\mu(\tau)+\frac{1}{2}\epsilon^A_\rho(p)\eta_{AB}
\frac{\partial\epsilon^B_\sigma(p)}{\partial
p_\mu}S^{\rho\sigma}(\tau),
 \label{5.10}
 \end{equation}

\noindent and the table (\ref{4.24}) with the following set of non
canonical variables

\begin{equation}
\begin{array}{|cc|c|}
\hline
q^\mu(\tau)&S^{\mu\nu}(\tau)&
\Sigma^{\check{r}}(\tau,\vec{\sigma}_o)\\
p^\mu(\tau)&b^\mu_{\check{A}}(\tau)&
K^{\check{r}}(\tau,\vec{\sigma}_o)\\
\hline
\end{array}.
 \label{5.11}
 \end{equation}

\medskip

The $q^\mu(\tau),p^\mu(\tau)$ and $\Sigma^{\check{r}}(\tau,
\vec{\sigma}_o), K^{\check{r}}(\tau,\vec{\sigma}_o)$ are still
canonical

\begin{eqnarray}
\{q^\mu(\tau ) ,q^\nu(\tau ) \} = 0,\qquad \{q^\mu(\tau )
,p^\nu(\tau ) \} = - \eta^{\mu\nu},\nonumber\\
 \{q^\mu(\tau ), \Sigma^{\check{r}}(\tau,\vec{\sigma}_o)\} = 0, \qquad
\{q^\mu(\tau ), K^{\check{r}}(\tau,\vec{\sigma}_o)\} = 0,
 \label{5.12}
 \end{eqnarray}

\noindent but for $S^{\mu\nu}(\tau),b^\mu_{\check A}(\tau)$ we
have the following non canonical brackets

\begin{equation}
\{q^\mu(\tau ), S^{\alpha\beta}(\tau ) \} \neq 0,\qquad
\{q^\mu(\tau ), b^\alpha_{\check{A}}(\tau ) \} \neq 0.
 \label{5.13}
 \end{equation}

Eq.(\ref{5.10}) also implies the following new decomposition for
the canonical generators of Lorentz transformations

\begin{equation}
J^{\mu\nu}(\tau)=q^\mu(\tau)p^\nu(\tau)-q^\nu(\tau)p^\mu(\tau)+
\Omega^{\mu\nu}(\tau),
 \label{5.14}
 \end{equation}

\noindent where

\begin{equation}
\Omega^{\mu\nu}(\tau)=
S^{\mu\nu}-\frac{1}{2}\epsilon^A_\alpha(p)\eta_{AB} \left(
\frac{\partial\epsilon^B_\beta(p)}{\partial p_\mu}p^\nu
-\frac{\partial\epsilon^B_\beta(p)}{\partial p_\nu}p^\mu
\right)S^{\alpha\beta}.
 \label{5.15}
 \end{equation}

Since we have

 \begin{equation}
\{q^\mu,\Omega^{hk}\}=0,\qquad
 \{q^\mu,\Omega^{ok}\}\neq0,
 \label{5.16}
 \end{equation}

\noindent we see that $q^\mu$ is not a true four-vector.

\hfill

If we restrict ourselves to configurations with $p^\mu(\tau)$
time-like ($p^2(\tau)>0$), there exists a family of space-like
hyper-planes orthogonal to $p^\mu(\tau)$ and we can select this
family with the gauge fixing \cite{8}

\begin{equation}
T^\mu_A(\tau)=b^\mu_{\check{A}=A}(\tau)-\epsilon^\mu_A(p)\approx
0,
 \label{5.17}
 \end{equation}

\noindent where the index $\check{r}$ is enforced to coincide with
$r$ with transformation property given by Eq.(\ref{5.9}). These
gauge fixings imply $\lambda_{\mu\nu}(\tau) \approx 0$
\footnote{So that Eqs.(\ref{4.31}) are satisfied.} and we have the
Dirac Hamiltonian

\begin{equation}
H_D=\lambda^\mu(\tau)H_\mu(\tau),
 \label{5.18}
 \end{equation}

\noindent where

\begin{equation}
H^\mu(\tau)= p^\mu-n^\mu{\cal M}(\tau) +\epsilon^\mu_r(p){\cal
P}^r(\tau)\approx 0,
 \label{5.19}
 \end{equation}

\noindent are the remaining four first class Dirac constraints
saying that $x^{\mu}(\tau )$ and terefore $q^{\mu}(\tau )$ are
gauge variables. Since we have ${\dot x}^{\mu}(\tau )\, {\buildrel
\circ \over =}\, \{ x^{\mu}(\tau ), H_D \} = - \lambda^{\mu}(\tau
)$, we see that the centroid has an arbitrary (gauge) acceleration
described by the remaining Dirac multipliers. It is useful to
observe that we can rewrite the constraints and the Dirac
Hamiltonian in the form (we choose the positive sheet of the mass
hyperboloid)

\begin{eqnarray}
{\cal H}(\tau)&=&n^\mu H_\mu(\tau)= \sqrt{p^2}-{\cal
M}(\tau)\approx 0,\nonumber\\
 &&\nonumber\\
  {\cal P}^r(\tau)&=&\epsilon^r_\mu(p)H^\mu(\tau) \approx
  0,\nonumber \\
  &&{}\nonumber \\
  H_D &=& \lambda (\tau )\, {\cal H} - \vec \lambda (\tau ) \cdot
  {\vec {\cal P}},
   \label{5.20}
 \end{eqnarray}

\noindent where, recalling the definitions (\ref{4.27}), the
quantity

\begin{eqnarray}
{\cal M}(\tau)&=&\int d^3\sigma_o\, \left[
\det\left(\frac{\partial\Sigma}{\partial\sigma_o}\right) \,\left(
\rho(X,s)-\frac{\partial\rho}{\partial X}\,X \right)+
\frac{n^2_o(\vec{\sigma}_o)}{
\det\left(\frac{\partial\Sigma}{\partial\sigma_o}\right)}
\,\frac{1}{X}\,\frac{\partial\rho}{\partial X}\, \right](\tau
,{\vec \sigma}_o) =\nonumber \\
 &&\nonumber\\
 &{\buildrel {def}\over =}& \int d^3\sigma_o\, \Delta (\tau
 ,{\vec \sigma}_o)
 \label{5.21}
 \end{eqnarray}

\noindent is the {\em invariant mass} and where

\begin{equation}
\vec{\cal P}(\tau)=\int d^3\sigma_o\,
\vec{K}(\tau,\vec{\sigma}_o)\approx 0,
 \label{5.22}
 \end{equation}

\noindent is a constraint on the total momentum of the fluid
inside the Wigner hyperplane: it gives the {\it rest-frame
condition}.

See Appendix B for the expression of the invariant mass ${\cal M}$
for the dust, the photon gas and the polytropic with $n = {1\over
2}$.

\hfill

After this gauge fixing the non canonical variables $S^{\mu\nu}$
can be everywhere substituted with the expression implied for them
by the constraints $H^{\mu\nu}(\tau ) \approx 0$ of
Eqs.(\ref{4.28}). By defining

\begin{eqnarray}
{\cal J}^{rs}(\tau)&=&\int d^3\sigma_o\,\left[
\Sigma^{r}(\tau,\vec{\sigma}_o)K^{s}(\tau,\vec{\sigma}_o)-
\Sigma^{s}(\tau,\vec{\sigma}_o)K^{r}(\tau,\vec{\sigma}_o)\right],
\nonumber\\
 &&\nonumber\\
  {\cal J}^{\tau r}(\tau)&=& {\cal K}^r = - \int
d^3\sigma_o\, \Sigma^{r}(\tau,\vec{\sigma}_o)\times\nonumber\\
 &&\nonumber\\
 && \left[
\det\left(\frac{\partial\Sigma}{\partial\sigma_o}\right)
\sqrt{\gamma(\Sigma)} \,\left(
\rho(X,s)-\frac{\partial\rho}{\partial X}\,X \right)+
\frac{n^2_o(\vec{\sigma}_o)}{\sqrt{\gamma(\Sigma)}
\det\left(\frac{\partial\Sigma}{\partial\sigma_o}\right)}
\,\frac{1}{X}\,\frac{\partial\rho}{\partial X}\, \right](\tau
,{\vec \sigma}_o) =\nonumber \\
 &&\nonumber\\
 &=& - \int d^3\sigma_o\, \Sigma^r(\tau ,{\vec \sigma}_o)\,
 \Delta (\tau ,{\vec \sigma}_o),
 \label{5.23}
 \end{eqnarray}

\noindent we get

\begin{equation}
S^{\mu\nu}(\tau)\approx \epsilon^\mu_A(p)\epsilon^\nu_B(p){\cal
J}^{AB}(\tau).
 \label{5.24}
 \end{equation}

\hfill

After the elimination of the variables $b^{\mu}_{\check A}$,
$S^{\mu\nu}$ with Eqs. (\ref{4.28}) and (\ref{5.17}), the reduced
phase space is spanned by the variables

\begin{equation}
\begin{array}{|c|c|}
\hline
q^\mu(\tau)&\Sigma^r(\tau,\vec{\sigma}_o)\\
p^\mu(\tau)&K^r(\tau,\vec{\sigma}_o)\\
\hline
\end{array}.
 \label{5.25}
 \end{equation}

\medskip

Its canonical structure is defined by the new Dirac brackets

\begin{eqnarray}
&&\{A,B\}^{**}= \{A,B\}^*-\left[
\{A,H^{\mu\nu}\}^*D^A_{\mu\nu\rho}(p)\{T^\rho_A,B\}^*+
\{A,T^\mu_A\}^*D^A_{\mu\nu\rho}(p)\{H^{\nu\rho},B\}^*\right],
 \label{5.26}
 \end{eqnarray}

\noindent where

\begin{equation}
D^A_{\mu\nu\rho}(p)=\frac{1}{4}[
\eta_{\mu\rho}\epsilon^A_\nu(p)-\eta_{\nu\rho}\epsilon^A_\mu(p)].
 \label{5.27}
 \end{equation}

The variables (\ref{5.25}) are canonical with respect to this
bracket

\begin{eqnarray}
\{\Sigma^r(\tau,\vec{\sigma}_o),
K^s(\tau,\vec{\sigma}'_o)\}^{**}&=&\delta^{rs}\,
\delta(\vec{\sigma}_o-\vec{\sigma}'_o),\nonumber\\
 \{q^\mu,q^\nu\}^{**}&=& 0,   \nonumber\\
 \{q^\mu,p^\nu\}^{**}&=&-\eta^{\mu\nu},\nonumber\\
 \{q^\mu,\Sigma^r(\tau,\vec{\sigma}_o)\}^{**}&=& 0,  \nonumber\\
 \{q^\mu,K^s(\tau,\vec{\sigma}_o)\}^{**}&=& 0.
 \label{5.28}
 \end{eqnarray}

On the contrary, with the new brackets the old centroid $x^\mu$ is
not canonical, because we have

 \begin{equation}
\{x^\mu,\Sigma^r(\tau,\vec{\sigma}_o)\}^{**} \neq 0, \qquad
\{x^\mu,K^r(\tau,\vec{\sigma}_o)\}^{**}\neq 0,
 \label{5.29}
 \end{equation}

\noindent and this explains why we introduced the centroid
$q^{\mu}(\tau )$.

\medskip

After the gauge fixing we have

\begin{equation}
\Omega^{ij}(\tau)\approx\delta^{ir}\delta^{js}{\cal J}^{rs}(\tau),
\qquad \Omega^{oi}(\tau)\approx-\frac{\delta^{ir}\delta^{js}p^j}
{p^o+\sqrt{p^2}}{\cal J}^{rs}(\tau),
 \label{5.30}
 \end{equation}

\noindent and the canonical generators of Lorentz transformations
become ($\sqrt{p^2} \approx {\cal M}$)

\begin{eqnarray}
J^{\mu\nu}(\tau)&=&q^\mu(\tau)p^\nu(\tau)-q^\nu(\tau)p^\mu(\tau)+
\nonumber\\
 &&\nonumber\\
 &+&\left[\eta^\mu_i\eta^\nu_j
\delta^{ir}\delta^{js}-(\eta^\mu_o\eta^\nu_j-\eta^\nu_o\eta^\mu_j)
\frac{\delta^{ir}\delta^{js}p^j}{p^o+\sqrt{p^2}}\right] {\cal
J}^{rs}(\tau).
 \label{5.31}
 \end{eqnarray}

\medskip

This form of the canonical generator of Lorentz transformations
tell us that on a function dependent on the fluid variables alone
$F(\vec{\Sigma}(\tau, \vec{\sigma}_o), \vec{K}(\tau,
\vec{\sigma}_o))$ a Lorentz transformation acts as a rotation
inside the hyperplane. This rotations is the Wigner rotation
associated to the infinitesimal Lorentz transformation

\begin{equation}
\delta F=\delta\omega_{\mu\nu}\{F,J^{\mu\nu}\}^{**}=
\delta\varphi^{rs}\{F,{\cal J}^{rs}\}^{**},
 \label{5.32}
 \end{equation}

\noindent where

\begin{equation}
\delta\varphi^{rs}= \left[\eta^\mu_i\eta^\nu_j
\delta^{ir}\delta^{js}-(\eta^\mu_o\eta^\nu_j-\eta^\nu_o\eta^\mu_j)
\frac{\delta^{ir}\delta^{js}p^j}{p^o+\sqrt{p^2}}\right]
\delta\omega_{\mu\nu}.
 \label{5.33}
 \end{equation}

\hfill

The canonical generators of Poincar\'e group $p^\mu(\tau)$ and
$J^{\mu\nu}(\tau)$ are called {\em external} and they realize the
{\em true} Poincar\'e  symmetry in the reduced phase space.
Because of property (\ref{5.32}),  the {\em Lorentz covariance} of
the theory is replaced with the {\em Wigner covariance} of the
3-dimensional instant form variables on the hyper-planes. In this
sense the only canonical variable with a non covariant
transformation property is the pseudo four-vector $q^\mu$. As said
in Ref.\cite{14}, the centroid $q^{\mu}$ is interpreted as a {\it
non-covariant, but canonical external 4-center of mass}. In
particular its spatial components are proportional to a
3-center-of-mass-like position that is the classical analogue of
the Newton-Wigner position operator, whose reduced covariance
corresponds to the little group of time-like Poincare' orbits.

For these particular properties the hyper-planes defined by the
condition (\ref{5.17}) have been called  {\em Wigner hyper-planes}
in \cite{8}.

\hfill

We can also construct another realization of the Poincar\'e Lie
algebra. This realization, called {\em internal}, is constructed
using only the fluid variables $\vec{\Sigma}(\tau,\vec{\sigma}_o),
\vec{K}(\tau,\vec{\sigma}_o)$ living inside the Wigner hyperplane.
In fact, from Eqs. (\ref{5.21}), (\ref{5.22}) and (\ref{5.23}) the
10 functions: $\vec{\cal P},{\cal M},\vec{\cal J},\vec{\cal K}$,
where

\begin{equation}
{\cal J}^r=\frac{1}{2}\epsilon^{ruv}{\cal J}^{uv},\qquad {\cal
K}^r={\cal J}^{\tau r},
 \label{5.34}
 \end{equation}

\noindent are the generators of the following canonical
realization of the Poincar\'e Lie algebra

\begin{eqnarray}
\{{\cal J}^r,{\cal J}^s\}&=&\epsilon^{rsu}{\cal J}^u,\nonumber\\
 \{{\cal K}^r,{\cal K}^s\}&=&-\epsilon^{rsu}{\cal J}^u,\nonumber\\
 \{{\cal K}^r,{\cal J}^s\}&=&\epsilon^{rsu}{\cal K}^u,\nonumber\\
\{{\cal P}^r,{\cal J}^s\}&=&\epsilon^{rsu}{\cal P}^u,\nonumber\\
\{{\cal P}^r,{\cal K}^s\}&=&-{\cal M}\;\delta^{rs},\nonumber\\
\{{\cal M},{\cal K}^s\}&=&-{\cal P}^s,\nonumber\\
 \{{\cal M},{\cal P}^s\}&=&\{{\cal P}^r,{\cal P}^s\}=0.
  \label{5.35}
 \end{eqnarray}

This internal realization is unfaithful due to the constraints
$\vec{\cal P}\approx 0$. These constraints says that three degrees
of freedom, playing the role of an {\it internal} 3-center of mass
of the fluid inside the Wigner hyperplane, are gauge variables. As
shown in Ref. \cite{14}, ${\vec {\cal K}} \approx 0$ are the
natural gauge fixings to eliminate the internal 3-center of mass
and to imply $\vec \lambda (\tau ) = 0$ in the Dirac Hamiltonian
of Eq.(\ref{5.20}). Only the invariant mass ${\cal M}$ and the
rotation canonical generator $\vec{\cal J}$ are non vanishing and
they appear in the external Poincar\'e generators (\ref{5.31}).

With the gauge fixings ${\vec {\cal K}} \approx 0$, the Dirac
Hamiltonian is reduced to $H_D = \lambda (\tau )\, {\cal H}$.

\hfill

To complete the definition of the {\em rest-frame instant form} we
identify the temporal parameter $\tau$ with the common Lorentz
scalar rest frame time of the centroids $q^{\mu}$ and $x^{\mu}$,
$T(\tau)=n_\mu x^\mu(\tau)=n_\mu q^\mu(\tau)$. This is realized by
introducing the gauge fixing

\begin{equation}
T(\tau)\approx \tau,
 \label{5.36}
 \end{equation}

\noindent implying  $\lambda(\tau)=-1$. By using the canonical
transformation

\begin{eqnarray}
 &&T(\tau)=n_\mu q^\mu(\tau),\qquad {\cal E}=\sqrt{p^2},\nonumber\\
 &&\nonumber\\
 &&\vec{z}(\tau)\equiv\sqrt{p^2}\cdot\vec{Q}(\tau) =\sqrt{p^2}\left[
\vec{q}(\tau)-\frac{\vec{p}}{p^o}q^o\right],\qquad
\vec{k}=\frac{\vec{p}}{p^o},
 \label{5.37}
 \end{eqnarray}

\noindent with inverse

 \begin{eqnarray}
&&q^o(\tau)=\sqrt{1+\vec{k}^2}\left(T(\tau)+
\frac{\vec{k}\cdot\vec{z}(\tau)}{\cal E}\right),\quad
\vec{q}(\tau)=\frac{\vec{z}}{\cal E}+\left(T(\tau)+
\frac{\vec{k}\cdot\vec{z}(\tau)}{\cal E}\right)\vec{k},\nonumber\\
 &&\nonumber\\
 &&p^o={\cal E}\sqrt{1+\vec{k}^2},\qquad \vec{p}={\cal E}\vec{k},
 \label{5.38}
 \end{eqnarray}

\medskip

\noindent we arrive at the following Darboux canonical basis
\footnote{ $\vec Q = \vec z / {\cal E}$ is the classical analogue
of the non-covariant Newton-Wigner position operator. However we
cannot replace $\vec z$ and $\vec k$ with $\vec Q$, $\vec p$,
because they do not have vanishing Poisson bracket with the
internal variables $\vec \Sigma (\tau , {\vec \sigma}_o)$ and
$\vec K(\tau , {\vec \sigma}_o)$. }

\begin{equation}
\begin{array}{|c|}
\hline
q^\mu(\tau)\\
p^\mu(\tau)\\
\hline
\end{array}
\longrightarrow
\begin{array}{|c|c|}
\hline
T(\tau)&\vec{z}(\tau)\\
{\cal E}(\tau)&\vec{k}(\tau)\\
\hline
\end{array}.
 \label{5.39}
 \end{equation}

\bigskip

In the rest frame instant form we have $T \approx \tau$, ${\cal E}
\approx {\cal M}$, $H_D = 0$ and the external canonical
non-covariant 4-center of mass $q^{\mu}$ is interpreted as a {\it
decoupled point particle clock}. However, since the gauge fixing
$T - \tau \approx 0$ is explicitly $\tau$-dependent we get that
the effective Hamiltonian to reproduce the Hamilton equations for
${\vec \Sigma}(\tau,\vec{\sigma}_o),\vec{K}(\tau,\vec{\sigma}_o)$
 when $\lambda (\tau ) = - 1$ is

\beq
 H = {\cal
M} - \vec \lambda (\tau ) \cdot {\vec {\cal P}}.
 \label{5.40}
 \eeq

In the new canonical variables (\ref{5.37}) we have
[$\Omega^i=(1/2)\epsilon^{ijk}\Omega^{jk}$ with $\Omega^{jk}$
 given by (\ref{5.30})]

\begin{equation}
\vec{J}(\tau)= \vec{z}(\tau)\times\vec{k}(\tau)+
\vec{\Omega}(\tau)= \vec{Q}(\tau)\times\vec{p}(\tau)+
\vec{\Omega}(\tau).
 \label{5.41}
 \end{equation}

The reduced phase space now is

\begin{equation}
\begin{array}{|c|c|}
\hline \vec{z}(\tau)&\vec{\Sigma}(\tau,\vec{\sigma}_o)\\
\vec{k}(\tau)&\vec{K}(\tau,\vec{\sigma}_o)\\ \hline
\end{array}.
 \label{5.42}
 \end{equation}

\bigskip

The final result is a new instant form of dynamic \cite{25}, the
{\em Wigner-covariant rest frame instant form}. In this form the
system is described by a reduced phase space (\ref{5.42}) formed
by two sectors:

i) the sector of the {\em external decoupled point particle clock}
described by the canonical non-covariant variables
$\vec{z}(\tau),\vec{k}(\tau)$;

ii) the sector of the {\em internal Wigner-covariant variables}
$\vec{\Sigma}(\tau,\vec{\sigma}_o), \vec{K}(\tau,\vec{\sigma}_o)$
(they are Wigner spin-1 3-vectors), living inside the Wigner
hyper-planes. Since they are restricted by the constraints ${\vec
{\cal P}} \approx 0$, ${\vec {\cal K}} \approx 0$, only variables
relative to an internal inessential 3-center of mass (see next
Section) are physical.

The external variables define the 4-unit vector
$n^\mu=(\sqrt{1+\vec{k}^2},\vec{k})$, which identifies the
direction of the total four-momentum with respect to an external
inertial observer.

\medskip

In the rest-frame instant form of dynamic the 10 canonical
generator of the external Poincar\'e group are

\begin{eqnarray}
p^o &=& {\cal M}\, \sqrt{1+\vec{k}^2} = \sqrt{{\cal M}^2 + {\vec
p}^2},\nonumber \\
 &&\nonumber\\
 p^i &=&{\cal M}k^i,\qquad \vec{J} = \vec{z} \times \vec{k} +
 \vec{\Omega}= \vec Q \times \vec p + \vec \Omega,\nonumber \\
&&\nonumber\\
 \vec{K} &=& - \vec{z}\, \sqrt{1+\vec{k}^2} -
\frac{\vec{k}\times\vec{\Omega}}{1+\sqrt{1+\vec{k}^2}}= \nonumber
\\ &&\nonumber\\
 &=& - \vec Q\, \sqrt{{\cal M}^2 + {\vec p}^2} - {{
 \vec p \times \vec \Omega}\over {
 {\cal M} + \sqrt{{\cal M}^2 + {\vec p}^2}}},
 \label{5.43}
 \end{eqnarray}

\noindent where

\begin{equation}
\Omega^{i}\equiv\epsilon^{ijk}\,\delta^{jr}\,\delta^{ks}\left(\frac{1}{2}
\epsilon^{rsu}{\cal J}^u\right),
 \label{5.44}
 \end{equation}

\noindent is the (interaction-free) spin with respect to the
external center of mass.

\medskip

The properties of the instant form are now shown explicitly: i)
$\vec p$, $\vec J$ do not depend on the interactions; ii) only the
4 generators $p^o$, $\vec K$ depend on the dynamics through ${\cal
M}$. Even if in a generic instant form the dynamics is determined
by four independent potentials (the dynamical $SU(2)$ of
Ref.\cite{26}), in the rest frame instant form there is a unique
function, the invariant mass ${\cal M}$, carrying the whole
dynamical information.

\medskip

Let us observe that the {\em external} boost generator
$K^r=J^{or}$ of Eqs.(\ref{5.31}) can be rewritten either in the
form

\begin{equation}
\vec{K}(\tau)=q^o(\tau)\vec{p}-\vec{q}(\tau)p^o-\frac{
\vec{p}\times\vec{\Omega}(\tau)}{p^o+\sqrt{p^2}},
 \label{5.45}
 \end{equation}

\noindent or in the form

\begin{equation}
\vec{K}(\tau)=q^o(\tau)\vec{p}+\vec{K}'(\tau).
 \label{5.46}
 \end{equation}

If $\vec{Q}(\tau)$ is defined by (\ref{5.37}), we have

\begin{equation}
\vec{Q}(\tau)=-\frac{\vec{K}(\tau)}{p^o}-\frac{
\vec{p}\times\vec{\Omega}(\tau)}{p^o(p^o+\sqrt{p^2})},
 \label{5.47}
 \end{equation}

\noindent and

\begin{equation}
\vec{q}(\tau)=-\frac{\vec{K}'(\tau)}{p^o}-\frac{
\vec{p}\times\vec{\Omega}(\tau)}{p^o(p^o+\sqrt{p^2})}.
 \label{5.48}
 \end{equation}

\medskip

In this way we recover the usual definitions of the canonical,
non-covariant relativistic 3-center of mass \cite{27}. The first
is given using the complete external boost generator, the second
using the boost generator on the hyperplane $q^o(\tau)=0$. Being
$n_\mu x^\mu(\tau)=n_\mu q^\mu(\tau)$, the 4-center of mass
$q^\mu(\tau)$ defines a point on the Wigner hyperplane different
from the centroid $x^\mu(\tau) = z^{\mu}(\tau , \vec 0)$, with
coordinates $\vec{\sigma}(q)\neq 0$ such that $q^\mu(\tau) =
z^\mu(\tau,\vec{\sigma}(q))$. Using only the canonical generators
of the external realization of the Poincar\'e group it is also
possible to define the external non-covariant, non-canonical {\em
M$\o$ller 3-center of energy } \cite{27}:

\begin{equation}
\vec{R}(\tau)=-\frac{\vec{K}(\tau)}{p^o},
 \label{5.49}
 \end{equation}

\noindent and the external covariant, non-canonical relativistic
{\em Fokker-Pryce 3-center of inertia}\cite{18}:

\begin{equation}
\vec{Y}(\tau)=-\frac{\vec{K}(\tau)}{p^o}-
\frac{\vec{p}\times\vec{\Omega}(\tau)}{p^o\sqrt{p^2}}.
 \label{5.50}
 \end{equation}

\medskip

In Ref.\cite{14} it is shown how to identify the position on the
Wigner hyperplane of the associated external (pseudo-vector)
4-center of energy $R^{\mu}$ and of the external 4-center of
inertia $Y^{\mu}$, which is a 4-vector by construction. If we put
$q^{\mu} = (q^o; \vec q)$, then we have $R^{\mu} = (q^o; \vec R +
q^o\, \vec p)$ and $Y^{\mu} = (q^o; \vec Y + q^o\, \vec p)$. In
Ref.\cite{14} it is also shown that all the possible
pseudo-vectors $q^{\mu}$ and $R^{\mu}$ fill a world-tube around
the 4-vector $Y^{\mu}$, whose radius is the {\it M$\o$ller radius}
\cite{28} $\rho = {{|\vec \Omega|}\over {{\cal M}}}$ of the fluid
configuration. This radius is defined by the Poincare' Casimirs of
the external Poincare' group ($p^2 = {\cal M}^2$, $W^2 = - {\cal
M}^2\, {\vec \Omega}^2$) and is a classical unit of length
determined by the Cauchy data of the configuration of the system.
See Ref.\cite{3} for the properties of this radius and for the
proposal of using it as a natural physical ultraviolet cutoff at
the quantum level for all rotating configurations of the isolated
system.

\bigskip

Let us conclude this Section with the Hamilton equations
associated with the Hamiltonian (\ref{5.40}) in a gauge where
$\vec \lambda (\tau ) = 0$

\bea
 &&{{\partial \vec \Sigma (\tau ,{\vec \sigma}_o)}\over {\partial
 \tau}}\, \cir\, \{ \vec \Sigma (\tau ,{\vec \sigma}_o), {\cal M}
 \},\nonumber \\
  &&{{\partial \vec K(\tau ,{\vec \sigma}_o)}\over {\partial
 \tau}}\, \cir\, \{ \vec K(\tau ,{\vec \sigma}_o), {\cal M} \}.
 \label{5.51}
 \eea

\medskip

\noindent For the dust Eq.(\ref{b6}) implies

\bea
  &&{{\partial \vec \Sigma (\tau ,{\vec \sigma}_o)}\over {\partial
 \tau}}\, \cir\, {{ \vec K(\tau ,{\vec \sigma}_o)}\over {\sqrt{
 [\mu\, n_o({\vec \sigma}_o)]^2 + {\vec K}^2(\tau ,{\vec
 \sigma}_o)}}},\qquad  {{\partial \vec K(\tau ,{\vec \sigma}_o)}\over {\partial
 \tau}}\, \cir\, 0,\nonumber \\
 &&{}\nonumber \\
 &&\vec \Sigma (\tau ,{\vec \sigma}_o)\, \cir\, {\vec \sigma}_o +
 {{ \vec K(\tau ,{\vec \sigma}_o)}\over {\sqrt{
 [\mu\, n_o({\vec \sigma}_o)]^2 + {\vec K}^2(\tau ,{\vec
 \sigma}_o)}}}\, \tau .
 \label{5.52}
 \eea

Let us remark that in this description these hyperbolic Hamilton
equations replace the hydrodynamical  Euler equations (\ref{2.26})
implied by the conservation (\ref{2.23}) of the stress-energy
tensor.

\vfill\eject

\section{Internal centers of mass and relative variables}

In the rest frame instant form defined in the previous Section the
3-dimensional  variables on the Wigner hyper-planes are not all
{\em physical} due to the first class Dirac constraint $\vec{\cal
P}\approx 0$. To select the physical degree of freedom is a
problem equivalent to determine the internal 3-center of mass and
relative variables on the Wigner hyperplane. As already said, the
internal 3-center of mass variable on the Wigner hyperplane is a
gauge variable, because the role of true center of mass is played
by the {\em external}, canonical non-covariant, 3-center of mass
$\vec{z}$. On the contrary the relative variables will be the
physical variables and they will describe the reduced phase space
after the final gauge fixings, whose natural form is ${\vec {\cal
K}} \approx 0$.

To justify the gauge fixing ${\vec {\cal K}} \approx 0$ we must
perform  two steps. First we select a naive internal center of
mass position $\vec{\cal X}$ canonical with respect to the total
momentum $\vec{\cal P}$ and the associated canonical relative
variables. The naive center of mass position $\vec{\cal X}$ allows
to define the gauge fixing $\vec{\cal X}\approx 0$, but this gauge
fixing has the unpleasant property that the arbitrary Dirac
multiplier $\vec{\lambda}(\tau )$ is fixed to a non null value
($\vec{\lambda}(\tau ) \neq 0$). Then we use the internal relative
variables obtained  as auxiliary variables in the first step for
defining the relative variable respect to a 3-center of mass
$\vec{\cal Q}$ such that the gauge fixings $\vec{\cal Q} \approx
0$ (identification of the internal 3-center of mass with the
centroid $x^{\mu}(\tau )$ origin of the 3-coordinates) imply
$\vec{\lambda}(\tau ) \approx 0$. This is done using the method of
the {\em Gartenhaus-Schwartz} transformation of  Ref.\cite{29}
(see also Appendix D).

\hfill

Let the total internal 3-momentum and an internal naive
3-center-of-mass position on the Wigner hyperplane $\vec{\cal X}$
be defined as

\begin{eqnarray}
{\cal P}^r(\tau)&=&\int d^3\sigma_o\;
K^r(\tau,\vec{\sigma}_o),\nonumber\\
 &&\nonumber\\
{\cal X}^s(\tau)&=&\frac{1}{\cal N}\int d^3\sigma_o\;
n_o(\vec{\sigma}_o) \Sigma^r(\tau,\vec{\sigma}_o),
 \label{6.1}
 \end{eqnarray}

\noindent with

\begin{equation}
{\cal N}=\int d^3\sigma_o\,n_o(\vec{\sigma}_o),
 \label{6.2}
 \end{equation}

\noindent and

\begin{equation}
\{{\cal X}^r(\tau),{\cal P}^s(\tau)\}=\delta^{rs}.
 \label{6.3}
 \end{equation}

\bigskip

Let the {\em internal relative canonical variables}
$\Re^r(\tau,\vec{\sigma}_o),\wp^s(\tau,\vec{\sigma}_o)$ be defined
in such a way that we get

\begin{eqnarray}
\Sigma^r(\tau,\vec{\sigma}_o)&=&{\cal X}^r(\tau)+\int
d^3\sigma'_o\,
\Gamma_\Sigma(\vec{\sigma}_o,\vec{\sigma}'_o)\Re^r(\tau,\vec{\sigma}'_o),
\nonumber\\
 &&\nonumber\\
 K^s(\tau,\vec{\sigma}_o)&=&\frac{n_o(\vec{\sigma}_o)}{\cal N}
{\cal P}^r(\tau)+\int d^3\sigma'_o\, \wp^r(\tau,\vec{\sigma}'_o)
\Gamma_K(\vec{\sigma}'_o,\vec{\sigma}_o).
 \label{6.4}
 \end{eqnarray}

\medskip

The kernels $\Gamma_K$, $\Gamma_\Sigma$ will be specified by some
conditions that we will analyze shortly. From the definitions
(\ref{6.1}) we obtain that

\begin{eqnarray}
\int d^3\sigma_o\,n_o(\vec{\sigma}_o)
\Gamma_\Sigma(\vec{\sigma}_o,\vec{\sigma}'_o)&=&0, \nonumber\\
 &&\nonumber\\
\int d^3\sigma_o\,\Gamma_K(\vec{\sigma}'_o,\vec{\sigma}_o)&=&0.
 \label{6.5}
\end{eqnarray}

\medskip

We impose the following canonical property

 \begin{equation}
\{\Re^r(\tau,\vec{\sigma}_o),\wp^s(\tau,\vec{\sigma}'_o)\}=
\delta^{rs}\delta(\vec{\sigma}_o-\vec{\sigma}'_o).
 \label{6.6}
\end{equation}

By using Eq.(\ref{6.3}), we can verify that from Eq.(\ref{6.4}) we
obtain the canonical property

\begin{equation}
\{\Sigma^r(\tau,\vec{\sigma}_o),K^s(\tau,\vec{\sigma}'_o)\}=
\delta^{rs}\,\delta(\vec{\sigma}_o-\vec{\sigma}'_o),
 \label{6.7}
 \end{equation}

\noindent if

\begin{equation}
\int d^3\sigma_o\,\Gamma_\Sigma(\vec{\sigma}_{o1},\vec{\sigma})
\Gamma_K(\vec{\sigma},\vec{\sigma}_{o2})=
-\frac{n_o(\vec{\sigma}_{o2})}{\cal N}+
\delta(\vec{\sigma}_{o1}-\vec{\sigma}_{o2}).
 \label{6.8}
 \end{equation}

\medskip

When Eqs.(\ref{6.8}) hold, Eq.(\ref{6.4}) is consistent with the
following definitions

\begin{eqnarray}
\Re^r(\tau,\vec{\sigma}_o)&=&\int d^3\sigma'_o\,
\Gamma_K(\vec{\sigma}_o,\vec{\sigma}'_o)\Sigma^r(\tau,\vec{\sigma}'_o),
\nonumber\\
 &&\nonumber\\
\wp^r(\tau,\vec{\sigma}_o)&=&\int d^3\sigma'_o\,
K^r(\tau,\vec{\sigma}'_o)\Gamma_\Sigma(\vec{\sigma}'_o,\vec{\sigma}_o).
 \label{6.9}
 \end{eqnarray}

In fact, if we substitute these expression in Eq.(\ref{6.4}),
using Eq.(\ref{6.8}) we have an identity. Again from
Eq.(\ref{6.9}) and using Eq.(\ref{6.7}), we can find
Eq.(\ref{6.6}) if the following condition is verified

\begin{equation}
\int d^3\sigma_o\, \Gamma_K(\vec{\sigma}_{o1},\vec{\sigma}_o)
\Gamma_\Sigma(\vec{\sigma}_o,\vec{\sigma}_{o2})=
\delta(\vec{\sigma}_{o1}-\vec{\sigma}_{o2}).
 \label{6.10}
 \end{equation}

Eqs. (\ref{6.5}),(\ref{6.8}) and (\ref{6.10}) are a set of
conditions that have to be satisfied by the kernels $\Gamma$.
These conditions are not independent: it can be proved that Eqs.
(\ref{6.8}) e (\ref{6.10}) imply Eq.(\ref{6.5}).

We can also verify that, using Eq.(\ref{6.5}), we have

\begin{eqnarray}
\vec{\cal J}(\tau) &=& \vec{\cal X}(\tau)\times\vec{\cal
P}(\tau)+\int d^3\sigma_o\,
\vec{\Re}(\tau,\vec{\sigma}_o)\times\vec{\wp}(\tau,\vec{\sigma}_o),
\nonumber \\
&&\nonumber\\
 {\vec {\cal K}}(\tau ) &=& - {\cal M}(\tau )\, {\vec {\cal
 X}}(\tau ) - \int d^3\sigma_o\, d^3\sigma_o^{'}\,
 \Gamma_{\Sigma}({\vec \sigma}_o, {\vec \sigma}_o^{'})\,
 \vec{\Re}(\tau ,{\vec \sigma}_o^{'})\, \Delta (\tau ,{\vec
 \sigma}_o).
 \label{6.11}
 \end{eqnarray}

\hfill

Let $\Phi_{\underline{n}}(\vec{\sigma}_o)$ be a base of
orthonormal functions on $R^3$ with $\underline{n}=(n_1,n_2,n_3)$
a set of multindices. Then we can consider the coefficients

\begin{eqnarray}
\vec{r}_{\underline{n}}(\tau)&=& \int
d^3\sigma_o\,\Phi_{\underline{n}}
(\vec{\sigma}_o)\vec{\Re}(\tau,\vec{\sigma}_o),\nonumber\\
 &&\nonumber\\
\vec{p}_{\underline{n}}(\tau)&=&\int
d^3\sigma_o\,\Phi_{\underline{n}}
(\vec{\sigma}_o)\vec{\wp}(\tau,\vec{\sigma}_o),
 \label{6.12}
 \end{eqnarray}

\noindent such that

\begin{eqnarray}
\vec{\Re}(\tau,\vec{\sigma}_o)&=&\sum_{\underline{n}}\vec{r}_{\underline{n}}(\tau)\,
\Phi_{\underline{n}}(\vec{\sigma}_o),\nonumber\\
 &&\nonumber\\
\vec{\wp}(\tau,\vec{\sigma}_o)&=&\sum_{\underline{n}}\vec{p}_{\underline{n}}(\tau)\,
\Phi_{\underline{n}}(\vec{\sigma}_o).
 \label{6.13}
 \end{eqnarray}

Moreover from Eq.(\ref{6.11}) we get

\begin{equation}
\int d^3\sigma_o\,
\vec{\Re}(\tau,\vec{\sigma}_o)\times\vec{\wp}(\tau,\vec{\sigma}_o)=
\sum_{\underline{n}}\,\vec{r}_{\underline{n}}(\tau)\times\vec{p}_{\underline{n}}(\tau),
 \label{6.14}
 \end{equation}

\noindent and

\begin{equation}
\{r^r_{\underline{n}}(\tau),p^s_{\underline{m}}(\tau)\}=
\delta^{rs}\delta_{\underline{n}\underline{m}}.
 \label{6.15}
 \end{equation}

\bigskip

In conclusion the coefficients $\vec{r}_{\underline{n}}(\tau)$,
$\vec{p}_{\underline{n}}(\tau)$ are a set of infinite canonical
variables that we can use as {\em internal relative variables}.
These variables are useful for defining the canonical
transformation that will realize the separation between rotational
and shape degree of freedom in the next Section. For the time
being we can use them to rewrite the definitions
(\ref{6.4}),(\ref{6.9}) in the form

\begin{eqnarray}
\vec{\Sigma}(\tau,\vec{\sigma}_o)&=&\vec{\cal
X}(\tau)+\sum_{\underline{n}}
\,\Gamma^{\Sigma}_{\underline{n}}(\vec{\sigma}_o)\,
\vec{r}_{\underline{n}}(\tau), \nonumber\\
 &&\nonumber\\
 \vec{K}(\tau,\vec{\sigma}_o)&=&\frac{n_o(\vec{\sigma}_o)}{\cal N}
\vec{\cal P}(\tau)+\sum_{\underline{n}}
\,\Gamma^{K}_{\underline{n}}(\vec{\sigma}_o)\,
\vec{p}_{\underline{n}}(\tau), \nonumber\\
 &&\nonumber\\
 \vec{r}_{\underline{n}}(\tau)&=&\int d^3\sigma_o\,
\Gamma^{K}_{\underline{n}}(\vec{\sigma}_o)\,
\vec{\Sigma}(\tau,\vec{\sigma}_o),\nonumber\\
 &&\nonumber\\
\vec{p}_{\underline{n}}(\tau)&=&\int d^3\sigma_o\,
\Gamma^{\Sigma}_{\underline{n}}(\vec{\sigma}_o)\,
\vec{K}(\tau,\vec{\sigma}_o),
 \label{6.16}
 \end{eqnarray}

\noindent where

\begin{eqnarray}
\Gamma^{\Sigma}_{\underline{n}}(\vec{\sigma}_o)&=& \int
d^3\sigma'_o\, \Gamma_\Sigma(\vec{\sigma}_o,\vec{\sigma}'_o)
\Phi_{\underline{n}}(\vec{\sigma}'_o),\nonumber\\
 &&\nonumber\\
\Gamma^{K}_{\underline{n}}(\vec{\sigma}_o)&=& \int
d^3\sigma'_o\,\Phi_{\underline{n}}(\vec{\sigma}'_o)
\Gamma_K(\vec{\sigma}'_o,\vec{\sigma}_o).
 \label{6.17}
 \end{eqnarray}

\medskip

The conditions (\ref{6.5}) and (\ref{6.8}), (\ref{6.10}) on the
kernel functions $\Gamma$  are rewritten in the form

\begin{eqnarray}
\int
d^3\sigma_o\,n_o(\vec{\sigma}_o)\,\Gamma^{\Sigma}_{\underline{n}}
(\vec{\sigma}_o)&=&0,\qquad
 \int d^3\sigma_o\,\Gamma^{K}_{\underline{n}}
(\vec{\sigma}_o)=0,\nonumber\\
 &&\nonumber\\
\sum_{\underline{n}}\,\Gamma^{\Sigma}_{\underline{n}}
(\vec{\sigma}_{1o})\Gamma^{K}_{\underline{n}}
(\vec{\sigma}_{2o})&=& -\frac{n_o(\vec{\sigma}_{2o})}{\cal N}+
\delta(\vec{\sigma}_{1o}-\vec{\sigma}_{2o}),\nonumber\\
 &&\nonumber\\
\int d^3\sigma_o\,\Gamma^{K}_{\underline{n}}
(\vec{\sigma}_o)\,\Gamma^{\Sigma}_{\underline{m}}
(\vec{\sigma}_o)&=&\delta_{\underline{n}\underline{m}}.
 \label{6.18}
 \end{eqnarray}

\hfill

Some possible solutions for the kernels $\Gamma$ are derived in
Appendix E.

\hfill

As said at the beginning of this Section, the internal 3-center of
mass-like position $\vec{\cal X}$ is such that the gauge fixing
$\vec{\cal X}\approx 0$ does not imply $\vec{\lambda}(\tau
)\approx 0$ as can be checked by using the Dirac Hamiltonian $H_D
= {\cal M} - \vec \lambda (\tau ) \cdot {\vec {\cal P}}$ in the
gauge $T \approx \tau$. We want to replace it with another
internal 3-center of mass $\vec{\cal Q}$ such that the conditions
$\vec{\cal Q}\approx 0$ imply $\vec{\lambda}(\tau ) \approx 0$.

\hfill

To this end we construct the {\em internal} 3-centers of mass,
energy and inertia in analogy to the {\em external} ones of the
previous Section, using the {\em internal} realization of the
Poincar\'e algebra instead of the external one

\begin{eqnarray}
\vec{\cal R}(\tau)&=&-\frac{\vec{\cal K}(\tau)}{\cal
M},\nonumber\\
 &&\nonumber\\
\vec{\cal Q}(\tau)&=&-\frac{\vec{\cal K}(\tau)}{\cal M}
-\frac{\vec{\cal P}\times\vec{\cal S}(\tau)} {{\cal M}({\cal
M}+\sqrt{{\cal M}^2-\vec{\cal P}^2)}} \approx \vec{\cal
R}(\tau),\nonumber\\
 &&\nonumber\\
 \vec{\cal Y}(\tau)&=&-\frac{\vec{\cal K}(\tau)}{\cal M}
-\frac{\vec{\cal P}\times\vec{\cal S}(\tau)} {{\cal M}\sqrt{{\cal
M}^2-\vec{\cal P}^2}} \approx \vec{\cal R}(\tau),
 \label{6.19}
 \end{eqnarray}

\noindent where

\begin{eqnarray}
\vec{\cal J}(\tau) &=& \vec{\cal Q}\times\vec{\cal P}+\vec{\cal
S}(\tau),\nonumber \\
&&\nonumber\\
 {\vec {\cal K}}(\tau ) &=& - {\cal M}(\tau )\, {\vec {\cal
 X}}(\tau ) - \sum_{\underline{n}}\, {\vec r}_{\underline{n}}(\tau
 )\, \int d^3\sigma_o\, \Gamma^{\Sigma}_{\underline{n}}({\vec
 \sigma}_o)\, \Delta (\tau ,{\vec \sigma}_o) =\nonumber \\
&&\nonumber\\
&=& - {\cal M}(\tau )\, {\vec {\cal R}}(\tau ) \approx - {\cal
 M}(\tau )\, {\vec {\cal Q}}(\tau ),\nonumber \\
&&\nonumber\\
&&\Downarrow \nonumber \\
&&\nonumber\\
{\vec {\cal X}}(\tau ) &=& {\vec {\cal R}}(\tau ) -
 \sum_{\underline{n}}\, {{ {\vec r}_{\underline{n}}(\tau )}\over
 {{\cal M}(\tau )}}\, \int d^3\sigma_o\,
 \Gamma^{\Sigma}_{\underline{n}}({\vec \sigma}_o)\, \Delta
 (\tau ,{\vec \sigma}_o).
 \label{6.20}
 \end{eqnarray}

These three centers are weakly equal due to the constraint
$\vec{\cal P}\approx 0$ and they are all canonically conjugate to
$\vec{\cal P}$

\begin{equation}
\{{\cal R}^r,{\cal P}^s\}^{**}= \{{\cal Q}^r,{\cal P}^s\}^{**}=
\{{\cal Y}^r,{\cal P}^s\}^{**}= \delta^{rs},
 \label{6.21}
 \end{equation}

\noindent and such that

\begin{equation}
\{{\cal R}^r,{\cal M}\}^{**}= \{{\cal Q}^r,{\cal M}\}^{**}=
\{{\cal Y}^r,{\cal M}\}^{**}=\frac{{\cal P}^r}{{\cal M}} \approx
0.
 \label{6.22}
 \end{equation}

But $\vec{\cal Q}$ is the only one such that

\begin{equation}
\{{\cal Q}^r,{\cal Q}^s\}^{**}=0,
 \label{6.23}
 \end{equation}

\noindent namely it is the real internal canonical 3-center of
mass.

\medskip

If we adopt the gauge fixings

\begin{equation}
\vec{\cal Q}(\tau)\approx \vec{\cal R}(\tau)\approx \vec{\cal
Y}(\tau)\approx 0,
 \label{6.24}
 \end{equation}

\noindent then we get

\begin{eqnarray}
 {\dot {\vec {\cal Q}}}(\tau )\, &{\buildrel \circ \over =}& \{
 {\vec {\cal Q}}(\tau ), {\cal M} - \vec \lambda (\tau ) \cdot {\vec {\cal P}} \} =
 - \vec{\lambda}(\tau) \approx 0,\nonumber \\
 &&\nonumber \\
 {\vec {\cal X}}(\tau ) &\approx& - \sum_{\underline{n}}\,  {{ {\vec r}_{\underline{n}}(\tau )}\over
 {{\cal M}(\tau )}}\, \int d^3\sigma_o\,
 \Gamma^{\Sigma}_{\underline{n}}({\vec \sigma}_o)\, \Delta
 (\tau ,{\vec \sigma}_o)
\nonumber \\
&&\nonumber\\
 \vec \Sigma (\tau , {\vec \sigma}_o) &\approx&
 \sum_{\underline{n}}\, {\vec r}_{\underline{n}}(\tau )\, \left[
 \Gamma^{\Sigma}_{\underline{n}}({\vec \sigma}_o) - {{ \int
 d^3\sigma_o^{'}\, \Gamma^{\Sigma}_{\underline{n}}({\vec
 \sigma}_o^{'})\,
 \Delta (\tau ,{\vec \sigma}_o^{'})}\over {{\cal M}(\tau
 )}}\right] =\nonumber \\
&&\nonumber\\
 &{\buildrel {def}\over =}&  \sum_{\underline{n}}\, {\vec r}_{\underline{n}}(\tau )\, \Big[
 \Gamma^{\Sigma}_{\underline{n}}({\vec \sigma}_o) -
 h_{\underline{n}}\Big].
 \label{6.25}
 \end{eqnarray}

Eq.(\ref{6.19}) shows that ${\vec {\cal Q}}(\tau ) \approx 0$ is
equivalent to the condition $\vec{\cal K}\approx 0$. After this
gauge fixing, in the internal unfaithful realization of the
Poincar\'e group there are only four non null functions: ${\cal
M}$ and $ \vec{\cal J}$.

\medskip

In Appendix D it is show that  with the relative variables
$\Re^r(\tau,\vec{\sigma}_o),\wp^s(\tau,\vec{\sigma}_o)$ we can
realize the {\em Garthenaus-Schwartz} canonical transformation

\begin{equation}
\begin{array}{|c|}
\hline
\vec{\Sigma}(\tau,\vec{\sigma}_o)\\
\vec{K}(\tau,\vec{\sigma}_o)\\
\hline
\end{array}
\longrightarrow
\begin{array}{|c|c|}
\hline
\vec{\cal Q}(\tau)&\Re^{\prime\, r}(\tau,\vec{\sigma}_o)\\
\vec{\cal P}(\tau)&\wp^{\prime\, s}(\tau,\vec{\sigma}_o)\\
\hline
\end{array}.
 \label{6.26}
 \end{equation}

In Appendix D it is also shown that, after having gone to Dirac
brackets with respect to the second class constraints ${\vec {\cal
P}} \approx 0$, ${\vec {\cal Q}} \approx 0$, we get

\begin{equation}
\vec{\cal P}\equiv\vec{\cal Q}\equiv 0 \;\; \Rightarrow \;\; \Re^{\prime\,
r}(\tau,\vec{\sigma}_o)\equiv \Re^{r}(\tau,\vec{\sigma}_o),\;\;
\wp^{\prime\, s}(\tau,\vec{\sigma}_o)\equiv
\wp^{s}(\tau,\vec{\sigma}_o).
 \label{6.27}
 \end{equation}

\medskip

Then the final reduced phase space is

\begin{equation}
\begin{array}{|c|c|}
\hline
\vec{z}(\tau)&\Re^{r}(\tau,\vec{\sigma}_o)\\
\vec{k}(\tau)&\wp^{r}(\tau,\vec{\sigma}_o)\\
\hline
\end{array},
 \label{6.28}
 \end{equation}

\noindent where

\begin{equation}
\vec{\cal J}(\tau)\approx\vec{\cal S}(\tau)= \int d^3\sigma_o\,
\vec{\Re}(\tau,\vec{\sigma}_o)
\times\vec{\wp}(\tau,\vec{\sigma}_o) =\sum_{\underline{n}}\,
\vec{r}_{\underline{n}}(\tau)\times \vec{p}_{\underline{n}}(\tau).
 \label{6.29}
 \end{equation}

On the Wigner hyper-planes the kinetic term ${\vec K}^2(\tau ,
{\vec \sigma}_o)$ appearing in the solution $X(\tau ,{\vec
\sigma}_o)$ of Eq.(\ref{4.28}) is

\begin{equation}
{\vec K}^2(\tau , {\vec \sigma}_o) \approx \sum_{
{\underline{n}}_1 {\underline{n}}_2 }\,
\Gamma^K_{{\underline{n}}_1}({\vec \sigma}_o)\,
\Gamma^K_{{\underline{n}}_2}({\vec \sigma}_o)\, {\vec
p}_{{\underline{n}}_1}(\tau ) \cdot {\vec
p}_{{\underline{n}}_2}(\tau ),
 \label{6.30}
 \end{equation}

\noindent while the dependence of $X(\tau ,{\vec \sigma}_o)$ on
the generalized Eulerian coordinates is concentrated in

\begin{equation}
 \det \, \left( {{\partial \Sigma}\over {\partial \sigma_o}}\right)
 = \det\, \left( \sum_{\underline{n}}\, {{\partial\,
 \Gamma^{\Sigma}_{\underline{n}}({\vec \sigma}_o)}\over {\partial
 \sigma_o^s}}\, r^r_{\underline{n}}(\tau ) \right).
 \label{6.31}
 \end{equation}

\vfill\eject

\section{Rotational Kinematics}

The internal relative canonical variables
$\vec{r}_{\underline{n}}(\tau)$, $\vec{p}_{\underline{n}}(\tau)$
are Wigner spin 1 vectors  under rotations. Then we can do on them
a canonical transformation that generalize the results obtained in
the $N$ particles case in Refs. \cite{13,14}. The
$\vec{r}_{\underline{n}}$ are interpreted as a set of infinite
relative position vectors; we want use them to construct a {\em
dynamical body frame} as done in Refs. \cite{13,14}. For this {\it
we have to select a pair of these vectors}. We assume to choose
the vector positions with multindices $\underline{u}_1,
\underline{u}_2$ as preferred vectors

\begin{equation}
\vec{r}_{\underline{u}_i}=\vec{R}_i,\;\;\;\;
\vec{p}_{\underline{u}_i}=\vec{\Pi}_i,\;\;\;i=1,2,
 \label{7.1}
 \end{equation}

\noindent and we use them to define the orthogonal vectors

\begin{equation}
\vec{N}=\frac{\hat{R}_1+\hat{R}_2}{2}, \;\;\;\;\;
\vec{\chi}=\frac{\hat{R}_1-\hat{R}_2}{2}, \;\;\;\;\;
\vec{N}\cdot\vec{\chi}=0.
 \label{7.2}
 \end{equation}

The most convenient choice of these two vectors will be dictated
by the spatial form of the initial density $n_o({\vec \sigma}_o)$.

\bigskip

Then a {\em dynamical body frame} is defined by the associated
unit vectors and their orthogonal complement

\begin{equation}
\hat{b}_{r}(\tau)=(\hat{\chi}(\tau),
\hat{N}(\tau)\times\hat{\chi}(\tau),\hat{N}(\tau)).
 \label{7.3}
 \end{equation}

By construction this frame is a orthonormal frame that rotates
with the motion of the fluid. In this sense it generalize the
concept of {\em body frame} of a non relativistic rigid body and
we can apply the same observations and interpretation done in
Refs. \cite{13,14}. Moreover we observe that

\begin{equation}
\{N^r,N^s\}=\{\chi^r,\chi^s\}=\{N^r,\chi^s\}=0.
 \label{7.4}
 \end{equation}

\medskip

All vectors can be projected on this {\em dynamical body frame} so
to obtain their components on it. In particular, for the relative
angular momentum, given by Eq. (\ref{6.29}), its components on the
{\em dynamical body frame} are

\begin{equation}
\overline{\cal S}^r(\tau)=\vec{\cal S}(\tau)\cdot\hat{b}_r(\tau),
 \label{7.5}
 \end{equation}

\noindent or more explicitly

\begin{equation}
\overline{\cal S}^1=\vec{\cal S}\cdot\hat{\chi} ;\;\;\;\;\;
\overline{\cal S}^2=\vec{\cal S}\cdot(\hat{N}\times\hat{\chi})
;\;\;\;\;\; \overline{\cal S}^3=\vec{\cal S}\cdot\hat{N}.
 \label{7.6}
 \end{equation}

\medskip

Using the results of Refs. \cite{13,14} we can construct the
following quantities

\begin{equation}
\vec{\cal
W}=\vec{R}_1\times\vec{\Pi}_{1}-\vec{R}_{2}\times\vec{\Pi}_{2},
 \label{7.7}
 \end{equation}

\begin{equation}
R_i=\mid\vec{R}_{i}\mid,\;\;\;\;\;\;\tilde{\Pi}_{i}=\vec{\Pi}_{i}
\cdot\hat{R}_{i},\;\;\;\;\;\;i=1,2,
 \label{a.7.8}
 \end{equation}

\noindent and, for $\underline{n}\neq \underline{u}_1,
\underline{u}_2$

\begin{eqnarray}
\overline{r}^1_{\underline{n}}=\vec{r}_{\underline{n}}\cdot\hat{\chi},\;\;\;&
\overline{r}^2_{\underline{n}}=\vec{r}_{\underline{n}}\cdot\hat{N}\times\hat{\chi},\;\;\;&
\overline{r}^3_{\underline{n}}=\vec{r}_{\underline{n}}\cdot\hat{N},\nonumber\\
 &&\nonumber\\
 \overline{p}^1_{\underline{n}}=\vec{p}_{\underline{n}}\cdot\hat{\chi},\;\;\;&
\overline{p}^2_{\underline{n}}=\vec{p}_{\underline{n}}\cdot\hat{N}\times\hat{\chi},\;\;\;&
\overline{p}^3_{\underline{n}}=\vec{p}_{\underline{n}}\cdot\hat{N}.
 \label{7.9}
 \end{eqnarray}

\medskip

 Then the transformation represented in the table

\begin{equation}
\begin{array}{|c|}
\hline
\vec{r}_{\underline{n}}\\
\vec{p}_{\underline{n}}\\
\hline
\end{array}
\longrightarrow
\begin{array}{|ccc|ccc|c|}
\hline
\mid\vec{\cal S}\mid&{\cal S}^3&\overline{\cal S}^3
&\mid\vec{N}\mid&R_1&R_2&\overline{r}^s_{\underline{n}\neq
\underline{u}_1,\underline{u}_2}\\
\alpha&\beta&\gamma&\xi&\tilde{\Pi}_1&\tilde{\Pi}_2&
\overline{p}^s_{\underline{n}\neq \underline{u}_1,\underline{u}_2}\\
\hline
\end{array},
 \label{7.10}
 \end{equation}

\noindent is a canonical transformation (the canonical pairs are
the variables on the same column) if we  define

\begin{eqnarray}
\alpha&=&\tan^{-1}\frac{(\hat{\cal S}\times\hat{N})^3} {[\hat{\cal
S}\times(\hat{\cal S}\times\hat{N})]^3}, \nonumber\\
 &&\nonumber\\
\beta&=&\tan^{-1}\frac{{\cal S}^2}{{\cal S}^1},\nonumber\\
 &&\nonumber\\
 \gamma&=&\tan^{-1}\frac{\overline{\cal S}^2}{\overline{\cal
 S}^1},
\nonumber\\
 &&\nonumber\\
\xi&=&\frac{\vec{\cal
W}\cdot(\hat{N}\times\hat{\chi})}{\sqrt{1-\vec{N}^2}}.
 \label{7.11}
 \end{eqnarray}

\hfill

The canonical variables in the final basis have been separated in
three sectors. The second and the third sector in the previous
table are constituted by canonical variables scalar under
rotations: these variables describe the {\em shape} of the fluid
and we call them {\em shape (or vibrational) variables} (see
Ref.\cite{30} for the original definition of shape variables). The
first sector is that of the {\em rotational (or orientational)
variables}; these variables describe the rotational motion of the
dynamical body frame.

\medskip

It is useful to analyze the variables using a new orthonormal
base: the {\em spin basis} of Refs. \cite{31,13,14}. This basis is
defined observing that there is  only a unit vector $\hat{\cal R}$
on the same plane of $\vec{N}$ and $\vec{\cal S}$, orthogonal to
$\vec{\cal S}$ such that

\begin{equation}
\alpha=\tan^{-1}\frac{(\hat{\cal S}\times\hat{\cal R})^3}
{[\hat{\cal S}\times(\hat{\cal S}\times\hat{\cal R})]^3}.
 \label{7.12}
 \end{equation}

\medskip

Then the three unit vectors $(\hat{\cal R},\hat{\cal S},\hat{\cal
S}\times\hat{\cal R})$ are a orthonormal basis, the {\em spin
basis}. By construction their components are given by the
following relations

\begin{equation}
\left\{
\begin{array}{l}
\hat{\cal S}^1= \frac{\sqrt{\vec{\cal S}^2-({\cal S}^3)^2}}
{\mid\vec{\cal S}\mid}\cos\beta,\\
\\
\hat{\cal S}^2= \frac{\sqrt{\vec{\cal S}^2-({\cal S}^3)^2}}
{\mid\vec{\cal S}\mid}\sin\beta,\\
\\
\hat{\cal S}^3=\frac{{\cal S}^3}{\mid\vec{\cal S}\mid},
\end{array}
\right.
 \label{7.13}
 \end{equation}

\hfill

\begin{equation}
\left\{
\begin{array}{l}
\hat{\cal R}^1=\sin\beta\sin\alpha- \frac{{\cal
S}^3}{\mid\vec{\cal S}\mid}\cos\beta\cos\alpha,
\\
\\
\hat{\cal R}^2=-\cos\beta\sin\alpha- \frac{{\cal
S}^3}{\mid\vec{\cal S}\mid}\sin\beta\cos\alpha,
\\
\\
\hat{\cal R}^3=\frac{\sqrt{\vec{\cal S}^2-({\cal S}^3)^2}}
{\mid\vec{\cal S}\mid}\cos\alpha,
\end{array}
\right.
 \label{7.14}
 \end{equation}

\hfill

\begin{equation}
\left\{
\begin{array}{l}
(\hat{\cal S}\times\hat{\cal R})^1=\sin\beta\sin\alpha+
\frac{{\cal S}^3}{\mid\vec{\cal S}\mid}\cos\beta\sin\alpha,
\\
\\
(\hat{\cal S}\times\hat{\cal R})^2=-\cos\beta\cos\alpha+
\frac{{\cal S}^3}{\mid\vec{\cal S}\mid}\sin\beta\sin\alpha,
\\
\\
(\hat{\cal S}\times\hat{\cal R})^3=\frac{\sqrt{\vec{\cal
S}^2-({\cal S}^3)^2}} {\mid\vec{\cal S}\mid}\sin\alpha.
\end{array}
\right.
 \label{7.15}
 \end{equation}

\medskip

We also define the angle $\psi$ such that

\begin{equation}
\cos\psi=\frac{\overline{\cal S}^3}{\mid\vec{\cal S}\mid}
;\;\;\;\; \sin\psi=\frac{\sqrt{\vec{\cal S}^2-(\overline{\cal
S}^3)^2}} {\mid\vec{\cal S}\mid}.
 \label{7.16}
 \end{equation}

By definition of $\hat{\cal R}$ we have

 \begin{equation}
\hat{N}=\cos\psi\;\hat{\cal S}+\sin\psi\;\hat{\cal R}.
 \label{7.17}
 \end{equation}

Moreover the definition of $\gamma$ implies that we have

\begin{eqnarray}
\hat{\cal S}\cdot\hat{\chi}&=&\sin\psi\cos\gamma,\nonumber\\
 \hat{\cal S}\cdot(\hat{\chi}\times\hat{N})
&=&\sin\psi\sin\gamma.
 \label{7.18}
 \end{eqnarray}

\medskip

We complete the conditions on $\hat{\chi},\hat{N}\times\hat{\chi}$
using the fact that the {\em dynamical body frame} and the {\em
spin basis} are connected by a (proper) rotation

\begin{eqnarray}
\hat{\chi}&=&\sin\psi\cos\gamma\;\hat{\cal S}-
           \cos\psi\cos\gamma\;\hat{\cal R}+
       \sin\gamma\;\hat{\cal S}\times\hat{\cal R},\nonumber\\
\hat{\chi}\times\hat{N}&=&\sin\psi\sin\gamma\;\hat{\cal S}-
           \cos\psi\sin\gamma\;\hat{\cal R}-
       \cos\gamma\;\hat{\cal S}\times\hat{\cal R}.
 \label{7.19}
 \end{eqnarray}

\medskip

Substituting in Eqs.(\ref{7.17}),(\ref{7.19}) the expression given
by Eqs.(\ref{7.13}),(\ref{7.14}),(\ref{7.15}),(\ref{7.16}) we
obtain the elements of the {\em dynamical body frame} expressed as
functions of the {\em rotational variables} alone. We can also
define the {\em Euler's angles} of the {\em dynamical body frame}
\footnote{As in Refs. \cite{13,14} we adopt the {y-convention} of
Ref. \cite{32}} $\varepsilon^1, \varepsilon^2, \varepsilon^3$

\begin{equation}
\cos\varepsilon^2 = \hat{N}^3;\;\; \cos\varepsilon^1 =
\frac{\hat{N}^1}{\sqrt{1-(\hat{N}^3)^2}} ;\;\; \cos\varepsilon^3 =
- \frac{\hat{\chi}^3}{\sqrt{1-(\hat{N}^3)^2}}.
 \label{7.20}
 \end{equation}

They are as functions of the {\em rotational variables} alone.

\medskip

The Euler's angles (\ref{7.20}) together with the relative angular
momentum components

\begin{equation}
\left\{
\begin{array}{l}
\overline{\cal S}^1= \sqrt{\vec{\cal S}^2-(\overline{\cal S}^3)^2}
\cos\gamma,\\
\\
\overline{\cal S}^2= \sqrt{\vec{\cal S}^2-(\overline{\cal S}^3)^2}
\sin\gamma,\\
\\
\overline{\cal S}^3.
\end{array}
\right.
 \label{7.21}
 \end{equation}

\noindent define a non canonical transformation for the rotational
sector. The corresponding canonical transformation is obtained
using the canonical momenta

\begin{eqnarray}
p_1 &=&-\sin\varepsilon^2 \cos\varepsilon^3\, \overline{\cal S}^1
+ \sin\varepsilon^2\, \sin\varepsilon^3\,\overline{\cal S}^2+
\cos\varepsilon^2\,\overline{\cal S}^3, \nonumber\\
 &&\nonumber\\
p_2&=&\sin\varepsilon^3\,\overline{\cal S}^1+
\cos\varepsilon^3\,\overline{\cal S}^2, \qquad p_3 =
\overline{\cal S}^3,\nonumber \\
 &&\nonumber \\
 &&\{ \varepsilon^r, \varepsilon^s \} = \{ p_r, p_s \} = 0,\qquad \{
 \varepsilon^r, p_s \} = \delta^r_s.
 \label{7.22}
 \end{eqnarray}

The inverses of the previous equations are

\begin{eqnarray}
\overline{\cal S}^1&=&\sin\varepsilon^3\,p_2 -
\frac{\cos\varepsilon^3}{\sin\varepsilon^2}\,p_1 +
\cos\varepsilon^3 \, \cot\varepsilon^2\,p_3,\nonumber\\
 &&\nonumber\\
\overline{\cal S}^2&=&\cos\varepsilon^3\, p_2+
\frac{\sin\varepsilon^3}{\sin\varepsilon^2}\, p_1 -
\sin\varepsilon^3\, \cot\varepsilon^2\, p_3,\nonumber\\
 &&\nonumber\\
\overline{\cal S}^3&=&p_3,
 \label{7.23}
 \end{eqnarray}

\noindent and we get

\begin{eqnarray}
 {\cal S}^3 &=& p_1 = - \sin\varepsilon^2\cos\varepsilon^3\,
\,\overline{\cal S}^1 + \sin\varepsilon^2\sin\varepsilon^3 \,\overline{\cal S}^2 +
 \cos\varepsilon^2\, \overline{\cal S}^3,\nonumber \\
&&\nonumber\\
 \beta &=& \left(\varepsilon^1-\frac{\pi}{2}\right)-\arctan
\frac{\sin\varepsilon^2\,(-\sin\varepsilon^2 \cos\varepsilon^3\, \overline{\cal S}^1
+ \sin\varepsilon^2\, \sin\varepsilon^3\,\overline{\cal S}^2+
\cos\varepsilon^2\,\overline{\cal S}^3)-\overline{\cal S}^3}
{\sin\varepsilon^2\,(\sin\varepsilon^3\,\overline{\cal S}^1+
\cos\varepsilon^3\,\overline{\cal S}^2)} \nonumber\\
&&\nonumber\\
 \alpha &=& \arctan\frac{\sin\varepsilon^2(\sin\varepsilon^3\,\overline{\cal S}^1+
\cos\varepsilon^3\,\overline{\cal S}^2)\,\mid\vec{\cal S}\mid}
 {\cos\varepsilon^2\, \mid\vec{\cal S}\mid^2-
 (-\sin\varepsilon^2 \cos\varepsilon^3\, \overline{\cal S}^1
+ \sin\varepsilon^2\, \sin\varepsilon^3\,\overline{\cal S}^2+
\cos\varepsilon^2\,\overline{\cal S}^3)\,\overline{\cal S}^3}\nonumber \\
&&\nonumber\\
\mid\vec{\cal S}\mid^2 &=& \sum_r\, \left({\overline{\cal S}}^r\right)^2.
 \label{7.24}
 \end{eqnarray}

\medskip

This chain of transformations can be represented with the table

\begin{equation}
\begin{array}{|ccc|}
\hline
\mid\vec{\cal S}\mid&{\cal S}^3&\overline{\cal S}^3
\\
\alpha&\beta&\gamma\\
\hline
\end{array}
\longrightarrow
\begin{array}{|ccc|}
\hline
\overline{\cal S}^1&\overline{\cal S}^2&\overline{\cal S}^3
\\
\varepsilon^1&\varepsilon^2&\varepsilon^3\\ \hline
\end{array}
\longrightarrow
\begin{array}{|ccc|}
\hline p_1&p_2&p_3
\\
\varepsilon^1&\varepsilon^2&\varepsilon^3\\ \hline
\end{array},
 \label{7.25}
 \end{equation}

\noindent and with the following Poisson brackets for the
non-canonical variables $\varepsilon^r$, $\overline{\cal S}^r$
\cite{29} ($\overline{f}$, $\overline{g}$ are functions only of these
variables)

\begin{eqnarray}
 && \{\varepsilon^r,\varepsilon^s\}=0,
\qquad
\{\overline{\cal S}^r,\overline{\cal S}^s\} = -\epsilon^{rsu}\,\overline{\cal S}^u,
\nonumber\\
&&\nonumber\\
 && \{\varepsilon^r,\overline{\cal S}_s\} ={\check X}^{(R)\, r}{}_s(\varepsilon^u),
\nonumber\\
&&\nonumber\\
 &&\{\overline{f},\overline{g}\}={\check X}^{(R)\, r}{}_s(\varepsilon^u)\,
\left(
 \frac{\partial \overline{f}}{\partial \varepsilon^r}\,
 \frac{\partial \overline{g}}{\partial \overline{\cal S}_s} -
 \frac{\partial \overline{f}}{\partial \overline{\cal S}_s}\,
 \frac{\partial \overline{g}}{\partial \varepsilon^r}
\right) -
\vec{\cal S}\cdot\left( \frac{\partial\overline{f}}{\partial \vec{\cal S}}\times
\frac{\partial \overline{g}}{\partial \vec{\cal S}}\right),
 \label{7.26}
 \end{eqnarray}

\noindent where ${\check X}^{(R)\, r}{}_s(\varepsilon^u)$ are the
components of the right invariant vector fields on the group
manifold of $SO(3)$. The components of the dual right invariant
one-forms are
\[\Lambda^{(R)\, r}{}_s = \left(
\begin{array}{ccc}
-\sin\varepsilon^2\,\cos\varepsilon^3 & \sin\varepsilon^3 & 0\\
\sin\varepsilon^2\,\sin\varepsilon^3 & \cos\varepsilon^3 & 0\\
\cos\varepsilon^2 & 0 & 1
\end{array}
\right) = \left[{\check X}^{(R)\,
-1}\right]^r{}_s,
\]
if the Euler angles are defined by the convention
$R(\varepsilon^r) = R_3(\varepsilon^1)\, R_2(\varepsilon^2)\,
R_3(\varepsilon^3)$.

These transformations stress the canonical equivalence between the
rotational variables and the canonical phase space of a non
relativistic rigid body \cite{32,33}; in other words, the
rotational variables are the rigid body-like variables, whereas
the {\em shape} variables describes the {\em non-rigidity} of the
system.

\hfill

Using the results obtained in the three body case in Refs.
\cite{13,14} it is easy to construct the inverse canonical
transformation and to express the original relative variables in
terms of the rotational and shape variables. In particular it is
trivial to observe that by using the rotation $R^r_{\,s}
(\varepsilon^1, \varepsilon^2, \varepsilon^3)$,  for
$\underline{n}\neq\underline{u}_1,\underline{u}_2$ we obtain
immediately

\begin{eqnarray}
r^r_{\underline{n}}&=&R^r_{\,s}
(\varepsilon^1,\varepsilon^2,\varepsilon^3)
\,\overline{r}^s_{\underline{n}},\nonumber\\
 &&\nonumber\\
p^r_{\underline{n}}&=& R^r_{\,s}
(\varepsilon^1,\varepsilon^2,\varepsilon^3)
\,\overline{p}^s_{\underline{n}}.
 \label{7.27}
 \end{eqnarray}

For $\underline{n}=\underline{u}_1,\underline{u}_2$ we can to use
the three body results of Refs.\cite{13,14}. In particular we have

\begin{eqnarray}
\overline{r}^1_{\underline{u}_i}
&=&\vec{r}_{\underline{u}_i}\cdot\hat{\chi}=(-)^{i+1}
R_i\sqrt{1-\vec{N}^2},\nonumber\\
 \overline{r}^2_{\underline{u}_i}&=&
\vec{r}_{\underline{u}_i}\cdot(\hat{\chi}\times\hat{N})=0,\nonumber\\
\overline{r}^3_{\underline{u}_i}&=&
\vec{r}_{\underline{u}_i}\cdot\hat{N}= R_i\mid\vec{N}\mid.
 \label{7.28}
 \end{eqnarray}

Then we obtain

\begin{equation}
r^r_{\underline{u}_i}= R^r_{\,s}
(\varepsilon^1,\varepsilon^2,\varepsilon^3)
\,\overline{r}^s_{\underline{u}_i}.
 \label{7.29}
 \end{equation}

Finally, if we define

\begin{equation}
\overline{\cal S}^r_{(12)}=\overline{\cal S}^r-
\sum_{\underline{n}\neq\underline{u}_1,\underline{u}_2}
\epsilon^{ruv}\overline{r}^u_{\underline{n}}
\overline{p}^s_{\underline{n}},
 \label{7.30}
 \end{equation}

\noindent we have

\begin{eqnarray}
\overline{p}_{\underline{u}_i}^1 &=&(-)^{i+1}\tilde{\Pi}_i
\sqrt{1-\vec{N}^2}+ \frac{\mid\vec{N}\mid}{2R_i}\left[
\overline{\cal S}^2_{(12)} +(-)^{i+1}\xi\sqrt{1-\vec{N}^2}
\right],\nonumber\\
 &&\nonumber\\
\overline{p}_{\underline{u}_i}^2 &=& (-)^{i+1}\frac{1}{2R_i}\left[
-(-)^{i+1} \frac{\overline{\cal S}^1_{(12)}}{\mid\vec{N}\mid}+
\frac{\overline{\cal S}^3_{(12)}}{\sqrt{1-\vec{N}^2}}
\right],\nonumber\\
 &&\nonumber\\
\overline{p}_{\underline{u}_i}^3 &=& \tilde{\Pi}_i
\mid\vec{N}\mid-(-)^{i+1} \frac{\sqrt{1-\vec{N}^2}}{2R_i}\left[
\overline{\cal S}_{(12)}^2 +(-)^{i+1}\xi\sqrt{1-\vec{N}^2}
\right],
 \label{7.31}
 \end{eqnarray}

\noindent and then

\begin{equation}
p^r_{\underline{u}_i}=R^r_{\,s}
(\varepsilon^1,\varepsilon^2,\varepsilon^3)
\,\overline{p}^s_{\underline{u}_i}.
 \label{7.32}
 \end{equation}

Therefore we get

\begin{eqnarray}
 \Sigma^r(\tau , {\vec \sigma}_o) &=& R^r{}_s(\varepsilon^u)\,
 \sum_{\underline{n}}\, \overline{r}^s_{\underline{n}}(\tau )\, \Big[
 \Gamma^{\Sigma}_{\underline{n}}({\vec \sigma}_o) -
 h_{\underline{n}}\Big]\, {\rightarrow}_{\tau \rightarrow 0}\,
 \sigma_o^r,\nonumber \\
&&\nonumber\\
 K^r(\tau ,{\vec \sigma}_o) &=&  R^r{}_s(\varepsilon^u)\,
 \sum_{\underline{n}}\, \overline{p}^s_{\underline{n}}(\tau )\,
 \Gamma^K_{\underline{n}}({\vec \sigma}_o).
 \label{7.33}
 \end{eqnarray}

\vfill\eject

\section{The Invariant Mass and the Equations of Motion.}

As we have seen, the invariant mass (Dixon's mass monopole as we
shall see in the next Section) ${\cal M} = \int d^3\sigma_o\,
\Delta (\tau , {\vec \sigma}_o)$ is the Hamiltonian on the
Wigner hyper-planes in the gauge $T \approx \tau$ and ${\vec {\cal
Q}} \approx 0$, where $\vec \Sigma (\tau ,{\vec \sigma}_o)$ and
$\vec K (\tau ,{\vec \sigma}_o)$ are given by Eqs.(\ref{7.33}).
Therefore we have ${d\over {d\tau }}\, {\cal M} = 0$, but it can
be shown that ${{\partial}\over {\partial \tau}}\, \Delta (\tau
,{\vec \sigma}_o) \not= 0$. Moreover, since $h_{\underline{n}}$ in
Eqs.(\ref{7.33}) depends on both configuration and momentum shape
variables, it can  be shown that we have ${d\over {d\tau}}\,
h_{\underline{n}} \not= 0$, namely

\begin{eqnarray}
 {{\partial \Sigma^r(\tau ,{\vec \sigma}_o)}\over {\partial \tau}}
&=& \Big(\vec \omega (\tau ) \times \vec \Sigma (\tau ,{\vec
\sigma}_o)\Big)^r + \nonumber\\
&&\nonumber\\
&+&R^r{}_s(\varepsilon^u(\tau ))\,
\sum_{\underline{n}}\, \left( {{d \overline{r}^s_{\underline{n}}(\tau
)}\over {d\tau}}\, \Big[\Gamma^{\Sigma}_{\underline{n}}(\vec{\sigma}_o) -
h_{\underline{n}}(\tau )\Big]
- \overline{r}^s_{\underline{n}}(\tau )\, {{d h_{\underline{n}}(\tau )}\over
{d\tau}}\right),
 \label{8.1}
 \end{eqnarray}

\noindent with the body frame components of the angular velocity
being $\overline{\omega}^r(\varepsilon^u(\tau ), {\dot \varepsilon}^u(\tau
)) = - {1\over 2}\, \epsilon^{ruv}\, \Big[R^T\, \dot
R\Big]^{uv}(\varepsilon^u(\tau ), {\dot \varepsilon}^u(\tau ))$ .

\medskip

As already said, for every equation of state  the mass density
$\Delta (\tau ,{\vec \sigma}_o)$ is a suitable function of
$n_o({\vec \sigma}_o)$, ${\vec K}^2(\tau ,{\vec \sigma}_o) /
n^2_o({\vec \sigma}_o)$ and $n_o({\vec \sigma}_o) / \det \Big(
{{\partial \Sigma}\over {\partial \sigma_o}} \Big)$. While the
last term (absent only in the case of dust, because $p = 0$) is
determined by Eq.(\ref{6.31}), from Eq.(\ref{6.30}) we get the
following expression of the second term

\begin{eqnarray}
 {{ {\vec K}^2(\tau ,{\vec \sigma}_o)}\over { n^2_o({\vec
\sigma}_o) }} &=& {1\over {n_o^2({\vec \sigma}_o)}}\,
\sum_{\underline{n}{}_1,\underline{n}{}_2}\,
\Gamma^K_{\underline{n}{}_1}({\vec \sigma}_o)\,
\Gamma^K_{\underline{n}{}_2}({\vec \sigma}_o)\, {\vec
p}_{\underline{n}{}_1}(\tau ) \cdot {\vec
p}_{\underline{n}{}_2}(\tau ) = \nonumber \\
 &=& {1\over {n_o^2({\vec \sigma}_o)}}\, \Big[ \sum_{i=1}^2\,
 \Big( \Gamma^K_{\underline{u}{}_i}({\vec \sigma}_o)\Big)^2\,
 {\vec p}^2_{\underline{u}{}_i}(\tau ) + 2\, \Gamma^K
_{\underline{u}{}_1}({\vec \sigma}_o)\, \Gamma^K
_{\underline{u}{}_2}({\vec \sigma}_o)\, {\vec
 p}_{\underline{u}{}_1}(\tau ) \cdot {\vec
 p}_{\underline{u}{}_2}(\tau ) +\nonumber \\
 &+& 2\, \sum_{i=1}^2\, \Gamma^K_{\underline{u}{}_i}({\vec
 \sigma}_o)\, \sum_{\underline{n} \not= \underline{u}{}_1,
 \underline{u}{}_2}\, \Gamma^K_{\underline{n}}({\vec \sigma}_o)\,
 {\vec p}_{\underline{u}{}_i}(\tau ) \cdot {\vec
 p}_{\underline{n}}(\tau ) +\nonumber \\
 &+& \sum_{\underline{n}{}_1 \not=
 \underline{u}{}_1,\underline{u}{}_2}\, \sum_{\underline{n}{}_2
 \not= \underline{u}{}_1, \underline{u}{}_2}\,
 \Gamma^K_{\underline{n}{}_1}({\vec \sigma}_o)\,
 \Gamma^K_{\underline{n}{}_2}({\vec \sigma}_o)\,
 {\vec p}_{\underline{n}{}_1}(\tau ) \cdot {\vec
 p}_{\underline{n}{}_2}(\tau ) \Big] .
 \label{8.2}
 \end{eqnarray}

\medskip

By using Eqs. (\ref{7.10}) and (\ref{7.31}) this term can be
expressed in the non-canonical basis $\varepsilon^r$, $\overline{\cal
S}^r$, $\mid\vec N\mid$, $\xi$, $R_1$, ${\tilde \Pi}_1$, $R_2$,
${\tilde \Pi}_2$, $\overline{r}^s_{\underline{n} \not=
 \underline{u}{}_1,\underline{u}_2}$, $\overline{p}^s_{\underline{n} \not=
 \underline{u}{}_1,\underline{u}_2}$. The result is that the
 invariant mass density

 i) is independent from the Euler angles $\varepsilon^r$;

 ii) contains terms bilinear and linear in the body frame
 components $\overline{\cal S}^r$ of the spin.

\medskip

However, $\Delta (\tau ,{\vec \sigma}_o)$ is a complicated
function of these three terms. The simplest expression is obtained
in the case of dust, where $\Delta (\tau ,{\vec \sigma}_o) =
\sqrt{[\mu\, n_o({\vec \sigma}_o)]^2 + {\vec K}^2(\tau ,{\vec
\sigma}_o)}$. Since the body frame components of the angular
velocity are defined \cite{13,14,22} as \footnote{From now on we
shall use the following notations: i) $q^{\mu}$, $p_{\mu}$ will
denote all the canonically conjugate shape variables $|\vec N|$,
$\xi$, $R_1$, ${\tilde \Pi}_1$, $R_2$, ${\tilde \Pi}_2$,
$\overline{ r}^s_{\underline{n} \not=
\underline{u}{}_1,\underline{u}_2}$,
$\overline{p}^s_{\underline{n} \not= \underline{u}{}_1,
\underline{u}_2}$; ii) $q^{\alpha}$, $p_{\alpha}$ will denote all
the canonically conjugate shape variables  $\overline{
r}^s_{\underline{n} \not= \underline{u}{}_1,\underline{u}_2}$,
$\overline{p}^s_{\underline{n} \not= \underline{u}{}_1,
\underline{u}_2}$ not including the first three pairs connected
with the choice of the body frame axes.}

\begin{equation}
 \overline{\omega}_r(\varepsilon^s) = {{\partial {\cal M}(\tau )}\over
 {\partial \overline{\cal S}^r}} = F_{rs}(\overline{\cal S}^u, q^{\mu},
 p_{\mu})\, \overline{\cal S}^s + G_r(\overline{\cal S}^u, q^{\mu},
 p_{\mu}),
 \label{8.3}
 \end{equation}

\noindent it turns out that there is no linear relation between
the spin and the angular velocity like in the non-relativistic
rigid body (this property is true also for non-relativistic
non-rigid bodies \cite{13}).

\bigskip

We can write the Hamilton equations for the orientational
variables $\varepsilon^r$, $\overline{\cal S}^r$ and for the shape
variables. From them we can deduce the equations of motion for the
orientational variables  $\alpha$, $\overline{\cal S}^3$, $\gamma$
in Eqs.(\ref{7.10}) (the other three $\mid\vec{\cal S}\mid$,
${\cal S}^3$, $\beta$ are Noether constants of motion). These
three variables are not constant of motion for deformable bodies:
they are coupled to the shape variables and {\it describe how the
dynamical body frame rotates when the body changes its shape}. In
particular $\alpha$, being conjugate to the constant of motion
$\mid\vec{\cal S}\mid$, is an ignorable variable (${\cal M}$,
expressed in the canonical basis (\ref{7.10}), does not depend on
it).

\medskip

Three types of configuration for the motion of the fluid are
interesting:

i) {\it Pure rotational motion} - It is defined by constant shape
configuration variables ${\dot q}^{\mu} = 0$. With this condition
the Hamilton equation ${\dot q}^{\mu}\, {\buildrel \circ \over
=}\, \{ q^{\mu}, {\cal M} \}$ become a system of algebraic
equations for the shape momenta $p^{(o)}_{\mu} = p_{\mu}{|}_{\dot
q = 0}$. Even if we cannot find the explicit solution, its form is
of the type $p_{\mu}^{(o)} = \sum_r\, \overline{\cal S}^r\, {\cal
C}^r_{\mu}(\overline{\cal S}^u, q^{\nu})$. The purely rotational
Hamiltonian is ${\cal M}^{(rot)} = {\cal M}{|}_{p_{\mu} =
p^{(o)}_{\mu}}$. Therefore for the dust invariant mass density we
get

\begin{equation}
\Delta^{(rot)}(\tau ,{\vec \sigma}_o) = \sqrt{ [\mu\,
n_o({\vec \sigma}_o)]^2 + \sum_{rs}\,A^{rs}({\vec \sigma}_o,
q^{\mu}(\tau ), \overline{\cal S}^u )\, \overline{\cal S}^r\,
\overline{\cal S}^s}.
\label{8.4}
\end{equation}

However Eqs.(\ref{8.1}) show that the {\it generalized Eulerian
coordinates} (i.e. the flux lines in adapted coordinates) {\it do
not perform a rigid motion}:

\begin{eqnarray}
{{\partial \Sigma^r(\tau ,{\vec
\sigma}_o)}\over {\partial \tau}}{|}_{\dot q = 0} &=& \Big(\vec
\omega (\tau ) \times \vec \Sigma (\tau ,{\vec \sigma}_o)\Big)^r -
R^r{}_s(\varepsilon^u(\tau ))\, \sum_{\underline{n}}\, \overline{
r}^s_{\underline{n}}(\tau )\, {{\partial h_{\underline{n}}(\tau
)}\over {\partial p^{(o)}_{\nu}}}\, {\dot p}^{(o)}_{\mu}(\tau )\, \neq\nonumber\\
&&\nonumber\\
&\not=&
\Big(\vec \omega (\tau ) \times \vec \Sigma (\tau ,{\vec
\sigma}_o)\Big)^r.
\label{8.5}
\end{eqnarray}

\medskip

ii) {\it Pure vibrational motion} - It is defined by the vanishing
of the angular velocity, so that Eq.(\ref{8.3}) determine the body
frame components of the spin in terms of the shape variables:
$\overline{\cal S}^r{|}_{\vec \omega = 0} = \overline{\cal
S}^r_{(o)}(q^{\mu}, p_{\mu}) \not= 0$. By putting this expression
in ${\cal M}$ gives a purely vibrational Hamiltonian ${\cal
M}^{(vib)}$.

\medskip

iii) {\it Small shape momenta} - We can study configurations in
which the shape momenta are very small. We can define the
following two approximations (we use the dust to illustrate them):

\medskip

a) $p_{\mu} \sim 0$ - The dust invariant mass density becomes
\begin{equation}
\Delta (\tau , {\vec \sigma}_o) \sim \sqrt{[\mu\, n_o({\vec
\sigma}_o)]^2 + \sum_{rs} C^{rs}({\vec \sigma}_o, q^{\mu})\, \overline
{\cal S}^r\, \overline{\cal S}^s}.
\label{8.6}
\end{equation}

If we make a Taylor expansion around the canonical center of mass
${\vec {\cal Q}} = 0$ of the function [see also Eq. (\ref{9.2})]

\begin{equation}
\widehat{T}^{\tau\tau}(\tau,\vec{\sigma}_o)=
{\det}^{-1}\left(\frac{\partial \Sigma}{\partial\sigma_o}\right)
\Delta(\tau,\vec{\sigma}_o)
\label{8.7}
\end{equation}

\noindent and if we define

\begin{equation}
D_o(\tau)={\det}^{-1}
\left(\frac{\partial \Sigma}{\partial\sigma_o}\right)_{\vec{\sigma}_o=0}
\label{8.8}
\end{equation}

\noindent we can define the following new $\Delta$-multipolar
expansion of the invariant mass ($V(\tau )$ is the volume of the
fluid)

\begin{eqnarray}
{\cal M} &\sim & V(\tau )\,D_o(\tau)\,\sqrt{[\mu\, n_o({\vec
\sigma}_o)]^2
+\sum_{rs} C^{rs}(\vec 0, q^{\mu})\, \overline{\cal
S}^r\, \overline{\cal S}^s} +\nonumber\\
&&\nonumber\\
&+& \sum_{n=1}^{\infty}\,
\sum_{r_1..r_n}\, \widehat{T}_{r_1..r_n}(q^{\mu}, \overline{\cal
S}^u)\, \int_{V(0)}\, d^3\sigma_o\,
\det\left(\frac{\partial \Sigma}{\partial\sigma_o}\right) \sigma_o^{r_1} ...
\sigma_o^{r_n},
\label{8.9}
\end{eqnarray}

\noindent where

\begin{equation}
\widehat{T}_{r_1..r_n}=
\left[\frac{\partial}{\partial\sigma_o^{r_1}...\partial\sigma_o^{r_n}}
\widehat{T}^{\tau\tau}(\tau,\vec{\sigma}_o)\right]_{\vec{\sigma}_o=0}.
\label{8.10}
\end{equation}

The first term in this multipolar expansion is just the {\it
relativistic rotator} defined in Eq.(5.12) of Ref. \cite{27} and $
C^{rs}(\vec 0, q^{\mu})$ plays the role  of the inverse of the
tensor of inertia.

\medskip

b) Only $p_{\alpha} \sim 0$ - If we denote $U^i$ the 3 shape
variables $|\vec N|$, $R_1$, $R_2$ and $V_i$ their conjugate
momenta $\xi$, ${\tilde \Pi}_1$, ${\tilde \Pi}_2$, the dust
invariant mass density becomes
\[
\Delta (\tau , {\vec \sigma}_o)
\sim \sqrt{[\mu\, n_o({\vec \sigma}_o)]^2 + \sum_{rs} C^{rs}({\vec
\sigma}_o, q^{\alpha},U^i)\, \overline{\cal S}^r\, \overline{\cal S}^s
+ \sum_r\, B^r({\vec \sigma}_o, q^{\alpha},U^i, V_i)\, \overline{\cal
S}^r + H_{(3)}(U^i, V_i)},
\]
where $H_{(3)}$
 is the
non-relativistic vibrational (namely only function of the shape
variables) Hamiltonian for the 3-body problem \cite{13}. If we
repeat the previous Taylor expansion ($\Delta$-multipolar expansion), the first
term will be
\[
V(\tau ) \,D_o(\tau)\, \sqrt{[\mu\, n_o({\vec \sigma}_o)]^2 +
\sum_{rs} C^{rs}(\vec 0, q^{\alpha},U^i)\, \overline{\cal S}^r\,
\overline{\cal S}^s + \sum_r\, B^r(\vec 0, q^{\alpha},U^i, V_i)\,
\overline{\cal S}^r + H_{(3)}(U^i, V_i)}
\]
It corresponds to a
generalized rotator interacting with a 3-body problem parametrized
by the two dipoles $R_1$, $R_2$, originally along two of the
chosen body frame axes, and by the angle between them, described
by $|\vec N|$.

\vfill\eject

\section{Dixon's Multipoles.}

In this Section, following the general treatment given in
Ref.\cite{17}, we shall give Dixon's multipoles \cite{16,17} for
the energy-momentum tensor of the perfect fluid in the rest-frame
instant form on the Wigner hyper-planes in the gauge $T \approx
\tau$ and ${\vec {\cal Q}} \approx 0$.

\bigskip

On the Wigner hyper-planes the energy-momentum tensor (\ref{2.22})
is rewritten in the form

\begin{equation}
 T^{AB}(\tau , \vec \sigma ) = \int d^3\sigma_o\, \det \left(
 {{\partial \Sigma}\over {\partial \sigma_o}}\right)\, \delta^3(\vec
 \sigma - \vec \Sigma (\tau ,{\vec \sigma}_o))\, {\hat T}^{AB}(\tau
 ,{\vec \sigma}_o),
 \label{9.1}
 \end{equation}

\noindent with ($\widehat{U}^A$ and ${\cal R}$ are defined in Eqs.
(\ref{3.16}) and (\ref{3.10}), respectively)

\begin{eqnarray}
 \widehat{T}^{AB}(\tau ,{\vec \sigma}_o) &=& \left[ \left( {{\partial
 \rho}\over {\partial \hat n}}\, \hat n\right)\, \widehat{U}^A\,
 \widehat{U}^B + \left( \rho - {{\partial \rho}\over {\partial
 \hat n}}\, \hat n\right)\, g^{AB}\right] (\tau ,{\vec
 \sigma}_o),\nonumber \\
 &&\nonumber \\
 \widehat{U}^A(\tau ,{\vec \sigma}_o) &=& {1\over {{\cal
 R}(\tau ,{\vec \sigma}_o)}}\, \Big( 1;\, {{\partial \vec \Sigma
 (\tau ,{\vec \sigma}_o)}\over {\partial \tau}}\Big),\nonumber \\
 &&\nonumber \\
 \widehat{T}^{\tau\tau}(\tau ,{\vec \sigma}_o) &=& \Delta (\tau
 ,{\vec \sigma}_o)\, {\det}^{-1} \left( {{\partial \Sigma}\over
 {\partial \sigma_o}}\right),\nonumber \\
&&\nonumber \\
 \widehat{T}^{\tau r}(\tau ,{\vec \sigma}_o) &=& K^r(\tau ,{\vec
 \sigma}_o)\,  {\det}^{-1} \left( {{\partial \Sigma}\over
 {\partial \sigma_o}}\right),\nonumber \\
&&\nonumber \\
 \widehat{T}^{rs}(\tau ,{\vec \sigma}_o) &=& {{
{\det}^{-1} \left( {{\partial \Sigma}\over
 {\partial \sigma_o}}\right)\, K^r(\tau ,{\vec
 \sigma}_o)\, K^s(\tau ,{\vec \sigma}_o)}\over {\sqrt{
 \left(n_o({\vec \sigma}_o)\, {{\partial \rho}\over {\partial
 X}}(X,\hat{s}_o)\right)^2 + {\vec K}^2(\tau, {\vec \sigma}_o) }}} + \hat p(X,\hat{s}_o)\,
 \delta^{rs}.
 \label{9.2}
 \end{eqnarray}

\bigskip

Let us consider a world-line $w^{\mu}(\tau ) = x^{\mu}(\tau ) +
\epsilon^{\mu}_r(p)\, \zeta^r(\tau )$, where $x^{\mu}(\tau )$
is the centroid, origin of the 3-coordinates on Wigner
hyper-planes. Dixon's multipoles of the energy-momentum tensor
with respect to this world-line  are defined as

\begin{eqnarray}
 t^{\mu_1...\mu_n\mu\nu}(w(\tau )) &=& \int d^3\sigma\, \left(
 z^{\mu_1}(\tau ,\vec \sigma ) - w^{\mu_1}(\tau )\right) ...
 \left( z^{\mu_n}(\tau ,\vec \sigma ) - w^{\mu_n}(\tau )\right)\,
 T^{\mu\nu}(z(\tau ,\vec \sigma )) =\nonumber \\
&&\nonumber\\
 &=& \epsilon^{\mu_1}_{r_1}(p) ...  \epsilon^{\mu_n}_{r_n}(p)\,
  \epsilon^{\mu}_A(p)\,  \epsilon^{\nu}_B(p)\, q^{r_1 .. r_n
  AB}({\vec \zeta}(\tau )).
  \label{9.3}
  \end{eqnarray}

\medskip

The rest-frame instant form multipoles are

\begin{eqnarray}
 &&q^{r_1 .. r_n AB}({\vec \zeta}(\tau )) = \int d^3\sigma\,
 \left( \sigma^{r_1} - \zeta^{r_1}(\tau )\right) ...
 \left( \sigma^{r_n} - \zeta^{r_n}(\tau )\right)\, T^{AB}(\tau
 ,\vec \sigma ) =\nonumber \\
&&\nonumber\\
 &=& \int d^3\sigma_o\, \left( \Sigma^{r_1}(\tau ,{\vec \sigma}_o)
 - \zeta^{r_1}(\tau )\right) ... \left( \Sigma^{r_1}(\tau ,{\vec \sigma}_o)
 - \zeta^{r_1}(\tau )\right) \det\, \left( {{\partial
 \Sigma}\over {\partial \sigma_o}}\right)\, \widehat{T}^{AB}(\tau ,{\vec
 \sigma}_o).
 \label{9.4}
 \end{eqnarray}

\medskip

The {\it monopoles} are $q^{AB}({\vec \zeta}(\tau ))$ with the
{\it mass monopole} $q^{\tau\tau}({\vec \zeta}(\tau )) = {\cal
M}$ and the vanishing (due to the rest frame condition) {\it
momentum monopole} $q^{\tau r}({\vec \zeta}(\tau )) = {\cal P}^r
\approx 0$. Then there are the {\it stress tensor monopole}
$q^{rs}({\vec \zeta}(\tau ))$ and the trace $q^A{}_A({\vec
\zeta}(\tau ))$.

\medskip

The {\it dipoles} are $q^{rAB}({\vec \zeta}(\tau ))$. If we ask
for the vanishing of the {\it mass dipole}
\[
q^{r\tau\tau}({\vec
\zeta}(\tau )) = \int d^3\sigma_o\, \left[ \Sigma^r(\tau ,{\vec
\sigma}_o) - \zeta^r(\tau )\right]\, \Delta (\tau ,{\vec
\sigma}_o),
\]
we find

\begin{equation}
 q^{r\tau\tau}({\vec \zeta}(\tau )) = 0\quad \Rightarrow
 {\vec \zeta}(\tau ) = {\vec {\cal R}} \approx {\vec {\cal Q}}
 \approx 0.
 \label{9.5}
 \end{equation}

\noindent This means that the vanishing of the mass dipole
identifies the world-line of the internal M$\o$ller center of
energy, namely the centroid $x^{\mu}(\tau )$, which, in the
rest-frame instant form in the gauge ${\vec {\cal Q}} \approx 0$,
is also both Tulczyjew and Pirani centroid, as shown in
Ref.\cite{17} in the case of particles.

\medskip

Therefore the multipoles with respect to the center of energy are

\begin{eqnarray}
 &&q^{r_1 .. r_n AB}(\tau ) =\nonumber\\
&&\nonumber\\
 &=& \int d^3\sigma_o\, \left( \Sigma^{r_1}(\tau ,{\vec \sigma}_o)
 - {\cal R}^{r_1}(\tau )\right) ... \left( \Sigma^{r_1}(\tau ,{\vec \sigma}_o)
 - {\cal R}^{r_1}(\tau )\right)\, \det\, \left( {{\partial
 \Sigma}\over {\partial \sigma_o}}\right)\, \widehat{T}^{AB}(\tau ,{\vec
 \sigma}_o) =\nonumber \\
 &&\nonumber \\
 &=& \sum_{\underline{m}{}_1,..,\underline{m}{}_n}\,
 r^{r_1}_{\underline{m}{}_1}(\tau ) ..  r^{r_n}_{\underline{m}{}_n}(\tau)
\int d^3\sigma_o\, \left( \Gamma^{\Sigma}_{\underline{m}{}_1}(\tau ,{\vec \sigma}_o)
  - h_{\underline{m}{}_1}(\tau ) \right) ...  \left(
  \Gamma^{\Sigma}_{\underline{m}{}_n}(\tau ,{\vec \sigma}_o)
  - h_{\underline{m}{}_n}(\tau ) \right)\times\nonumber\\
&&\nonumber\\
&\times&
\det\, \left( \sum_{\underline{n}}\, {{\partial\,
 \Gamma^{\Sigma}_{\underline{n}}({\vec \sigma}_o)}\over {\partial
 \sigma_o^s}}\, r^r_{\underline{n}}(\tau ) \right)\,
  \widehat{T}^{AB}(\tau ,{\vec \sigma}_o).
  \label{9.6}
  \end{eqnarray}

\noindent The last lines give the expression of Dixon's multipoles
in terms of the relative variables, once the energy momentum
tensor (\ref{9.2}) is rewritten in terms of them. Then we could
re-express the multipoles in terms of the  orientational and shape
variables.

\medskip

Then the {\it momentum dipole} takes the following form

\begin{eqnarray}
 q^{r\, s\tau}(\tau ) &=& \int d^3\sigma_o\, \Big( \Sigma^{r}(\tau ,{\vec \sigma}_o)
 - {\cal R}^{r}(\tau )\Big)\, K^s(\tau ,{\vec \sigma}_o)
 =\nonumber \\
&&\nonumber\\
 &=& \int d^3\sigma_o\, \sum_{\underline{m}}\,
 r^r_{\underline{m}}(\tau )\,  \Big( \Gamma^{\Sigma}_{\underline{m}}(\tau ,{\vec \sigma}_o)
  - h_{\underline{m}}(\tau ) \Big)\, K^s(\tau ,{\vec \sigma}_o)
  \approx\nonumber \\
&&\nonumber\\
  &\approx& \sum_{\underline{n}}\, r^r_{\underline{n}}(\tau )\,
  p^s_{\underline{n}}(\tau ),
  \label{9.7}
  \end{eqnarray}

 \noindent and the angular momentum with respect to the internal center
of energy, coinciding with the internal spin ${\vec {\cal S}}$, is

\begin{eqnarray}
 {\cal S}^u &=& \frac{1}{2}\epsilon^{urs}\Big[ q^{r\,
 s\tau}(\tau ) - q^{s\, r\tau}(\tau )\Big] \approx\nonumber \\
&&\nonumber\\
 &\approx& \frac{1}{2}\epsilon^{urs}
\sum_{\underline{n}}\, \Big( r^r_{\underline{n}}(\tau
 )\, p^s_{\underline{n}}(\tau ) -  r^s_{\underline{n}}(\tau
 )\, p^r_{\underline{n}}(\tau )\Big).
 \label{9.8}
 \end{eqnarray}

\medskip

As shown in Ref.\cite{17} the {\it quadrupoles}  $q^{r_1r_2\,
AB}(\tau )$ allow to introduce two definitions of {\it barycentric
tensor of inertia}:

i) Dixon's one using the mass quadrupole, $I^{r_1r_2}_{dixon}(\tau
) = \delta^{r_1r_2}\, \sum_u\, q^{uu \tau\tau}(\tau ) - q^{r_1r_2
\tau\tau}(\tau )$;

ii) Thorne's one, $I^{r_1r_2}_{thorne}(\tau ) = \delta^{r_1r_2}\,
\sum_u\, q^{uu A}{}_A(\tau ) - q^{r_1r_2 A}{}_A(\tau )$;

\noindent both definitions give the standard tensor of inertia in
the non-relativistic limit, since their difference is at the
post-Newtonian order.

\medskip

In Ref.\cite{17} there is the study of the non-relativistic limit
of Dixon's multipoles by means of the Gartenhaus-Schwartz
transformation, which are then compared with the non-relativistic
multipoles defined in Appendix A of that paper.

\vfill\eject

\section{Conclusions.}

We have developed a new Lagrangian and Hamiltonian formulation of
perfect fluids in terms of generalized Eulerian coordinates $\vec
\Sigma (\tau ,{\vec \sigma}_o)$ , which not only produces less
complicated invariant masses, but also allows to define dynamical
body frames, spin frames, orientational and shape variables by a
natural extension of the techniques developed for N-body systems.
On the contrary, it is too difficult to determine these quantities
in the description based on the Lagrangian comoving coordinates.
Since the Lagrangian contains the fluid density at the initial
time, the choice of the two first axes of the dynamical body frame
can be adapted to the initial form of the fluid. Then the
dynamical body frame will evolve in time according to the changes
in the form of the fluid according to the initial data and to the
equation of state.

\medskip

We have also evaluated Dixon's multipoles in the rest-frame
instant form. While they allow a description  of the mean motion
of the extended body, the fluid in this case, the orientational
and shape variables give a complete information about the real
motion with its changes of shape, moreover adapted to all the
generic Noether constants of motion. The invariant mass of the
fluid, i.e. the Hamiltonian governing the real motion, can in turn
be expressed with a $\Delta$-multipolar expansion, as shown in
Section VIII, which may be more useful than Dixon's multipoles
when only a finite number of shape variables is relevant (the
others may be treated as perturbations).

\medskip

We hope that this description of the fluid as an extended
deformable relativistic body will help to treat with numerical
simulations and/or approximations any type of system from droplet
models of the proton, to heavy ion fireballs, plasmas and rotating
stars. Regarding the simulation of stars we still need to make the
coupling to tetrad gravity and then to go to a completely fixed
3-orthogonal gauge with the technique developed in Ref.\cite{20}
in absence of matter. This will give a new starting point for
Hamiltonian numerical gravity and for the simulation of the
properties of rotating stars. In particular in the case of
incompressible fluids it will be interesting to try to recover the
ellipsoidal equilibrium configurations in the non-relativistic
limit. In the meantime in Ref.\cite{22} it will be shown that the
non-relativistic limit of the formalism of this paper allows to
describe such configuration after the addition of the Newton
gravitational potential.

\vfill\eject

\appendix

\section{Definitions and relations on the space-like hyper-surface}

The {\em vierbeins} $z^{\mu}_{\check A}(\tau ,\vec \sigma ) =
{{\partial z^{\mu}(\tau ,\vec \sigma )}\over {\partial
\sigma^{\check A}}}$ satisfy the {\em Gauss-Codazzi-Weingarten}
integrability relation

\begin{equation}
\frac{\partial}{\partial\sigma^{\check{B}}}z^\mu_{\check{A}}(\tau,\vec{\sigma})-
\frac{\partial}{\partial\sigma^{\check{A}}}z^\mu_{\check{B}}(\tau,\vec{\sigma})=0.
 \label{a.1}
 \end{equation}

\medskip

By construction the three 4-vectors
$z^\mu_{\check{r}}(\tau,\vec{\sigma})$ ($\check{r}=1,2,3$) define
a {\em non-orthonormal} basis for the tangent space to
hyper-surface $\Sigma(\tau)$; then it is possible to define the
induced metric on the hyper-surface

\begin{equation}
g_{\check{r}\check{s}}(\tau,\vec{\sigma})=
z^\mu_{\check{r}}(\tau,\vec{\sigma})\eta_{\mu\nu}z^\nu_{\check{s}}(\tau,\vec{\sigma}),
 \label{a.2}
 \end{equation}

\noindent and  the {\em normal unit vector}

\begin{equation}
l^\mu(\tau,\vec{\sigma})=\frac{1}{\sqrt{\gamma(\tau,\vec{\sigma})}}
\epsilon^{\mu\alpha\beta\gamma} z_{1\alpha}(\tau,\vec{\sigma})
z_{2\beta}(\tau,\vec{\sigma}) z_{3\gamma}(\tau,\vec{\sigma}),
 \label{a.3}
 \end{equation}

\noindent where

\begin{equation}
\gamma(\tau,\vec{\sigma})=-\det(g_{\check{r}\check{s}}(\tau,\vec{\sigma})),
 \label{a.4}
 \end{equation}

\noindent so that

\begin{equation}
l^\mu(\tau,\vec{\sigma})l_\mu(\tau,\vec{\sigma})=1.
 \label{a.5}
 \end{equation}

\medskip

Equally if we define

\begin{equation}
g(\tau,\vec{\sigma})=-\det(g_{\check{A}\check{B}}(\tau,\vec{\sigma})),
 \label{a.6}
 \end{equation}

\noindent we have

\begin{equation}
l_\mu(\tau,\vec{\sigma})z^\mu_\tau(\tau,\vec{\sigma})=
\sqrt{\frac{g(\tau,\vec{\sigma})}{\gamma(\tau,\vec{\sigma})}}.
 \label{a.7}
 \end{equation}

\hfill

We can define the inverse vierbeins
$z^{\check{A}}_\mu(\tau,\vec{\sigma})$ such that

\begin{eqnarray}
 &&z^{\check{A}}_\mu(\tau,\vec{\sigma})z^\mu_{\check{B}}(\tau,\vec{\sigma})
=\delta^{\check{A}}_{\check{B}},
\nonumber \\
 &&z^\mu_{\check{A}}(\tau,\vec{\sigma})z^{\check{A}}_\nu(\tau,\vec{\sigma})=\eta^\mu_\nu.
 \label{a.8}
 \end{eqnarray}

Then the inverse metric
$g^{\check{A}\check{B}}(\tau,\vec{\sigma})$, such that

\begin{equation}
g^{\check{A}\check{B}}(\tau,\vec{\sigma})g_{\check{B}\check{C}}(\tau,
\vec{\sigma})=\delta^{\check{A}}_{\check{C}},
 \label{a.9}
 \end{equation}

\noindent is defined by

\begin{equation}
g^{\check{A}\check{B}}(\tau,\vec{\sigma})=
z^{\check{A}}_\mu(\tau,\vec{\sigma})\eta^{\mu\nu}z^{\check{B}}_\nu(\tau,\vec{\sigma}).
 \label{a.10}
 \end{equation}

We have also

\begin{equation}
\eta^{\mu\nu}=
z^\mu_{\check{A}}(\tau,\vec{\sigma})g^{\check{A}\check{B}}(\tau,\vec{\sigma})
z^\nu_{\check{B}}(\tau,\vec{\sigma}).
 \label{a.11}
 \end{equation}

\medskip

It is useful to  consider the inverse 3-dimensional metric
$\gamma^{\check{r}\check{s}}(\tau,\vec{\sigma})$ such that

\begin{equation}
\gamma^{\check{r}\check{s}}(\tau,\vec{\sigma})g_{\check{r}\check{s}}(\tau,
\vec{\sigma})=\delta^{\check{r}}_{\check{s}}.
 \label{a.12}
 \end{equation}

The following relation holds

\begin{equation}
\eta^{\mu\nu}=l^\mu(\tau,\vec{\sigma})l^\nu(\tau,\vec{\sigma})+
\gamma^{\check{r}\check{s}}(\tau,\vec{\sigma})
z^\mu_{\check{r}}(\tau,\vec{\sigma})z^\nu_{\check{s}}(\tau,\vec{\sigma}).
 \label{a.13}
\end{equation}

From this expression we can find

\begin{equation}
z^\mu_\tau(\tau,\vec{\sigma})=
\sqrt{\frac{g(\tau,\vec{\sigma})}{\gamma(\tau,\vec{\sigma})}}
l^\mu(\tau,\vec{\sigma})+
g_{\tau\check{r}}(\tau,\vec{\sigma})\gamma^{\check{r}\check{s}}(\tau,\vec{\sigma})
z^\mu_{\check{s}}(\tau,\vec{\sigma}).
 \label{a.14}
 \end{equation}

\medskip

If we define the {\em lapse} function

\begin{equation}
N(\tau,\vec{\sigma})=
\sqrt{\frac{g(\tau,\vec{\sigma})}{\gamma(\tau,\vec{\sigma})}},
 \label{a.15}
 \end{equation}

\noindent and the {\em shift} function

\begin{equation}
N^{\check{r}}(\tau,\vec{\sigma})=
g_{\tau
\check{u}}(\tau,\vec{\sigma})\gamma^{\check{u}\check{r}}(\tau,\vec{\sigma}),
 \label{a.16}
 \end{equation}

\noindent we can write the following expression for the inverse
metric

\begin{eqnarray}
g^{\tau\tau}(\tau,\vec{\sigma})&=&
\frac{1}{N^2(\tau,\vec{\sigma})},\nonumber\\
 &&\nonumber\\
g^{\check{r}\tau}(\tau,\vec{\sigma})&=&
-\frac{1}{N^2(\tau,\vec{\sigma})}
N^{\check{r}}(\tau,\vec{\sigma}),\nonumber\\
 &&\nonumber\\
g^{\check{r}\check{s}}(\tau,\vec{\sigma})&=&
\gamma^{\check{r}\check{s}}(\tau,\vec{\sigma})+
\frac{1}{N(\tau,\vec{\sigma})}
N^{\check{r}}(\tau,\vec{\sigma})N^{\check{s}}(\tau,\vec{\sigma}).
 \label{a.17}
 \end{eqnarray}

\noindent and we have

\begin{equation}
z^\mu_\tau(\tau,\vec{\sigma})= N(\tau,\vec{\sigma})
l^\mu(\tau,\vec{\sigma})+ N^{\check{s}}(\tau,\vec{\sigma})
z^\mu_{\check{s}}(\tau,\vec{\sigma}).
 \label{a.18}
 \end{equation}

Moreover from the definition

\begin{equation}
g_{\tau\tau}(\tau,\vec{\sigma})=z^\mu_\tau(\tau,\vec{\sigma})
\eta_{\mu\nu}z^\nu_\tau(\tau,\vec{\sigma}),
 \label{a.19}
 \end{equation}

\noindent we have

\begin{equation}
g_{\tau\tau}(\tau,\vec{\sigma})
=N^2(\tau,\vec{\sigma})+N^{\check{r}}(\tau,\vec{\sigma}) g_{\tau
\check{r}}(\tau,\vec{\sigma}).
 \label{a.20}
\end{equation}

\vfill\eject

\section{Perfect Fluids Admitting a Closed Form of the Invariant
Mass.}

Let us rewrite Eq.(\ref{4.8}) in the following form

\begin{eqnarray}
 &&\left( {{\partial \rho(X, s)}\over {\partial
 X}} \right)^2(\tau ,{\vec \sigma}_o)\, \left[ X^2 - B^2\right] (\tau ,{\vec \sigma}_o)  =
A(\tau ,{\vec \sigma}_o)\, X^2(\tau ,{\vec \sigma}_o),\nonumber \\
 &&{}\nonumber \\
 &&B(\tau ,{\vec \sigma}_o) = {{n_o({\vec \sigma}_o)}\over {
\det\left( {{\partial \Sigma}\over {\partial \sigma_o}} \right)\,
 \sqrt{\gamma (\Sigma )} }},\nonumber \\
&&\nonumber\\
 &&A(\tau ,{\vec \sigma}_o) =
{{\gamma^{\check{r}\check{s}}(\Sigma)K_{\check{r}}(\tau,\vec{\sigma}_o)
K_{\check{s}}(\tau,\vec{\sigma}_o)}\over {n_o^2({\vec
\sigma}_o)}}.
 \label{b1}
 \end{eqnarray}

\medskip

For each equation of state $\rho = \rho (\hat n,{\hat s}_o)$ with
$\hat n = X$ \footnote{In Ref.\cite{6} $X = \sqrt{\gamma}\, n$ was
used as unknown.} the solution of this equation allows to get the
explicit phase space form of the constraint ${\cal H}_{\perp}(\tau
, \vec \sigma ) \approx 0$ in Eqs.(\ref{4.9}). Referring to
Section 5 of Ref.\cite{6} for the determination of the equations
of state $\rho = \rho (\hat n,{\hat s}_o)$, in this Section we
will show the few cases in which the solution for $X(\tau ,{\vec
\sigma}_o)$ can be obtained in closed form.

1) As shown in Ref.\cite{6}, for the {\it dust} we have $p = 0$
and $\rho (\hat n) = \mu\, \hat n$ with $\mu = const.$ By using
Eqs. (\ref{3.10}) and (\ref{3.11}), the action (\ref{3.15})
becomes

\begin{eqnarray}
 S &=& - \mu\, \int d\tau d^3\sigma_o\, n_o({\vec \sigma}_o)\,
 {\cal R}(\tau ,{\vec \sigma}_o),\nonumber \\
 &&{}\nonumber \\
 {\cal R}(\tau ,{\vec \sigma}_o) &=& \sqrt{
 \Big(g_{\tau\tau}(\tau,\vec{\Sigma}) +
2\, g_{\tau \check{r}}(\tau,\vec{\Sigma})\,
\frac{\partial\Sigma^{\check{r}}}{\partial\tau} +
g_{\check{r}\check{s}}(\tau,\vec{\Sigma}) \,
\frac{\partial\Sigma^{\check{r}}}{\partial\tau} \,
\frac{\partial\Sigma^{\check{s}}}{\partial\tau}\Big)(\tau ,{\vec
\sigma}_o)}.
 \label{b2}
 \end{eqnarray}

Let us remark that with the positions $n_o({\vec \sigma}_o) =
\sum_{i=1}^N\, \delta^3({\vec \sigma}_o - {\vec \eta}_i(0))$ and
$\vec \Sigma (\tau , {\vec \eta}_i(0)) = {\vec \eta}_i(\tau )$ [so
that consistently $\vec \Sigma (0, {\vec \eta}_i(0)) = {\vec
\eta}_i(0)$] we get the action of N free particles of equal mass
$\mu$:

\begin{eqnarray}
&&S = - \mu\, \sum_{i=1}^N\, \int d\tau\, \sqrt{g_{\tau\tau}(\tau
,{\vec \eta}_i(\tau )) + 2\, g_{\tau r}(\tau ,{\vec \eta}_i(\tau
))\, {\dot \eta}^r_i(\tau ) + g_{rs}(\tau ,{\vec \eta}_i(\tau ))\,
{\dot \eta}^r_i(\tau )\, {\dot \eta}^s_i(\tau )}.\nonumber \\
 &&{}
 \label{b3}
 \end{eqnarray}

Since we have $\rho (X) - {{\partial \rho (X)}\over {\partial
X}}\, X = 0$, Eq.(\ref{4.8}) has the solution

\begin{equation}
 X(\tau ,{\vec \sigma}_o) =
 {{ \mu\, n^2_o({\vec \sigma}_o) }\over { \sqrt{\gamma
(\tau , {\vec \sigma}_o)}\, \det\left( {{\partial \Sigma}\over
{\partial \sigma_o}} \right)\,  \sqrt{[\mu\, n_o({\vec \sigma}_o)]^2
- \gamma^{rs}(\tau ,\vec \Sigma
 (\tau , {\vec \sigma}_o))\, K_r(\tau , {\vec \sigma}_o)\, K_s(\tau ,
  {\vec \sigma}_o)} }}
 \label{b4}
 \end{equation}

\noindent and the second of Eqs.(\ref{4.9}) becomes

\begin{eqnarray}
 &&{\cal H}_{\perp}(\tau , \vec \sigma ) = \rho_{\mu}(\tau ,\vec
 \sigma )\, l^{\mu}(\tau ,\vec \sigma) +\nonumber \\
&&\nonumber\\
&-&
\int d^3\sigma_o\,
 \delta^3(\vec \sigma - \vec \Sigma (\tau , {\vec
 \sigma}_o))
\sqrt{[\mu\, n_o({\vec \sigma}_o)]^2 - \gamma^{rs}(\tau ,\vec \Sigma
 (\tau , {\vec \sigma}_o))\, K_r(\tau , {\vec \sigma}_o)\, K_s(\tau ,
  {\vec \sigma}_o)}.
 \label{b5}
 \end{eqnarray}

On space-like hyper-planes  and on Wigner hyper-planes the
invariant mass and the internal boost of Eqs.(\ref{4.27}),
(\ref{5.21}), (\ref{5.23}) become

\begin{eqnarray}
 {\cal M} &=& \int d^3\sigma_o\, \sqrt{[\mu\, n_o({\vec \sigma}_o)]^2 +
 {\vec K}^2(\tau , {\vec \sigma}_o)},\nonumber \\
 &&{}\nonumber \\
 {\vec {\cal K}} &=& -\int d^3\sigma_o\, \vec \Sigma (\tau , {\vec
 \sigma}_o)\,  \sqrt{[\mu\, n_o({\vec \sigma}_o)]^2 +
 {\vec K}^2(\tau , {\vec \sigma}_o)}.
 \label{b6}
 \end{eqnarray}

\medskip

2) Let us now consider some cases of barotropic, $p = p(\rho (\hat
n,{\hat s}_o))$, and isentropic, $p = p(\rho (\hat n))$, fluids.
Let us remember that the dominant energy condition on the
energy-momentum tensor requires $|p| \leq \rho$.

2a) $p = k\, \rho (\hat n)$ ($k \not= -1$), whose equation of
state is $\rho (\hat n) = \mu\, {\hat n}^{k+1}$. With $X = \hat n$
we get ${{\partial \rho}\over {\partial X}} = (k+1)\, \mu\, X^k$
and Eq.(\ref{b1}) becomes

\begin{equation}
 [(k+1)\, \mu]^2\, X^{2(k-1)}\, (X^2 - B^2) = A.
 \label{b7}
 \end{equation}

This equation can be solved in various cases, in particular for
$k=1$ and $k={1\over 3}$
\medskip

A) $k=1$, $p = \rho$, $\rho = \mu\, {\hat n}^2$, we have:

\beq
 X = \sqrt{B^2 + {{A}\over {4 \mu^2}}},
 \label{b8}
 \eeq

\noindent and then

\begin{eqnarray}
 {\cal H}_{\perp}(\tau , \vec \sigma ) &=&\rho_{\mu}(\tau ,\vec
 \sigma )\, l^{\mu}(\tau ,\vec \sigma) -\int d^3\sigma_o\,
 \delta^3(\vec \sigma - \vec \Sigma (\tau , {\vec \sigma}_o))
\left[
\frac{2n_o(\vec{\sigma}_o)\,\mu}{\sqrt{\gamma(\Sigma)}
\det\left(\frac{\partial\Sigma}{\partial\sigma_o}\right)}\right.+\nonumber\\
\nonumber\\
&+&\left.
\sqrt{\gamma(\Sigma)}\det\left(\frac{\partial\Sigma}{\partial\sigma_o}\right)\mu\left(
\frac{\gamma^{rs}(\Sigma)K_r(\tau,\vec{\sigma}_o)K_s(\tau,\vec{\sigma}_o)}{4\mu^2n_o(\vec{\sigma}_o)}
+\frac{n_o^2(\vec{\sigma}_o)\,\mu}{\sqrt{\gamma(\Sigma)}
{\det}^2\left(\frac{\partial\Sigma}{\partial\sigma_o}\right)}\right)\right]
\approx 0,
\nonumber \\
&&\nonumber\\
&&\nonumber \\
 {\cal M} &=&\int d^3\sigma_o\,
\left[
\frac{2n_o(\vec{\sigma}_o)\,\mu}{
\det\left(\frac{\partial\Sigma}{\partial\sigma_o}\right)}
+\det\left(\frac{\partial\Sigma}{\partial\sigma_o}\right)\mu\left(
\frac{\vec{K}^2(\tau,\vec{\sigma}_o)}{4\mu^2n_o(\vec{\sigma}_o)}
+\frac{n_o^2(\vec{\sigma}_o)\,\mu}{
{\det}^2\left(\frac{\partial\Sigma}{\partial\sigma_o}\right)}\right)\right]
 \label{b9}
 \end{eqnarray}

\medskip

B) {\it Photon gas}, $k= {1\over 3}$, $p = {1\over 3} \rho$, $\rho
= \mu\, {\hat n}^{4/3}$. Eq.(\ref{b1}) is the following cubic
equation in $Y = X^2$

\beq
 \left(\frac{4\mu}{3}\right)^6\, (Y - B^2)^3 - A^3\,Y^2 = 0.
 \label{b10}
 \eeq

If we define

\beq
Z=Y-\frac{1}{3}\left(3B^2+\left(\frac{3}{4\mu}\right)^6\,A^3\right),
 \label{b11}
 \eeq

\noindent to find the solution of the previous equation is
equivalent to solve the cubic equation

\beq
 Z^3+C_1\,Z-C_o\,=0,
 \label{b12}
 \eeq

\noindent where

\begin{eqnarray}
C_1&=&-2\left(\frac{3}{4\mu}\right)^6A^3B^2-\frac{1}{3}\left(\frac{3}{4\mu}\right)^{12}A^6
\nonumber \\
 &&{}\nonumber \\
C_o&=&\frac{2}{3}\left(\frac{3}{4\mu}\right)^{12}A^6B^2+\frac{2}{27}
\left(\frac{3}{4\mu}\right)^{18}A^9+\left(\frac{3}{4\mu}\right)^6A^3B^4,
 \label{b13}
 \end{eqnarray}

Using the {\em Cardano solution} we obtain:

\begin{eqnarray}
D&=&\frac{C_o^2}{4}+\frac{C_1^3}{27}= \frac{1}{4}
\left(\frac{3}{4\mu}\right)^{12}A^6B^8+\frac{1}{27}
\left(\frac{3}{4\mu}\right)^{18}A^9B^6, \nonumber \\
 &&{}\nonumber \\
Z&=&\left(\frac{C_o}{2}+\sqrt{D}\right)^{\frac{1}{3}}+
  \left(\frac{C_o}{2}-\sqrt{D}\right)^{\frac{1}{3}},
 \label{b14}
 \end{eqnarray}

\noindent and then:

\begin{eqnarray}
Y=X^2=+\frac{1}{3}\left(3B^2+\left(\frac{3}{4\mu}\right)^6\,A^3\right)+
\left(\frac{C_o}{2}+\sqrt{D}\right)^{\frac{1}{3}}+
  \left(\frac{C_o}{2}-\sqrt{D}\right)^{\frac{1}{3}}.
 \label{b15}
 \end{eqnarray}

\medskip

We can obtain the explicit expression of the constraint ${\cal
H}_\perp(\tau,\vec{\sigma})\approx 0$ and of the invariant mass
$\cal M$ using the solution (\ref{b15}) in the eqs.(\ref{4.9}) and
(\ref{4.15}). Since we have

\begin{eqnarray}
&&\rho=\mu\,X^{\frac{4}{3}}=\mu\,Y^{\frac{2}{3}},\;\;\;\;
\frac{1}{X}\frac{\partial \rho}{\partial
X}=\frac{4}{3}\mu\,X^{-\frac{2}{3}}=
\frac{4}{3}\mu\,Y^{-\frac{1}{3}},\nonumber\\
 &&\nonumber\\
 p(X)&=&X\frac{\partial \rho}{\partial
X}-\rho=\frac{1}{3}\mu\,X^{\frac{4}{3}}=
\frac{1}{3}\mu\,Y^{\frac{2}{3}},
 \label{b16}
 \end{eqnarray}

we obtain ($Y = X^2$)

\begin{equation}
{\cal M}=\int d^3\sigma_o\,\frac{\mu}{3}\,Y^{-\frac{1}{3}}(\tau
,{\vec \sigma}_o)\,
\det\left(\frac{\partial\Sigma}{\partial\sigma_o}\right)\left[
4B^2-Y\right](\tau ,{\vec \sigma}_o).
 \label{b17}
 \end{equation}

\medskip

2b) A variant is the equation of state $\rho (\hat n) = m\, \hat n
+ {k\over {\gamma - 1}}\, (m\, \hat n)^{\gamma}$ ($\gamma \not=
1$) with $p = k\, (m\, \hat n)^{\gamma} = (\gamma - 1)\, (\rho -
m\, \hat n)$. It is called a polytropic fluid ($\gamma = 1 +
{1\over n}$) by some authors and we can have $k=k({\hat s}_o)$ in
the non-isentropic case. Since we have

\beq
 {{\partial \rho}\over {\partial X}} = m\, \Big[ 1 +
{{\gamma}\over {\gamma - 1}}\, k\, (m\, X)^{\gamma - 1}\Big],
 \label{b18}
 \eeq

\noindent Eq.(\ref{b1}) becomes

\beq
 m^2\, (X^2 -
B^2)\, \Big[ 1 + {{\gamma\, k\, m^{\gamma - 1}}\over {\gamma -
1}}\, X^{\gamma - 1} \Big]^2 = A\, X^2.
 \label{b20}
 \eeq

\medskip

A) For $\gamma = 2$ we have the fourth order equation in $X$:

\beq
 m^2\, (X^2 -
B^2)\, \Big[ 1 + 2\, k\, m\, X \Big]^2 = A\, X^2.
 \label{b21}
 \eeq

\medskip

B) For $\gamma = 3$ we have a third order equation in $Y=X^2$:

\beq
 m^2\, (Y -
B^2)\, \Big[ 1 + \frac{3\, k\, m^2}{2}\, Y \Big]^2 = A\, Y.
 \label{b22}
 \eeq

\bigskip

3) Standard polytropic perfect fluids have $p = k\,
\rho^{\gamma}(\hat n)$ ($\gamma = 1 + {1\over n} \not= 1$) and

\beq
 \rho (\hat n) = m\, \hat n\, \Big[ 1 - k\, (m\, \hat
n)^{\gamma - 1} \Big]^{- {1\over {\gamma - 1}}}.
 \label{b23}
 \eeq

 Since we have

\beq
 {{\partial \rho}\over {\partial X}} = m\, \Big[ 1 - k\, (m\, \hat
n)^{\gamma - 1} \Big]^{- {{\gamma}\over {\gamma - 1}}},
 \label{b24}
 \eeq

\noindent Eq.(\ref{b1}) becomes

\beq
 m^2\, \Big(
1 - k\, m^{\gamma -1}\, X^{\gamma -1} \Big)^{-{{2\, \gamma}\over
{\gamma -1}}}\, (X^2 - B^2) = A\, X^2.
 \label{b25}
 \eeq

 For $\gamma = 3$ ($n =
{1\over 2}$) it is a fourth order equation in $Y=X^2$:

\beq
 m^2\, \Big(
1 - k\, m^2\, Y \Big)^3\, (Y - B^2) = A\,Y.
 \label{b26}
 \eeq

\vfill\eject

\section{On the Poisson Bracket}

In Section III we  observed that the constraints of the
Hamiltonian formulation are not explicitly known as function of
the canonical variables. Nevertheless it is possible to calculate
their Poisson Bracket with a another functional on the phase
space. To see this, we define the following short notations

\begin{eqnarray}
A&=&\frac{
\gamma^{\check{r}\check{s}}(\Sigma)K_{\check{r}}(\tau,\vec{\sigma}_o)
K_{\check{s}}(\tau,\vec{\sigma}_o)}{n^2_o(\vec{\sigma}_o)},
\nonumber\\
 &&\nonumber\\
B&=&
\frac{n_o(\vec{\sigma}_o)}{\det\left(\frac{\partial\Sigma}{\partial\sigma_o}\right)
\,\sqrt{\gamma(\Sigma)}}, \nonumber\\
 &&\nonumber\\
Q&=&\frac{1}{B}\left(\rho-X\, \frac{\partial\rho}{\partial
X}\right)+ \frac{B}{X}\,\frac{\partial\rho}{\partial X},
\nonumber\\
 &&\nonumber\\
P&=&\frac{1}{B}\left(\rho-X\, \frac{\partial\rho}{\partial
X}\right), \nonumber\\
 &&\nonumber\\
R&=&\frac{X}{2B}\,\left(\frac{\partial\rho}{\partial
X}\right)^{-1}.
 \label{c1}
 \end{eqnarray}

\medskip

With this notation the implicit definition of $X$, Eq.(\ref{4.9})
can be rewritten in the form

\begin{equation}
A=\left(\frac{\partial\rho}{\partial
X}\right)^2\left[-\frac{B^2}{X^2}+1\right],
 \label{c2}
 \end{equation}

\noindent and the second constraint of Eqs.(\ref{4.10}) is

\begin{equation}
{\cal H}_\perp(\tau,\vec{\sigma})=
\rho_\mu(\tau,\vec{\sigma})l^\mu(\tau,\vec{\sigma})- {\cal
F}(\tau,\vec{\sigma}),
 \label{c3}
 \end{equation}

\noindent where

\begin{equation}
{\cal F}(\tau,\vec{\sigma})=\int d^3\sigma_o\,
n_o(\vec{\sigma}_o)\,\delta^3(\vec{\sigma}-
\vec{\Sigma}(\tau,\vec{\sigma}_o))Q.
 \label{c4}
 \end{equation}

\medskip

From Eq.(\ref{c2}), using the rules on the Poisson Bracket, we can
see that

\begin{eqnarray}
\{.,A\}&=&-2B\left(\frac{\partial\rho}{\partial
X}\right)^2\frac{1}{X^2}\,\{.,B\}+\nonumber\\
 &&\nonumber\\
 &+&2\left(\frac{\partial\rho}{\partial
X}\right)\left[-\frac{\partial^2\rho}{\partial
X^2}\frac{B^2}{X^2}+\frac{\partial^2\rho}{\partial
X^2}+\frac{\partial\rho}{\partial
X}\frac{B^2}{X^3}\right]\,\{.,X\}.
 \label{c5}
 \end{eqnarray}

Equally we can see that

\begin{eqnarray}
\{.,{\cal F}(\tau,\vec{\sigma})\}&=&-\int d^3\sigma_o\,
n_o(\vec{\sigma}_o)\,\frac{\partial}{\partial\sigma^{\check{r}}}
\delta^3(\vec{\sigma}- \vec{\Sigma}(\tau,\vec{\sigma}_o))\,Q\,
\{.,\Sigma^{\check{r}}(\tau,\vec{\sigma}_o)\}+ \nonumber\\
 &&\nonumber\\
 &+& \int d^3\sigma_o\,
n_o(\vec{\sigma}_o)\,\delta^3(\vec{\sigma}-
\vec{\Sigma}(\tau,\vec{\sigma}_o))\times \nonumber\\
 &&\nonumber\\
&\times& \left[ -\frac{1}{B^2}\left(\rho-X\,
\frac{\partial\rho}{\partial X}\right)+
\frac{1}{X}\,\frac{\partial\rho}{\partial X}\right]\{.,B\}+
\nonumber\\
 &&\nonumber\\
  &+&\int d^3\sigma_o\,
n_o(\vec{\sigma}_o)\,\delta^3(\vec{\sigma}-
\vec{\Sigma}(\tau,\vec{\sigma}_o))\times \nonumber\\
 &&\nonumber\\
&\times&\left[ \frac{1}{B}\left(-X\,\frac{\partial^2\rho}{\partial
X^2}\right)+ \frac{B}{X}\,\frac{\partial^2\rho}{\partial
X^2}-B\,\frac{1}{X^2}\,\frac{\partial\rho}{\partial X}\right]
\{.,X\}.
 \label{c6}
 \end{eqnarray}

\medskip

Using the results (\ref{c5}) we can  write

\begin{eqnarray}
&&\left[ \frac{1}{B}\left(-X\,\frac{\partial^2\rho}{\partial
X^2}\right)+ \frac{B}{X}\,\frac{\partial^2\rho}{\partial
X^2}-B\,\frac{1}{X^2}\,\frac{\partial\rho}{\partial X}\right]
\{.,X\}=\nonumber\\
 &&\nonumber\\
 &=&-\frac{X}{2B}\left(
\frac{\partial \rho}{\partial X}\right)^{-1}
\left[\{.,A\}+2B\left(\frac{\partial\rho}{\partial X}\right)^2\,
\frac{1}{X^2}\{.,B\}\right].
 \label{c7}
 \end{eqnarray}

If we substitute this expression in Eq. (\ref{c6}) and using the
short notation (\ref{c1}) we obtain

\begin{eqnarray}
\{.,{\cal F}(\tau,\vec{\sigma})\}&=&-\int d^3\sigma_o\,
n_o(\vec{\sigma}_o)\,\frac{\partial}{\partial\sigma^{\check{r}}}
\delta^3(\vec{\sigma}- \vec{\Sigma}(\tau,\vec{\sigma}_o))\,Q\,
\{.,\Sigma^{\check{r}}(\tau,\vec{\sigma}_o)\}+ \nonumber\\
 &&\nonumber\\
 &+&\int d^3\sigma_o\,
\delta^3(\vec{\sigma}- \vec{\Sigma}(\tau,\vec{\sigma}_o))\,
n_o(\vec{\sigma}_o)\,\left[-\frac{P}{B} \{.,B\}-R\{.,A\}\right].
 \label{c8}
 \end{eqnarray}

Being $A$ and $B$ known functions of the canonical variables, the
previous equation permits to calculate the Poisson bracket with
the constraint ${\cal H}_\perp(\tau,\vec{\sigma})$ although the
$X$ is unknown explicitly. The rule (\ref{c8}) can be used, for
example, to verify the algebra (\ref{4.13}).

\hfill

On the hyper-planes, the previous observations are again valid; in
this case we have

\begin{equation}
 A = -\frac{ {\vec
K}^2(\tau,\vec{\sigma}_o)}{n^2_o(\vec{\sigma}_o)}, \qquad B =
\frac{n_o(\vec{\sigma}_o)}
{\det\left(\frac{\partial\Sigma}{\partial\sigma_o}\right)}.
 \label{c9}
 \end{equation}

From Eq.(\ref{4.27}) the invariant mass and the canonical
generator of the internal boost are defined in term of the density

\begin{eqnarray}
\Delta(\tau,\vec{\sigma_o})&=&\left[
\det\left(\frac{\partial\Sigma}{\partial\sigma_o}\right) \,\left(
\rho(X,s)-\frac{\partial\rho}{\partial X}\,X \right)+
\frac{n^2_o(\vec{\sigma}_o)}{
\det\left(\frac{\partial\Sigma}{\partial\sigma_o}\right)}
\,\frac{1}{X}\,\frac{\partial\rho}{\partial X}\, \right](\tau
,{\vec \sigma}_o)=\nonumber\\
 &&\nonumber\\
 &=&n_o(\vec{\sigma}_o)\,Q(\tau ,{\vec \sigma}_o),
  \label{c10}
 \end{eqnarray}

\noindent such that

\begin{eqnarray}
{\cal M}&=&\int
d^3\sigma_o\,\Delta(\tau,\vec{\sigma_o}),\nonumber\\
 &&\nonumber\\
\vec{\cal K}&=&\int
d^3\sigma_o\,\vec{\Sigma}(\tau,\vec{\sigma_o})\,
\Delta(\tau,\vec{\sigma_o}).
 \label{c11}
 \end{eqnarray}

On them the previous rule becomes

\begin{equation}
\{.,\Delta(\tau,\vec{\sigma_o})\}=n_o(\vec{\sigma}_o)\,\left[-\frac{P}{B}
\{.,B\}-R\{.,A\}\right].
 \label{c12}
 \end{equation}

This rule allows us to calculate  the Poisson brackets of a
functional on the phase space with the constraint also in the case
of hyper-planes or Wigner hyper-planes. This is useful for example
for getting the equations of motion from  the Hamilton-Dirac
equations.

\vfill\eject

\section{Garthenaus-Schwartz transformations}

Using the notations of Section VI, let ${\cal G}=\vec{\cal
P}\cdot\vec{\cal Q}$ be the canonical generator of a
transformation, called {\em Garthenaus-Schwartz canonical
transformations} \cite{29}. If $F$ is an arbitrary functions on
the phase space, we have that its infinitesimal transformation is

\begin{equation}
\delta F=\delta\alpha\cdot\{F,\vec{\cal P}\cdot\vec{\cal Q}\}.
 \label{d1}
 \end{equation}

For finite values of the parameter $\alpha$  the transformation
can be written as

\begin{equation}
F(\alpha)=F+\int^\alpha_0 d\overline{\alpha}\:\{
F(\overline{\alpha}),\vec{\cal P}(\overline{\alpha})\cdot
\vec{\cal Q}(\overline{\alpha})\}.
 \label{d2}
 \end{equation}

We are interested to the singular limit $\alpha\rightarrow\infty$
and we use the notation

\begin{equation}
F'=\lim_{\alpha\rightarrow\infty}F(\alpha).
 \label{d3}
 \end{equation}

\medskip

Deriving both sides of (\ref{d2}) we can see that to realize the
canonical transformation is equivalent to solve the differential
equation

\begin{equation}
\left\{
\begin{array}{ccl}
\frac{dF}{d\alpha}(\alpha)&=&\{F(\alpha),\vec{\cal
P}(\alpha)\cdot\vec{\cal Q} (\alpha)\},\\ &&\\ F(0)&=&F.
\end{array}
\right.
 \label{d4}
 \end{equation}

It is trivial to verify that

\begin{eqnarray}
\vec{\cal P}(\alpha)&=&e^{-\alpha}\vec{\cal
P}\,\Rightarrow\,\vec{\cal P}'=0, \nonumber\\
 &&\nonumber\\
\vec{\cal Q}(\alpha)&=&e^{+\alpha}\vec{\cal
Q}\,\Rightarrow\,\vec{\cal Q}'\rightarrow\infty.
 \label{d5}
 \end{eqnarray}

\bigskip

The usefulness of the singular limits is shown by the following
observation \cite{29}. Let $F$ be a function on the phase space
such that

\begin{equation}
\{\vec{\cal P},F\}\equiv\{\vec{\cal P}(\alpha),F(\alpha)\}=0.
 \label{d6}
 \end{equation}

\noindent For this function the singular limit

\begin{equation}
\lim_{\alpha\rightarrow\infty}F(\alpha)=F',
 \label{d7}
 \end{equation}

\noindent exists and  is well defined. Moreover, if we define

\begin{equation}
\vec{G}=\{\vec{\cal Q},F\},
 \label{d8}
 \end{equation}

\noindent the Jacoby identity implies that

\begin{equation}
\{\vec{\cal P},\vec{G}\}=0,
 \label{d9}
\end{equation}

\noindent and the limit

\begin{equation}
\lim_{\alpha\rightarrow\infty}\vec{G}(\alpha)=\vec{G}',
 \label{d10}
 \end{equation}

\noindent exists and  is well defined. In conclusion

\begin{equation}
\{\vec{\cal P},F'\}=\lim_{\alpha\rightarrow\infty} \{\vec{\cal
P},F(\alpha)\}=\lim_{\alpha\rightarrow\infty}
e^{+\alpha}\{\vec{\cal P}(\alpha),F(\alpha)\}=0,
 \label{d11}
 \end{equation}

\noindent because $\{\vec{\cal P}(\alpha),F(\alpha)\}\equiv 0$,
and

\begin{equation}
\{\vec{\cal Q},F'\}=\lim_{\alpha\rightarrow\infty} \{\vec{\cal
Q},F(\alpha)\}=\lim_{\alpha\rightarrow\infty}
e^{-\alpha}\{\vec{\cal Q}(\alpha),F(\alpha)\}=0,
 \label{d12}
 \end{equation}

\noindent because the singular limit of $\{\vec{\cal
Q}(\alpha),F(\alpha)\}$ is the well defined quantity $\vec{G}'$.

\medskip

This observation can be applied to the relative variables
$\Re^r(\tau,\vec{\sigma}_o),\wp^s(\tau,\vec{\sigma}_o)$. These
variables are by construction such that

\begin{equation}
\{\vec{\cal P},\Re^r(\tau,\vec{\sigma}_o)\}= \{\vec{\cal
P},\wp^s(\tau,\vec{\sigma}_o)\}=0,
 \label{d13}
 \end{equation}

\noindent and then their singular limits

\begin{eqnarray}
\lim_{\alpha\rightarrow\infty}
\Re^r(\alpha\mid\tau,\vec{\sigma}_o)&=&
\Re^{\prime\,r}(\tau,\vec{\sigma}_o), \nonumber\\
 &&\nonumber\\
\lim_{\alpha\rightarrow\infty}
\wp^s(\alpha\mid\tau,\vec{\sigma}_o)&=&
\wp^{\prime\,s}(\tau,\vec{\sigma}_o),
 \label{d14}
 \end{eqnarray}

\noindent exist and are well defined. In particular we get

\begin{eqnarray}
\{\vec{\cal P},\Re^{\prime\,r}(\tau,\vec{\sigma}_o)\}&=&
\{\vec{\cal P},\wp^{\prime\,r}(\tau,\vec{\sigma}_o)\}=0,
\nonumber\\
 &&\nonumber\\
\{\vec{\cal Q},\Re^{\prime\,r}(\tau,\vec{\sigma}_o)\}&=&
\{\vec{\cal
Q},\wp^{\prime\,r}(\tau,\vec{\sigma}_o)\}=0,\nonumber\\
 &&\nonumber\\
\{\Re^{\prime\,r}(\tau,\vec{\sigma}_o),
\wp^{\prime\,s}(\tau,\vec{\sigma}'_o)\}&=&\delta^{rs}\,
\delta^3(\vec{\sigma}_o-\vec{\sigma}'_o).
 \label{d15}
 \end{eqnarray}

\medskip

In conclusion the coordinates of the table

\begin{equation}
\begin{array}{|c|c|}
\hline
\vec{\cal Q}&\Re^{\prime\,r}(\tau,\vec{\sigma}_o)\\
\vec{\cal P}&\wp^{\prime\,s}(\tau,\vec{\sigma}_o)\\
\hline
\end{array},
 \label{d16}
 \end{equation}

\noindent are canonical coordinates. Moreover we can observe that

\begin{eqnarray}
\Re^{\prime\,r}(\tau,\vec{\sigma}_o)&=&
\Re^r(\tau,\vec{\sigma}_o)+\int_0^\infty d\alpha\,\,
\{\Re^r(\alpha\mid\tau,\vec{\sigma}_o)
,\vec{\cal P}(\alpha)\cdot\vec{\cal Q}(\alpha)\}=
\nonumber\\
 &&\nonumber\\
&=&\Re^r(\tau,\vec{\sigma}_o)
+{\cal P}^s\,\int_0^\infty d\alpha\,\,e^{-\alpha}
\{\Re^r(\alpha\mid\tau,\vec{\sigma}_o),\vec{\cal Q}^s(\alpha)\}=\nonumber\\
 &&\nonumber\\
&=&\Re^r(\tau,\vec{\sigma}_o)+{\cal P}^s\,I^{rs}_{\Re},\nonumber\\
 &&\nonumber\\
\wp^{\prime\,r}(\tau,\vec{\sigma}_o)&=&\wp^r(\tau,\vec{\sigma}_o)+\int_0^\infty d\alpha\,\,
\{\wp^r(\alpha\mid\tau,\vec{\sigma}_o)
,\vec{\cal P}(\alpha)\cdot\vec{\cal Q}(\alpha)\}=
\nonumber\\
 &&\nonumber\\
&=&\wp^r(\tau,\vec{\sigma}_o)+{\cal P}^s\,\int_0^\infty d\alpha\,\,e^{-\alpha}
\{\wp^r(\alpha\mid\tau,\vec{\sigma}_o),\vec{\cal Q}^s(\alpha)\}=\nonumber\\
 &&\nonumber\\
&=&\wp^r(\tau,\vec{\sigma}_o)+{\cal P}^s\,I^{rs}_{\wp}.
 \label{d17}
 \end{eqnarray}

Due to the previous considerations, the
$\{\Re^r(\alpha\mid\tau,\vec{\sigma}_o),\vec{\cal Q}^s(\alpha)\}$
and the $\{\wp^r(\alpha\mid\tau,\vec{\sigma}_o),\vec{\cal
Q}^s(\alpha)\}$ are well defined constants in the singular limit.
Then the integral $I_\Re,I_\wp$ in Eq.(\ref{d17})  are well
defined for the presence of $e^{-\alpha}$ factor. Then if we use
explicitly the condition $\vec{\cal P}\approx 0$ we get

\begin{equation}
\Re^{\prime\,r}(\tau,\vec{\sigma}_o)\approx
\Re^r(\tau,\vec{\sigma}_o)\;\;\;\;\;
\wp^{\prime\,r}(\tau,\vec{\sigma}_o)\approx
\wp^r(\tau,\vec{\sigma}_o).
 \label{d18}
 \end{equation}

This is the case of the gauge fixing (\ref{6.27}).

\vfill\eject

\section{Some solutions for  the kernel $\Gamma$}

In this Appendix we construct some kernels $\Gamma$ that satisfy
the conditions (\ref{6.5}),(\ref{6.8}) and (\ref{6.10}) or Eq.
(\ref{6.18}).

\medskip

The first solution is based on the possibility to read the kernel
as distributions. For example, we can define in this case

\begin{equation}
\Gamma_K(\vec{\sigma}_o-\vec{\sigma}'_o)=
\nabla^2_{\sigma_o}\,\delta^3(\vec{\sigma}_o-\vec{\sigma}'_o),
 \label{e1}
 \end{equation}

\noindent and the second of Eqs. (\ref{6.5}) is satisfied being
reduced to

\begin{equation}
\nabla^2_{\sigma_o}\, 1=0.
 \label{e2}
 \end{equation}

\medskip

Let us define the usual symmetric Green function
$c(\vec{\sigma}_o-\vec{\sigma}'_o)$

\begin{equation}
\nabla^2_{\sigma_o}\,c(\vec{\sigma}_o-\vec{\sigma}'_o)=
\delta^3(\vec{\sigma}_o-\vec{\sigma}'_o),
 \label{e3}
 \end{equation}

\noindent and let us make the following  {\em ansatz} on
$\Gamma_\Sigma$

\begin{equation}
\Gamma_\Sigma(\vec{\sigma}_o,\vec{\sigma}'_o)=
c(\vec{\sigma}_o-\vec{\sigma}'_o)+f(\vec{\sigma}'_o).
 \label{e4}
 \end{equation}

Then Eq. (\ref{6.10}) is satisfied

\begin{equation}
\nabla^2_{\sigma_o}\,\Gamma_\Sigma(\vec{\sigma}_o,\vec{\sigma}'_o)=
\nabla^2_{\sigma_o}\,c(\vec{\sigma}_o-\vec{\sigma}'_o)=
\delta^3(\vec{\sigma}_o-\vec{\sigma}'_o).
 \label{e5}
 \end{equation}

\medskip

The function $f$ is determined imposing Eq.(\ref{6.5}) and we get

\begin{equation}
f(\vec{\sigma}'_o)=-\frac{1}{\cal N}\int d^3\sigma_{o1}\,
n_o(\vec{\sigma}_{o1})c(\vec{\sigma}_{o1}-\vec{\sigma}'_o).
 \label{e6}
 \end{equation}

Automatically also Eq.(\ref{6.8}) is satisfied. In conclusion

\begin{eqnarray}
\Gamma_K(\vec{\sigma}_o,\vec{\sigma}'_o)&=&
\nabla^2_{\sigma_o}\,\delta^3(\vec{\sigma}_o-\vec{\sigma}'_o),
\nonumber\\
 &&\nonumber\\
\Gamma_\Sigma(\vec{\sigma}_o,\vec{\sigma}'_o)&=&
c(\vec{\sigma}_o-\vec{\sigma}'_o)-\frac{1}{\cal N}\int
d^3\sigma_{o1}\,
n_o(\vec{\sigma}_{o1})c(\vec{\sigma}_{o1}-\vec{\sigma}_o),
 \label{e7}
 \end{eqnarray}

\noindent is a distribution-like solution for the kernels
$\Gamma$.

\bigskip

Another class of possible solutions is obtained if we use the
representation $\Gamma^{K}_{\underline{n}}(\vec{\sigma}_o),
\Gamma^{\Sigma}_{\underline{n}}(\vec{\sigma}_o)$ given by Eq.
(\ref{6.17}). Then we can consider a second base of orthonormal
functions $\Psi_{\underline{n}}(\vec{\sigma}_o)$ such that

\begin{equation}
\int d^3\sigma_o\,n_o(\vec{\sigma}_o)\,
\Psi_{\underline{o}}(\vec{\sigma}_o)\neq 0.
 \label{e8}
 \end{equation}

\medskip

The base $n_o(\vec{\sigma}_o), \Psi_{\underline{n}\neq
\underline{o}}(\vec{\sigma}_o)$ is  a complete, non orthonormal
base of functions. Using the {\em Gram-Schmidt procedure}
\cite{34} we can  construct the orthonormal base
$\Psi'_{\underline{n}}(\vec{\sigma}_o)$ such that in particular

\begin{equation}
\Psi'_{\underline{o}}(\vec{\sigma}_o)=
\frac{n_o(\vec{\sigma}_o)}{R},
 \label{e9}
 \end{equation}

\noindent with the normalization constant

\begin{equation}
R=\int d^3\sigma_o\,n^2_o(\vec{\sigma}_o).
 \label{e10}
 \end{equation}

\medskip

The other elements of
$\Psi'_{\underline{n}\neq\underline{o}}(\vec{\sigma}_o)$ are given
by the recurrence formula of the Gram-Schmidt's algorithm. With
these definitions the first of the conditions (\ref{6.18}) is
satisfied if we choose

\begin{eqnarray}
\Gamma^{\Sigma}_{\underline{o}}(\vec{\sigma}_o)&=&0,\nonumber\\
 &&\nonumber\\
\Gamma^{\Sigma}_{\underline{n}}(\vec{\sigma}_o)&=&
\Psi'_{\underline{n}}(\vec{\sigma}_o)\;\;\;\;if
\;\;\;\;\underline{n}\neq\underline{o}.
 \label{e11}
 \end{eqnarray}

\medskip

The fourth of Eqs.(\ref{6.18}) is satisfied if the $\Gamma^{K}$'s
have the following form

\begin{eqnarray}
\Gamma^{K}_{\underline{o}}(\vec{\sigma}_o)&=&0,\nonumber\\
 &&\nonumber\\
\Gamma^{K}_{\underline{n}}(\vec{\sigma}_o)&=&
\Psi'_{\underline{n}}(\vec{\sigma}_o)-
c_{\underline{n}}\Psi'_{\underline{o}}(\vec{\sigma}_o) \;\;\;\;if
\;\;\;\;\underline{n}\neq\underline{o}.
 \label{e12}
 \end{eqnarray}

We use the second of Eqs.(\ref{6.18}) for fixing the values of the
coefficients $c_{\underline{n}}$

\begin{equation}
c_{\underline{n}}=-\frac{R}{\cal N} \int
d^3\sigma_o\,\Psi'_{\underline{n}}(\vec{\sigma}_o)
\;\;\;\;\underline{n}\neq\underline{o}.
 \label{e13}
 \end{equation}

\medskip

With this choice also the third of Eqs.(\ref{6.18}) is satisfied.
In fact we can  calculate explicitly the sum in the left-hand side
of this conditions using the completeness of the basis
$\Psi'_{\underline{n}}(\vec{\sigma}_o)$

\begin{eqnarray}
\sum_{\underline{n}}\,\Gamma^{\Sigma}_{\underline{n}}
(\vec{\sigma}_{1o})\Gamma^{K}_{\underline{n}}
(\vec{\sigma}_{2o})&=&\sum_{\underline{n}\neq\underline{o}}
\,\Psi'_{\underline{n}}(\vec{\sigma}_{1o})
\Psi'_{\underline{n}}(\vec{\sigma}_{2o})+
\sum_{\underline{n}\neq\underline{o}} c_{\underline{n}}
\Psi'_{\underline{o}}(\vec{\sigma}_{2o})
\Psi'_{\underline{n}}(\vec{\sigma}_{1o})=\nonumber\\
 &&\nonumber\\
 &=&\delta^3(\vec{\sigma}_{1o}-\vec{\sigma}_{2o})-
\Psi'_{\underline{o}}(\vec{\sigma}_{2o})
\Psi'_{\underline{o}}(\vec{\sigma}_{1o})+
\sum_{\underline{n}\neq\underline{o}} c_{\underline{n}}
\Psi'_{\underline{o}}(\vec{\sigma}_{2o})
\Psi'_{\underline{n}}(\vec{\sigma}_{1o})=\nonumber\\
 &&\nonumber\\
 &=&\delta^3(\vec{\sigma}_{1o}-\vec{\sigma}_{2o})+
\Psi'_{\underline{o}}(\vec{\sigma}_{2o})
\sum_{\underline{n}}\,c_{\underline{n}}
 \Psi'_{\underline{n}}(\vec{\sigma}_{1o}),
  \label{e14}
 \end{eqnarray}

\noindent with $c_{\underline{o}}=-1$ in the last line . Finally
we can observe that, in the sum,  the $c_ {\underline{n}}$ are the
components on the base $\Psi'_{\underline{n}}(\vec{\sigma}_o)$ of
the $(-R/{\cal N})$ constant function and then we have in accord
with Eqs.(\ref{6.18})

\begin{equation}
\Psi'_{\underline{o}}(\vec{\sigma}_{2o})
\sum_{\underline{n}}\,c_{\underline{n}}
\Psi'_{\underline{n}}(\vec{\sigma}_{1o})=
-\Psi'_{\underline{o}}(\vec{\sigma}_{2o})\frac{R}{\cal N}=
-\frac{n_o(\vec{\sigma}_{2o})}{\cal N}.
 \label{e15}
 \end{equation}

\medskip

In conclusion

\begin{eqnarray}
\Gamma_\Sigma(\vec{\sigma}_{1o},\vec{\sigma}_{2o})&=&
\sum_{\underline{n}\neq\underline{o}}
\Psi'_{\underline{n}}(\vec{\sigma}_{1o})
\Phi_{\underline{n}}(\vec{\sigma}_{2o}),\nonumber\\
 &&\nonumber\\
\Gamma_K(\vec{\sigma}_{1o},\vec{\sigma}_{2o})&=&
\sum_{\underline{n}\neq\underline{o}}
\Phi_{\underline{n}}(\vec{\sigma}_{1o})
\Psi'_{\underline{n}}(\vec{\sigma}_{2o})-
\frac{n_o(\vec{\sigma}_{2o})}{\cal N}
\sum_{\underline{n}\neq\underline{o}}\,c_{\underline{n}}\,
\Phi_{\underline{n}}(\vec{\sigma}_{1o}).
 \label{e16}
 \end{eqnarray}

\vfill\eject


\begin{thebibliography}{99}





\bibitem{1}J.L.Tassoul, {\it Theory of Rotating Stars} (Princeton
Univ. Press, Princeton, 1978).

\bibitem{2}P.A.M. Dirac, {\it Lectures on Quantum Mechanics},
Belfer Graduate School of Science (Yeshiva University, New York,
N.Y., 1964).

\bibitem{3}L.Lusanna, {\it Towards a Unified Description of the Four
Interactions in Terms of Dirac-Bergmann Observables}, invited
contribution to the book {\it Quantum Field Theory: a 20th Century
Profile} of the Indian National Science Academy, ed. A.N.Mitra
(Hindustan Book Agency, New Delhi, 2000) (HEP-TH/9907081).
Gen.Rel.Grav. {\bf 33}, 1579 (2001)(gr-qc/0101048).

\bibitem{4}S.Adler and T.Buchert, Astron.Astrophys. {\bf 343}, 317 (1999)
(ASTRO-PH/9806320).


\bibitem{5}J. D. Brown Class. Quantum Grav. {\bf 10}, 1579
(1993).

\bibitem{6}L. Lusanna and D. Nowak-Szczepaniak, Int. J. Mod. Phys. {\bf A15}, 4943
(2000).



\bibitem{7}D.Bao, J.Marsden and R.Walton, Commun.Math.Phys. {\bf 99}, 319
(1985).\hfill\break D.D.Holm, {\it Hamiltonian Techniques for
Relativistic Fluid Dynamics and Stability Theory}, in {\it
Relativistic Fluid Dynamics}, eds. A.Anile and Y.Choquet-Bruhat
(Springer, Berlin, 1989).


\bibitem{8}L. Lusanna, Int. J. Mod. Phys. {\bf A12}, 645
(1997).

\bibitem{9}D. Alba and L. Lusanna, Int. J. Mod. Phys. {\bf A13}, 2791
(1998).

\bibitem{10}H. Crater and L. Lusanna, Ann. Phys. {\bf 289}, 87 (2001)
(hep-th/0001046).

\bibitem{11} D. Alba, H. Crater and L. Lusanna, Int. J. Mod.Phys. {\bf A16}, 3365 (2001)
(hep-th/0103109).

\bibitem{12} D. Alba and L. Lusanna, Int. J. Mod. Phys. {\bf A13}, 3275
(1998).

\bibitem{13}D. Alba, L. Lusanna and M. Pauri,
 J. Math. Phys. {\bf 43}, 373 (2002) (hep-th/0011014).

\bibitem{14}D. Alba, L. Lusanna and M. Pauri,
J. Math. Phys. {\bf 43}, 1677 (2002) (hep-th/0102087).

\bibitem{15}D. Alba, L. Lusanna, M. Pauri, {\it Multipolar Expansions
for the Relativistic N-Body Problem in the Rest-Frame Instant
Form}(hep-th/0103092).

\bibitem{16}W.G. Dixon, J. Math. Phys. {\bf 8}, 1591 (1967).


\bibitem{17}W. G. Dixon, {\it Exended Bodies in General Relativity:
Their Description and Motion}, Ed. J. Ehlers (North Holland,
Amsterdam, 1979).



\bibitem{18}L.Lusanna, {\it The Rest-Frame Instant Form of Metric Gravity},
Gen.Rel.Grav. {\bf 33}, 1579 (2001)(gr-qc/0101048).


\bibitem{19}L.Lusanna and S.Russo, {\it A New Parametrization for Tetrad Gravity},
Gen.Rel.Grav. {\bf 34}, 189 (2002)(gr-qc/0102074).


\bibitem{20}R.De Pietri, L.Lusanna, L.Martucci and S.Russo, {\it Dirac's
Observables for the Rest-Frame Instant Form of Tetrad Gravity in a
Completely Fixed 3-Orthogonal Gauge}, to appear in Gen.Rel.Grav.
(gr-qc/0105084).

\bibitem{21}J.Agresti, R.DePietri, L.Lusanna and L.Martucci, {\it
Hamiltonian Linearization of the Rest-Frame Instant Form of Tetrad
Gravity in a Completely Fixed 3-Orthogonal Gauge: a Radiation
Gauge for Background-Independent Gravitational Waves in a
Post-Minkowskian Einstein Space-Time}, in preparation.


\bibitem{22}D.Alba, {\it Eulerian Coordinates for Non-Relativistic
Perfect Fluids and the Ellipsoidal Equilibrium Configurations of
Self-Gravitating Incompressible Fluids}, in preparation.

\bibitem{23} S. Chandrasekhar, {\it Ellipsoidal Figures of
Equilibrium}, Dover Publications Inc., New York (1968).


\bibitem{24}A. J. Hanson and T. Regge, Ann. Phys. {\bf 87}, 498
(1974).


\bibitem{25}P.A.M. Dirac, Rev. Mod. Phys. {\bf 21}, 392
(1949).

\bibitem{26}H.Leutwyler and J.Stern, Ann.Phys.(N.Y.) {\bf 112}, 94
(1978).


\bibitem{27}M. Pauri and G. M. Prosperi, J. Math. Phys. {\bf 16}, 1503
(1975).

\bibitem{28}C. M$\o$ller, Ann. Inst. Henri Poincar\'e, {\bf 11},251 (1949);
{\it The theory of Relativity} (Oxford University Press, Oxford,
1957).

\bibitem{29}H. Osborn, Phys. Rev. {\bf 176}, 1514 (1968).

\bibitem{30}R. G. Littlejohn and M. Reinsch, Rev. Mod. Phys. {\bf 69}, 213
(1997).

\bibitem{31}A. Lucenti, L.Lusanna and M. Pauri, J. Phys. {A31}, 1633
(1998).



\bibitem{32}H. Goldstein, {\it Classical Mechanics}, Addison-Wesley
(1959).

\bibitem{33}L. D. Landau and E. M. Lifsits, {\it Meccanica}, Editori Riuniti
(1976).

\bibitem{34}H. Hochstadt, {\it The functions of the mathematical physics},
Wiley-Interscience (1971).















\bibitem{ADM}R. Arnowitt, S. Deser, C. W. Misner, {\it The dynamics of General
relativity} in L. Witten, {\it Gravitation: an introduction to
modern research}, Ed. Wiley (1962)

\end{thebibliography}
\end{document}